\documentclass[10pt]{article}
\usepackage{geometry}
\usepackage{amssymb,amsfonts, amsmath}
\usepackage{hyperref}

\usepackage{longtable}
\usepackage{graphicx}
\usepackage[table,xcdraw]{xcolor}
\usepackage{amsthm, subcaption, multicol, enumerate, bm, textcomp, listings, mathtools, multirow, framed, soul, subcaption}
\usepackage[makeroom]{cancel}

\usepackage[utf8]{inputenx}
\usepackage{rotating}
\setstcolor{red}

\newcommand{\Tau}{\mathcal{T}}
\newcommand{\Sau}{\mathcal{S}}
\newcommand{\dx}{~\text{d}\mathbf{x}}

\newcommand{\ds}{~\text{d}s}

\newcommand{\corr}{}

\newcommand{\stkout}[1]{\ifmmode\text{\sout{\ensuremath{#1}}}\else\sout{#1}\fi}

\theoremstyle{definition}

\newtheorem{proposition}{Proposition}[section]
\newtheorem{theorem}{Theorem}[section]

\begin{document}

\begin{center}
    
{\textbf{\Large A COUPLED BULK-SURFACE MODEL FOR \\\vspace{0.15cm}CELL POLARISATION}}
\vspace{1cm}

\textsc{\large
  Davide Cusseddu 
  \footnote{\label{UOS}Dept of Mathematics, School of Mathematical and Physical Sciences, \textsc{University of Sussex}, Brighton, UK\\ \textit{E-mail contacts:} \textbf{d.cusseddu@sussex.ac.uk} (D. Cusseddu), \textbf{a.madzvamuse@sussex.ac.uk} (A. Madzvamuse)},
  Leah Edelstein-Keshet \footnote{Dept  of Mathematics, \textsc{University of British Columbia}, Vancouver, Canada},
  John A. Mackenzie \footnote{Dept of Mathematics and Statistics, \textsc{University of Strathclyde}, Glasgow, UK},\\
  St\'ephanie Portet \footnote{Dept of Mathematics, \textsc{University of Manitoba}, Winnipeg, Canada},
  Anotida Madzvamuse \textsuperscript{\ref{UOS}}
}
\vspace{1cm}
\end{center}

\begin{minipage}{0.92\textwidth}
\textbf{Abstract}: Several cellular activities, such as directed cell migration, are coordinated by an intricate network of biochemical reactions which lead to a polarised state of the cell, in which cellular symmetry is broken, causing the cell to have a well defined front and back. Recent work on balancing biological complexity with mathematical tractability resulted in the proposal and formulation of a famous minimal model for cell polarisation, known as the {\it wave pinning} model. In this study, we present a three-dimensional generalisation of this mathematical framework through the maturing theory of coupled bulk-surface semilinear partial differential equations in which protein compartmentalisation becomes natural. We show how a local perturbation over the surface can trigger propagating reactions, eventually stopped in a stable profile by the interplay with the bulk component. We describe the behavior of the model through asymptotic and local perturbation analysis, in which the role of the geometry is investigated. The bulk-surface finite element method is used to generate numerical simulations over simple and complex geometries, which confirm our analysis, showing pattern formation due to propagation and pinning dynamics. The generality of our mathematical and computational framework allows to study more complex biochemical reactions and biomechanical properties associated with cell polarisation in multi-dimensions.

\vspace{0.5cm}

\vspace{0.5cm}
\textbf{Note.} This article will be published in a forthcoming issue of the \textit{Journal of Theoretical Biology}. The article appears here in its accepted, peer-reviewed form, as it was provided by the submitting author. It has not been copyedited, proofread, or formatted by the publisher. 
\textbf{doi}: \href{https://doi.org/10.1016/j.jtbi.2018.09.008}{10.1016/j.jtbi.2018.09.008}

\vspace{0.5cm}
\textcopyright  2018. This manuscript version is made available under the CC-BY-NC-ND 4.0 license \url{https://creativecommons.org/licenses/by-nc-nd/4.0/}
\end{minipage}

\section{Introduction}
Cell polarity is a complex process by which cells lose  symmetry. However, its precise definition is still not very clear \cite{Frankel2017}. Polarity appears in single-cell organisms and multi-cell tissues. 
Many common basic polarisation mechanisms are shared and adapted by many different kinds of cells \cite{Nelson2003}. 
Roughly speaking, by breaking symmetry, cells define their front and rear and this process is characterised and driven by molecular chemical processes. Cell polarity is mediated and coordinated by a huge number of molecules and proteins and their  interactions \cite{Drubin1996, Guilluy2011}. The polarisation process, which can be caused by some external stimuli or can be spontaneous \cite{Andrew2007,Graessl2017},  is  necessary for many cellular activities, such as morphogenesis, and directed cell migration \cite{Ladoux2016,StJohnston2010}. Studies have identified the main directors of this phenomenon in the  Rho family small guanosine triphosphate (GTP)-binding proteins (Rho GTPases). They behave like molecular switches, cycling between active (GTP-bound) and inactive forms (GDP-bound). Activation and inactivation are regulated by guanine nucleotide exchange factors (GEFs) and GTPase-activating proteins (GAPs). Moreover, the inactive Rho GTPases are sequestered in the cytosol by guanine nucleotide dissociation inhibitors (GDIs), that prevent the association of Rho GTPases with the plasma membrane \cite{DerMardirossian2005, Hodge2016}. Among the Rho GTPase family, RhoA, Rac and Cdc42 are the most well known representatives in initiating the polarisation of migrating cells
 \cite{Etienne-Manneville2008,Ridley2003,Sadok2014}. During cell migration, Rac and Cdc42 tend to concentrate their activities at the front, controlling the protrusive actin network, while RhoA is mostly active at the rear and regulates large focal adhesions and stress fibres \cite{Mayor2010, Ohashi2017}. Microtubules and intermediate filaments are also involved in the process, for example binding the RhoA-effectors GEF-H1 and Solo \cite{Chang2008, Fujiwara2016}. 
 
\begin{figure}[ht]
    \centering
    \includegraphics[width=.4\linewidth]{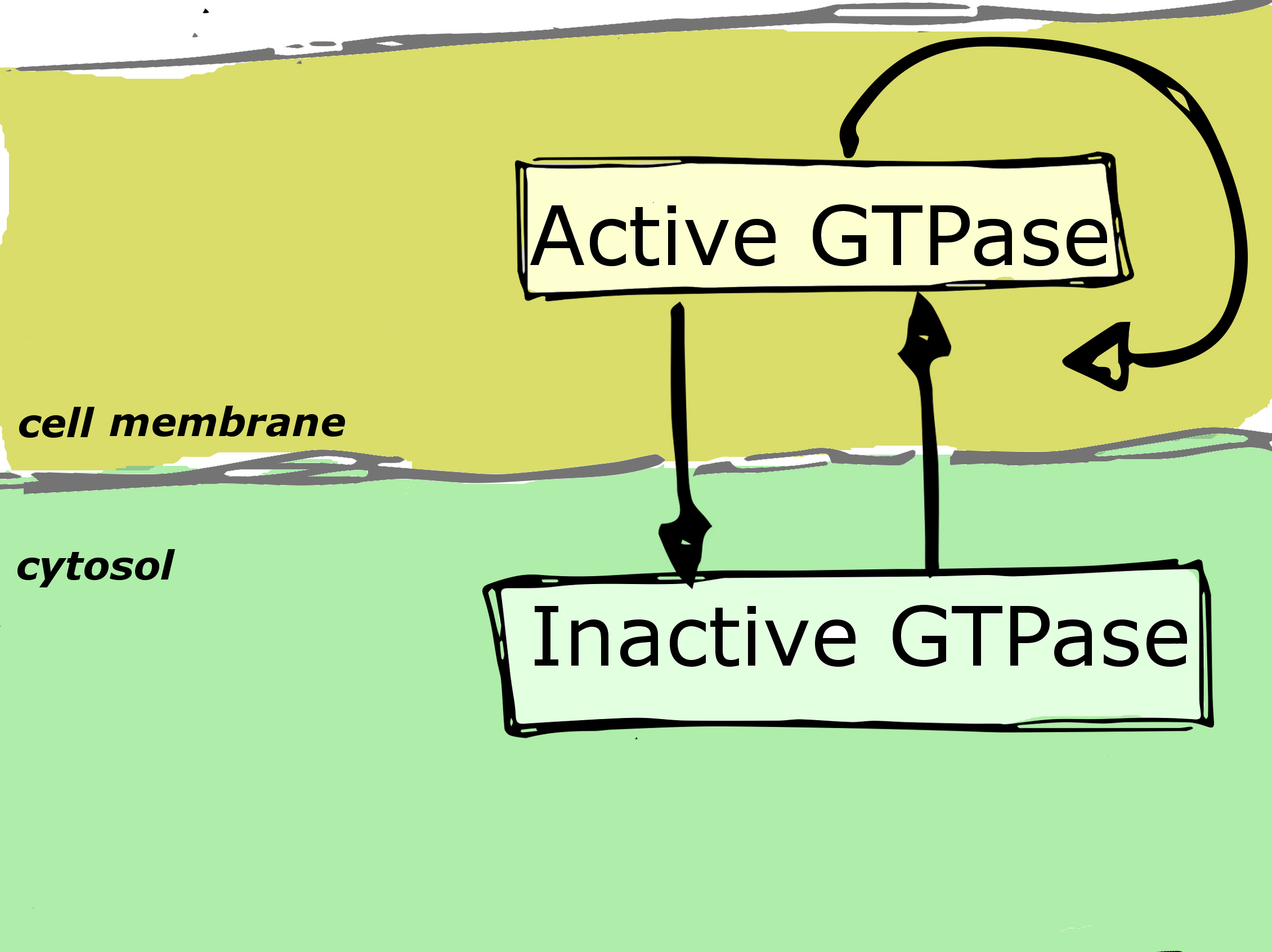}
    \caption{The minimal GTPase circuit with positive feedback for the activation (\cite{Mori2008, Altschuler2008}). Active GTPase is bounded to the membrane, while inactive GTPase moves in the cytosol.}
    \label{fig:minimal_circuit}
\end{figure}

In recent years, Rho GTPases and cell polarisation have attracted the attention of many modellers \cite{Goryachev2017, Rappel2017}.  {\corr Mar\'ee {\it et al.} \cite{Maree2006}
were able to simulate polarisation on a two-dimensional domain, in which the crosstalk between RhoA, Rac and Cdc42 in their active and inactive forms could generate the expected patterns.} However, despite the fact that good computational results were obtained, a rigorous mathematical analysis of the biochemical system comprising six partial differential equations (PDEs), remained out of reach \cite{Edelstein2013}, until two years later, when Mori {\it et al.} \cite{Mori2008} proposed a significant mathematical simplification of this modelling framework for cell polarisation, which became very popular and {\corr can be considered as the starting point of our study}. The work in \cite{Mori2008} focused on a conceptual minimal model of a single Rho GTPase and its switch between active and inactive forms, in which activation was supported by a positive feedback of the active GTPase in its own activation (see Figure \ref{fig:minimal_circuit} for a schematic representation). 
Their model consisted of the following pair of reaction-diffusion equations posed on a one dimensional domain
\begin{align}
    & \frac{\partial a}{\partial t} = D_a \frac{\partial^2 a}{\partial^2 x} + f(a,b) &x \in (0, L), ~ t>0,
    \label{eq:model_Mori_a}\\ 
    & \frac{\partial b}{\partial t} = D_b \frac{\partial^2 b}{\partial^2 x} - f(a,b) &x \in (0, L), ~ t>0,
    \label{eq:model_Mori_b}
\end{align}
with 
\begin{equation}
     f(a,b) = \Big(k_0+\frac{\gamma a^2}{K^2+a^2}\Big)b-\beta a, \label{eq:model_Mori_f(a,b)}
\end{equation}
and boundary conditions
\begin{align}
    &\frac{\partial a}{\partial x} = \frac{\partial b}{\partial x} = 0 & x = 0, L, ~ t>0,
\end{align}
where $a (x,t)$ and $b(x,t)$ denote the active and inactive forms, respectively. Here, $k_0$ represents the basal rate of activation and $\beta$ is the rate of inactivation.
The maximal rate for the positive feedback is indicated by $\gamma$ and $K$ is the parameter representing the quantity of $a$ needed to achieve a feedback-induced activation rate of $\gamma/2$ in the reaction. 

The mathematical model was based on three key properties: (1) a large difference in diffusivities between active and inactive forms ($D_a/D_b \ll 1$); (2) conservation in time of the total mass $\int_0^L (a+b) dx$; and (3) bistability in the reaction term $f(a,b)$ with respect to $a$. Bistable reaction-diffusion equations are known to produce travelling waves for certain initial conditions \cite{Fife1977}. In this work, \cite{Mori2008}, a local narrow peak of active GTPase was able to generate a travelling wave of active GTPase which is eventually stopped due to the interplay with the inactive GTPase, where conservation of total mass and fast cytosolic diffusion were key ingredients. An asymptotic analysis of the model, known as wave pinning (WP) phenomena, was later carried out in \cite{Mori2011}. 

{\corr Over the years, the need for a mathematical understanding of cell polarity led to the reduction of different mathematical models of polarisation to minimal conceptual models, revealing different underlying mechanisms, not necessarily based on wave pinning. 
Some of them, however, share common features, for example positive feedback is still the key to achieve cell polarity in the work by Altschuler \emph{et al.} \cite{Altschuler2008}, in which one ordinary and partial differential equation (ODE-PDE) system and one stochastic model are proposed for the interactions between an active and inactive GTPase component. Reaction-diffusion systems have also been used by Otsuji \emph{et al.} \cite{Otsuji2007}. They derive conceptual models of two components based on mass conservation and difference in diffusivity, which they show to be fundamental properties to achieve polarisation. 
In addition, Goryachev and Pokhilko \cite{Goryachev2008} proposed a reaction-diffusion model for Cdc42 clustering in budding yeast, which was based on the Turing pattern formation mechanism.
In \cite{Edelstein2013,Jilkine2011} some of these models are described and compared.
}

{\corr An important biological aspect of cell polarisation is the compartmentalisation of the membrane-bound and cytosolic proteins, which has inspired several works:}
Novak \emph{et al.} presented a computational approach for three-dimensional modelling of Rac proteins cycling between cell membrane and cytosol, using reaction diffusion equations \cite{Novak2007}. 
In a more recent paper, \cite{Xu2018}, a one-dimensional model for Cdc42 and its GEFs in budding yeast is proposed. The cytosolic components purely diffuse over the line domain while slow membrane diffusion motivates the use of ordinary differential equations (ODEs) to model the membrane-bound species at the two ends. Interactions between the two occur through the flux conditions of the cytosolic components and the ODE reactions. 
A three-dimensional bulk-surface model showing Turing pattern formation is proposed in \cite{Ratz2014}. The GDI-bound inactive GTPase diffuses freely in the cell interior (the bulk) and, through an appropriate coupling boundary conditions, it binds to the cell membrane (surface of the domain), on which, its membrane-bound counterpart interacts with the active form. Both species were modelled by reaction-diffusion equations.
Another three-dimensional bulk-surface model is also proposed in \cite{Spill2016}. This model is more detailed as all the three GTPases Cdc42, Rac, RhoA (in the cytosolic, membrane-bound active and membrane-bound inactive forms) and phosphatidylinositols (PIPs) are taken into account. The model results in a system of twelve reaction-diffusion equations. 

The wave pinning model has seen its bulk-surface extension in two works \cite{Giese2015,Ramirez2015} and very recently in \cite{Diegmiller2018}. The first one by Ramirez \emph{et al.} \cite{Ramirez2015} adapts the WP model to GTPases in dendritic spines in neurons. The cytosolic GTPase is assumed spatially homogeneous, while the membrane-bound active form is subject to a surface reaction-diffusion equation. The interesting result is that the pinning mechanism can be induced only by the geometry of the domain: the smaller the neck of the spine, the easier is the confinement of the active GTPase. Confinement is also facilitated by higher diffusion, which however is in contrast with other models for cell polarisation based on slow membrane diffusivity. 
The second work, by Giese \emph{et al.} \cite{Giese2015}, presents a natural extension of the wave pinning model in the bulk-surface setting (see the following equations \eqref{eq:model_b}-\eqref{eq:model_a}), where the molecular interactions between the bulk and  surface chemical components are mediated through an appropriate coupling boundary condition on the surface. In their work they investigate the role of shape, internal organelles and inhomogeneities in polarisation processes.  
Diegmiller \textit{et al.} \cite{Diegmiller2018} have recently presented a three-dimensional analysis of the steady state of the wave pinning model in the bulk-surface setting on a sphere. They were able to show pattern formation {\corr in} the surface component, after having shown analytically that spatial variation of the bulk component is negligible.  

Inspired by these previous works, we study the extension of the wave pinning model in more general three-dimensional stationary convex and non-convex domains. Indeed in the work by \cite{Ramirez2015} the geometry naturally reduces the model to a single one-dimensional reaction-diffusion equation and the cytosolic component is assumed constant, while \cite{Giese2015} is an entirely two-dimensional work. Finally the work by Diegmiller \textit{et al.} \cite{Diegmiller2018} reveals very important insights, however it is restricted to a sphere.
The novelty of our work lies in that we {\corr mathematically quantify} the role of the three-dimensional geometry in the wave pinning process, yielding new insights into this minimal model for wave pinning. For simplicity throughout the paper, we will refer to the reformulated WP model as the bulk-surface wave pinning (BSWP) model. 

{\corr We present new three-dimensional results on regular and irregular geometries, exhibiting the wave pinning process on complex geometries. A key part of our study involves the numerical simulation of the BSWP model in three-dimensional geometries using a recently developed bulk-surface finite element method (BS-FEM) \cite{Dziuk2013,Elliott2013,Macdonald2016,Madzvamuse2016b,Madzvamuse2016a,Madzvamuse2015}. This numerical framework allows to compute the solutions of the BSWP model on complex convex and non-convex geometries.}

To put into context our computational framework with respect to the current-state-of-the-art, throughout this paper we confirm previous works based on the wave pinning model \eqref{eq:model_Mori_a}-\eqref{eq:model_Mori_f(a,b)} and show analogies with our results. For example, we show the evolution of the solutions of the model at very large times which display interesting spatial effects.
{\corr Our results reveal that certain geometries induce a metastable behavior of the model, in which the apparently stable active patch undergoes a very slow shifting on the surface towards more rounded areas of the domain.  This was also shown in previous published results for the two-dimensional wave pinning model presented by \cite{Vanderlei2011}. }
In addition, the BSWP model shows competition between active regions, as recently shown in the classical WP model \cite{Chiou2018}. {\corr We also show how the geometry of the domain plays a crucial role in the pattern formation for the special case of spatial homogeneous initial conditions. This was interestingly reported in the two-dimensional case by Giese \emph{et al.} in \cite{Giese2015}. Hence, our work through mathematical and numerical analysis, aims to extend the current knowledge of the wave pinning model to realistic three-dimensional settings and to provide a satisfactory understanding of the influence of the geometry, showing the role that cell shape plays in the polarisation mechanism.
}

The structure of this work is therefore as follows:
In Section \ref{sec:BS_model} we describe the model and its parameters as well as discussing its fundamental properties. The polarisation mechanism of the BSWP model is explained in Section  \ref{sec:asymptotic_analysis} by an asymptotic analysis on a simple geometry.
In Section \ref{sec:bistabiliy_LPA} we present the parameter regions for bistability and polarisation. Analysis of the steady states for the well-mixed system provide a bistability region, whereas spatial effects were studied using the local perturbation analysis (LPA) \cite{Holmes2014,Holmes2015}. This latter tool is able to identify parameter spaces in which a local and narrow perturbation of the spatially homogeneous slow-diffusing component can generate spatial effects on the system. In our work we present a novel application of the LPA in a bulk-surface setting, which provides a natural way to investigate the effect of the ratio between surface area and bulk volume on the system. 
In Section \ref{sec:BSFEM} we present the bulk-surface finite element method (BS-FEM) \cite{Macdonald2016,Madzvamuse2016a}, used to simulate the model on various geometries. Numerical results are then presented in Section \ref{sec:Results} to confirm and validate theoretical findings. A summary of the main results and a discussion follow in Section \ref{sec:conclusions}, with suggestions on future extensions and applications of the BSWP model.

\begin{figure}[ht]
\begin{subfigure}{.5\textwidth}
\centering
	\includegraphics[width=.8\linewidth]{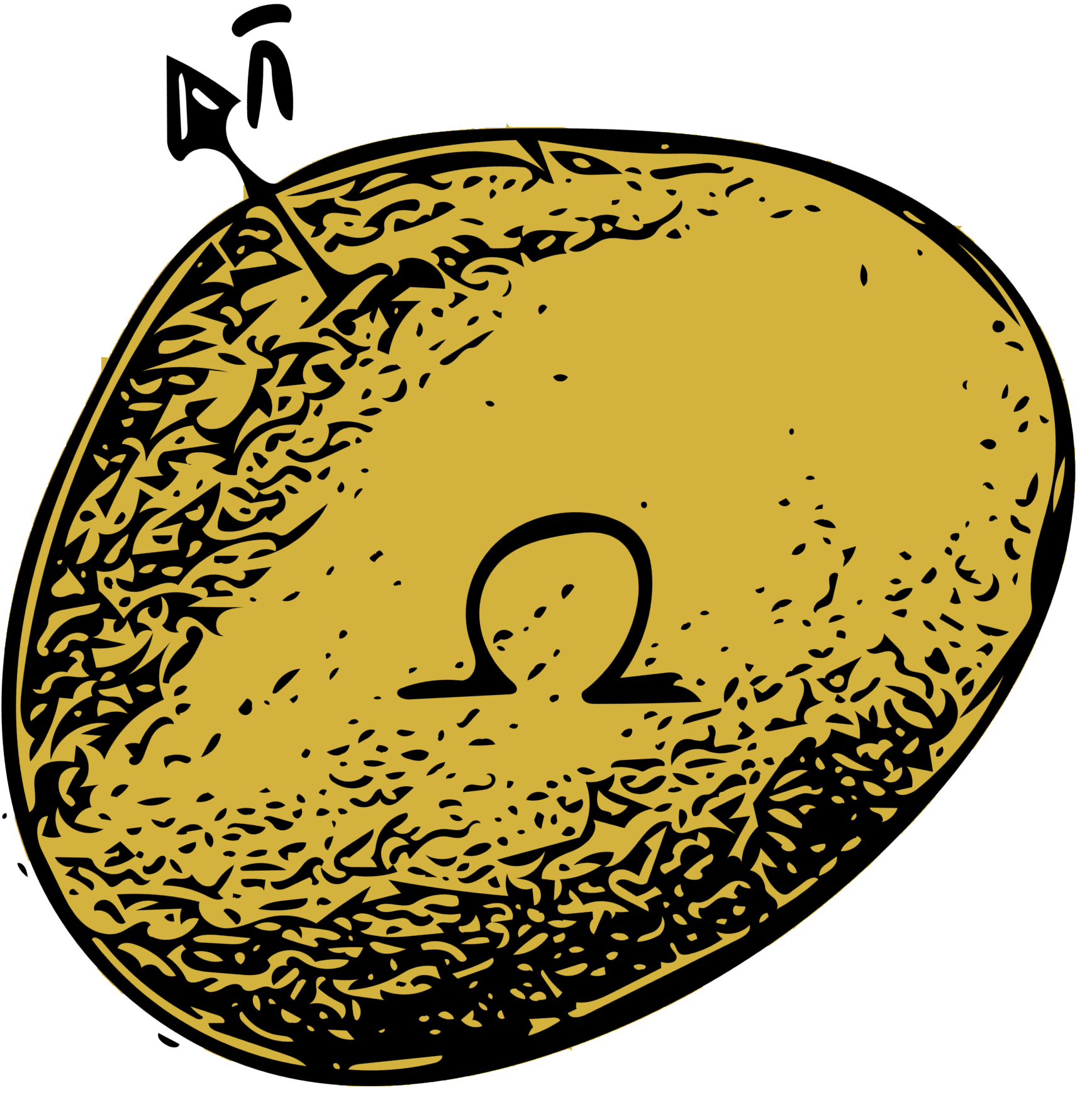}
	\caption{}
	\label{fig:Omega}
\end{subfigure}
\begin{subfigure}{.5\textwidth}
\centering
     \includegraphics[width=.8\linewidth]{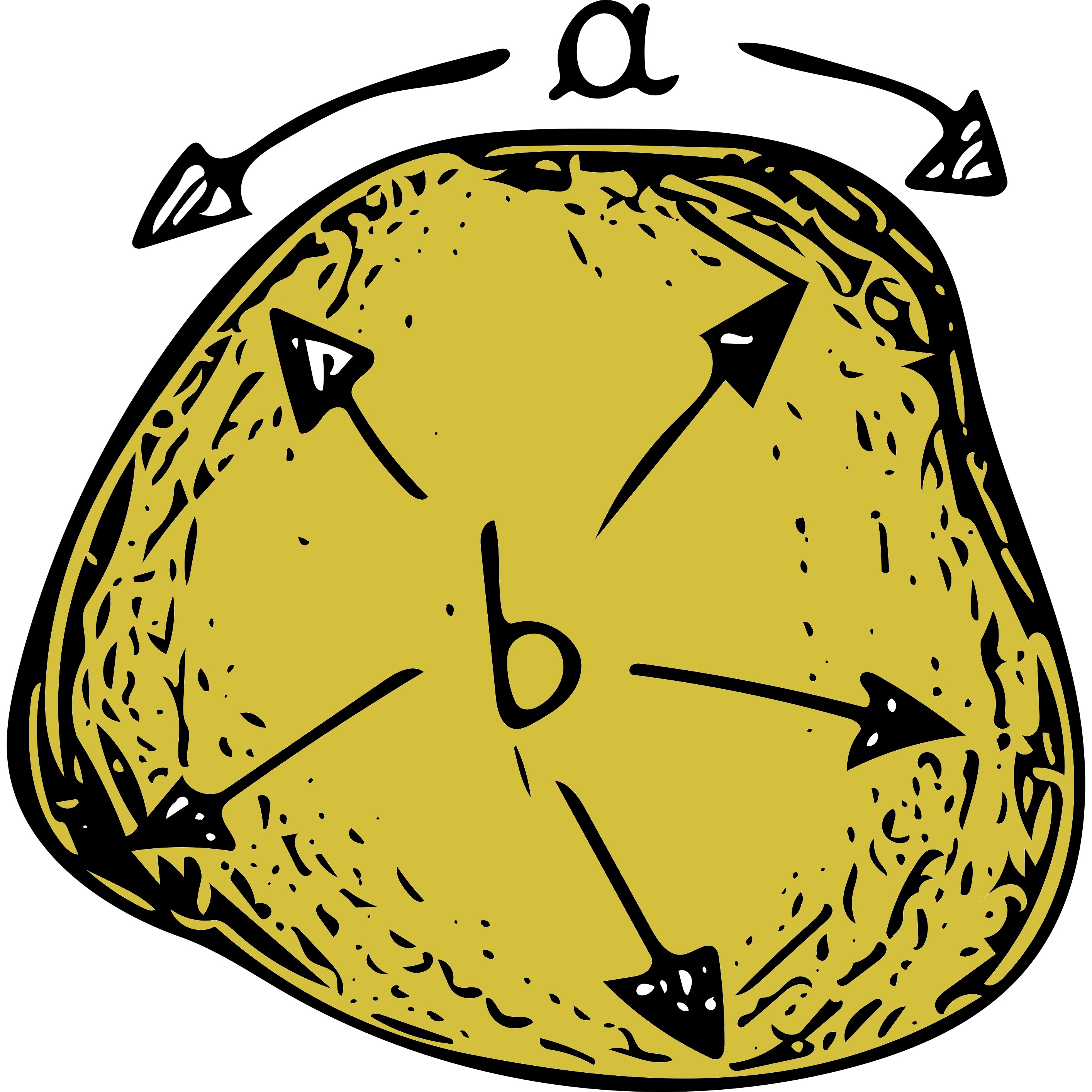}
     \caption{}
     \label{fig:SpeciesOnDomain}
\end{subfigure}
\caption{(a) A three-dimensional domain $\Omega$ representing the cell. (b) Activation of the bulk species occurs through the boundary conditions and propagates over the surface $\Gamma$ of the domain.}
\end{figure}

\section{The bulk-surface wave pinning model}\label{sec:BS_model}
The model is derived for a single stationary cell (studies on migrating cells are deferred to future work) whose shape is described by a smooth closed surface $\Gamma\subset\mathbb{R}^3$, hence with no boundary, which encloses the bulk geometry $\Omega\subset\mathbb{R}^3$ such that $\Gamma=\partial\Omega$. In biological terms, $\Gamma$ represents the cell membrane and $\Omega$ the cell interior.
Let $a$ be the active membrane-bound GTPase and $b$ the inactive GTPase. 
For model consistency we need to require our solutions to be smooth enough, so we look for {\corr \textit{classical}} solutions {\corr $a\in C(\Gamma\times[0, T])\cap C^{2,1}(\Gamma\times(0, T])$ and $b\in C(\overline{\Omega}\times[0, T])\cap C^{2,1}(\Omega\times(0, T])\cap C^{1,0}(\overline{\Omega}\times(0, T])$,
where $C^{k,h}$ indicates the set of functions $k$ times differentiable in space and $h$ times in time.} 
We will assume pure diffusion for the cytosolic form and impose Robin-type boundary conditions on $\Gamma$, which take into account the switching between active and inactive species. From conservation principles we get
\begin{align}
\frac{\partial b}{\partial t} &=D_b\Delta b, &\mathbf{x}\in\Omega,{\corr ~t\in(0,T],}  \label{eq:model_b}\\
-D_b(\mathbf{n}\cdot\nabla b) &=f(a,b), &\mathbf{x}\in\Gamma,{\corr ~t\in(0,T],} \label{eq:model_b_Gamma} 
\end{align}
where $D_b$ represents the diffusion coefficient, $\mathbf{n}$ is the outward unit vector to $\Omega$ and $f$ is a function which depends on both $a$ and $b$ and represents the variation of the bulk variable $b$ due to activation and inactivation of the GTPase on the cell membrane. 
One key property of the model is that the reactions for the bulk species are incorporated into the boundary condition, while no reactions occur inside the cell. 
We use a relatively simple nonlinear reaction function $f(a,b)$, the same as in  \cite{Mori2008}. Nonlinearity is achieved through a Hill function, commonly used in biochemistry to represent what is called a \emph{cooperative binding} \cite{Lehninger2005}. One can work with a generalised function of \eqref{eq:model_Mori_f(a,b)}  given by
\begin{align}\label{eq:f(a,b)}
f(a,b)&=\omega\Big(k_0+\frac{\gamma a^n}{{\corr K^n}+a^n}\Big)b-\beta a, 
\end{align}
where the Hill coefficient $n=2$ is sufficient to achieve bistability \cite{Mori2008, Mori2011}, It must be noted that other choices for $n$ have been presented \cite{Diegmiller2018,Holmes2016}.  Following \cite{Ratz2012} we define $\omega :=|\Omega|/ |\Gamma|$ as the ratio between bulk volume and surface area; it characterises the geometric effects in the reaction function {\corr and can be seen as a parameter describing the protein \emph{binding} to the cell membrane. The length unit dimension of $\omega$ is needed to reduce the dimensionality of the bulk protein to the two-dimensional surface, where activation occurs.
For} a fixed volume, $\omega$ is maximal when $\Omega$ is spherical, so activation is enhanced in resting cells which generally have, at least in two-dimensions, a rounded shape \cite{Kozlov2007}. 
 \begin{table}
\centering
\begin{tabular}{lll}
\rowcolor[HTML]{CACACA} 
\textbf{Param.} & \textbf{Value/Units}       & \textbf{Description}  \\ \hline
\rowcolor[HTML]{DDDDDD} 
$a$             & mol $\mu$m$^{-2}$  & concentration of active GTPase \\
\rowcolor[HTML]{CACACA} 
$b$             & mol $\mu$m$^{-3}$       & concentration of inactive GTPase  \\
\rowcolor[HTML]{DDDDDD} 
$D_a$           & 0.1 $\mu$m$^{2} \; s^{-1}$ & diffusion coefficient of $a$   \\
\rowcolor[HTML]{CACACA} 
$D_b$           & 10 {\corr $\mu$m$^{2} \; s^{-1}$ }      & diffusion coefficient of $b$ \\
\rowcolor[HTML]{DDDDDD} 
$k_0$           & 0.067 $s^{-1}$              & basal activation rate          \\
\rowcolor[HTML]{CACACA} 
$\beta$         & 1 $s^{-1}$            & deactivation rate              \\
\rowcolor[HTML]{DDDDDD} 
$\gamma$        & 1 $s^{-1}$                  & feedback activation rate       \\
\rowcolor[HTML]{CACACA} 
$K$             & 1 mol $\mu$m$^{-2}$         & saturation parameter      \\
\rowcolor[HTML]{DDDDDD} 
$n$        & 2                 &Hill coefficient       \\
\rowcolor[HTML]{CACACA} 
$\omega$             & $\mu$m         & {\corr volume to surface ratio membrane binding parameter}   
\end{tabular}
\caption{Parameters used in the bulk-surface model; the diffusion coefficients are taken as in \cite{Postma2004}, and the kinetic parameter values as in \cite{Mori2008}.} 
\label{tab:parameters}
\end{table}

The spatio-temporal dynamics of the membrane-bound active form $a$ on the cell membrane are described by the following surface reaction-diffusion equation 
\begin{align}
\frac{\partial a}{\partial t} &=D_a\Delta_\Gamma a+f(a,b), &\mathbf{x}\in\Gamma,{\corr ~t\in(0,T],} \label{eq:model_a}
\end{align}
where $D_a$ is the diffusion coefficient and $\Delta_\Gamma$ is the Laplace-Beltrami operator, which generalises the Laplacian over manifolds \cite{Dziuk2013} and it is here used to describe lateral diffusion of  membrane proteins. Since we are considering a closed system in which $a$ and $b$ are different forms of the same component, it makes sense to link entirely the reaction in $a$ with the boundary condition for $b$, meaning that there is full inter-conversion between the two forms. Therefore the reaction in the equation for $a$ is the same function $f$ defined in \eqref{eq:f(a,b)}. 
The parameters used in the bulk-surface model are listed in Table \ref{tab:parameters}.

{\corr It should be noted that the well-posedness and the global existence of solutions for the general bulk-surface reaction-diffusion system  of $k$ bulk and $m$ surface variables was studied by Sharma and Morgan in \cite{Sharma2016}. Hence, the following theorem holds:

\begin{theorem}
The BSWP model \eqref{eq:model_b}-\eqref{eq:model_a} admits a unique and non-negative classical solution 
$a(t,\mathbf{x})$ and $b(t,\mathbf{x})$ at any time $t>0$, for any suitable non-negative initial condition $a_{in}(\mathbf{x})$ and  $b_{in}(\mathbf{x})$.
\end{theorem}
\begin{proof}
See \cite{Sharma2016} and in particular Corollary 3.4.
\end{proof}
}
\subsection{Fundamental properties of the BSWP model}\label{sec:fundamental_properties}
We now briefly present some fundamental properties of the BSWP model \eqref{eq:model_b}-\eqref{eq:model_a}
 as follows.
\begin{enumerate}
	\item \emph{Conservation of total species}. Integrating (\ref{eq:model_b}) and (\ref{eq:model_a}) and applying the divergence theorem on a manifold with boundary condition (\ref{eq:model_b_Gamma}) {\corr and the divergence theorem on manifolds without boundary}, it is easy to prove the following.
	\begin{proposition}
		Let $a$ and $b$ be solutions of  (\ref{eq:model_b})-(\ref{eq:model_a}). Then
		\begin{equation}\label{eq:conservation_mass}
		M(t) := \int_{\Omega}b(\mathbf{x},t)\dx + \int_{\Gamma}a(\mathbf{x},t)\ds = M_0, \quad \forall t\geq 0
		\end{equation} 
for a certain fixed value $M_0\in\mathbb{R}$, defined by the initial conditions: indeed, $M_0$ represents the total amount of substance $a+b$ in the cell.
    \end{proposition}
	
	\item \emph{Difference in diffusivities}. As protein diffusion over the membrane is known to occur much slower than in the cytosol, we consider $D_a\ll D_b$  \cite{Postma2004}.
	
	\item \emph{Bistability}.  The following proposition holds 
	\begin{proposition} \label{prop:roots_f(a,b)}
	Consider a fixed value of $b$, denoted $\overline{b}$. Therefore,
	there exist two positive values $b_1$ and $b_2$ such that if $\overline{b}\in(b_1,b_2)$, the function $f(a,\overline{b})$, as defined in \eqref{eq:f(a,b)}, can have up to three zeros $a_1(\overline{b})<a_2(\overline{b})<a_3(\overline{b})$. In particular, when this occurs we have \cite{Mori2011}
    \begin{equation}\label{eq:f(a,b)_bistability}
	       \frac{\partial f}{\partial a}(a_1(\overline{b}), \overline{b})<0, ~\frac{\partial f}{\partial a}(a_2(\overline{b}), \overline{b})>0,~\frac{\partial f}{\partial a}(a_3(\overline{b}), \overline{b})<0.
    \end{equation}
	\end{proposition}
	See also Figure \ref{fig:f(a,b)=0} for a schematic representation. In particular, a necessary condition for bistability \cite{Mori2011} is that 
	\begin{equation}
	8k_0<\gamma. \label{eq:8k<gamma}
	\end{equation}
	This latter condition represents and highlights the important role of the feedback-induced activation rate in the model. 
\end{enumerate}

\section{Asymptotic analysis on a disk}\label{sec:asymptotic_analysis}
The basic mechanisms of the BSWP model \eqref{eq:model_b}-\eqref{eq:model_a} can be understood through an asymptotic analysis which is here presented in order to highlight the main steps of the spatio-temporal evolution of certain classes of initial conditions. Since the core of the analysis is based on the crucial difference of protein diffusivity between cell membrane and cytosol, a convenient setting to stress this relationship is the use of a nondimensional version of the model. Therefore, in this section we consider the following coupled system of bulk-surface reaction-diffusion equations, where diffusion on the surface is very slow relative to diffusion in the bulk
\begin{align}
\varepsilon\frac{\partial {b}}{\partial {t}} &={\Delta}{b},\quad &\mathbf{x}\in\Omega,  \label{eq:modelnondim_b}\\
-(\mathbf{n}\cdot{\nabla}{b}) &=f\;({a}, {b}),\quad &\mathbf{x}\in\Gamma, \label{eq:modelnondim_b_Gamma}\\
\varepsilon\frac{\partial {a}}{\partial {t}} &=\varepsilon^2{\Delta_\Gamma} {a}+ \; f(a, b),\quad &\mathbf{x}\in\Gamma, \label{eq:modelnondim_a}
\end{align}
with 
\begin{equation}\label{eq:f(a,b)_nondimensional}
f({a}, {b}):= \left({k_0}+\frac{{\gamma}{a}^2}{1+{a}^2}\right){b}-{a},
\end{equation}
where $a$ and $b$ are now nondimensional quantities and {\corr $\varepsilon^2={\frac{D_a}{D_b}}$} is a  small parameter. Details of the nondimensionalisation can be found in the Appendix. 

Provided condition (\ref{eq:8k<gamma}) is satisfied, for $b$ within a certain range $(b_1, b_2)$, the function $f(a,b)$ has three distinct and positive roots $a_1(b) < a_2(b) < a_3(b)$ and (\ref{eq:f(a,b)_bistability}) is satisfied{\corr , i.e. $a_1(b)$ and $a_3(b)$ are stable steady states for the ODE corresponding to the equation \eqref{eq:modelnondim_a} with zero diffusion.
Bistable reaction-diffusion equations are known to produce travelling wave solutions \cite{Fife1977} and this is a crucial aspect of the wave pinning mechanism.}
Figure \ref{fig:f(a,b)=0} shows the zero level set of $z=f(a,b)$, which also represents the nullcline of the ordinary differential system. 
\begin{figure}[ht]
	\centering
	\includegraphics[scale=0.4]{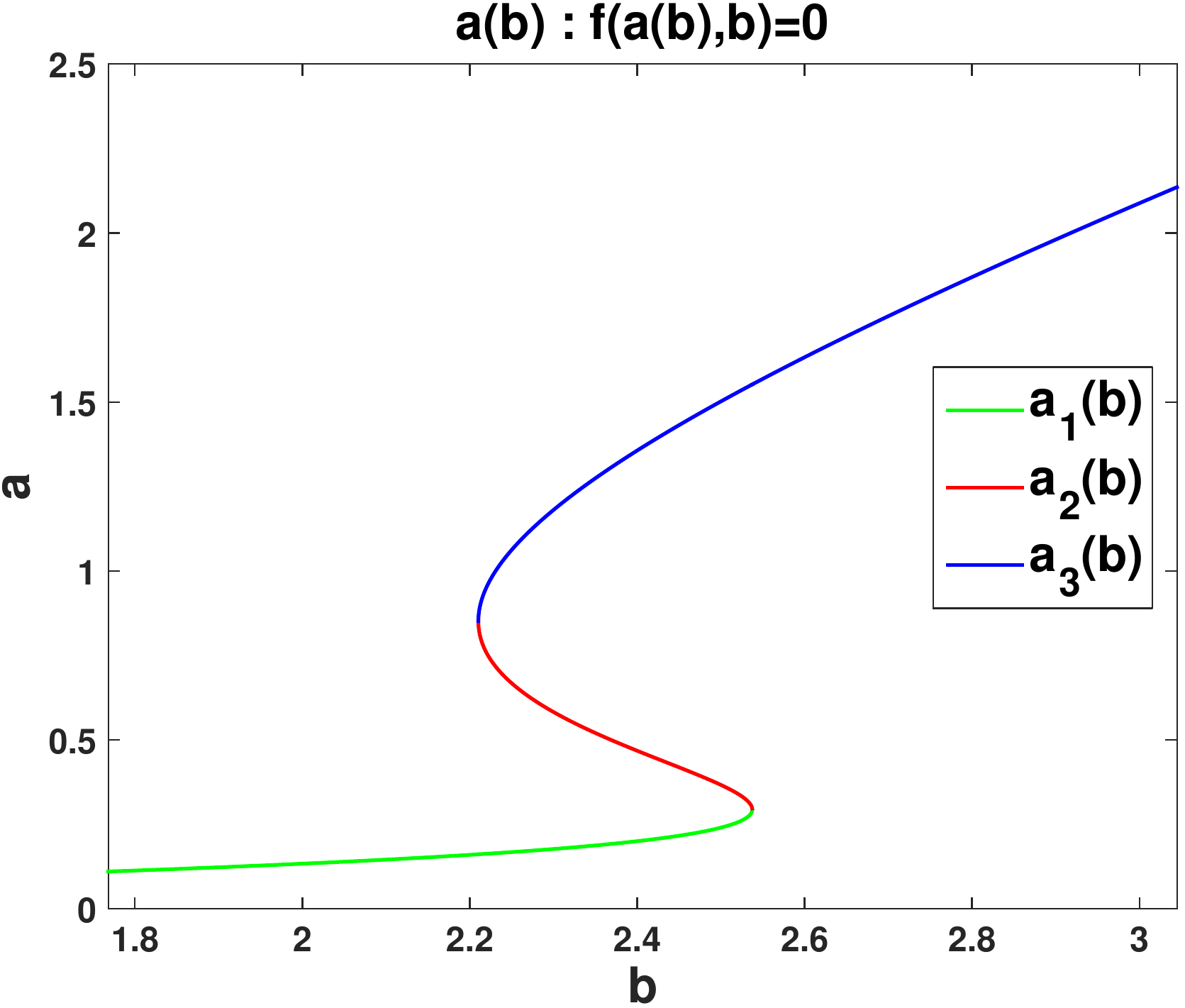}
	\caption{The solutions $(a,b)$ solving $f(a,b)=0$ as defined in (\ref{eq:f(a,b)_nondimensional}) with parameters $k_0=0.05$ and $\gamma=0.79$.}
	\label{fig:f(a,b)=0}
\end{figure}

\begin{figure}[ht!]
\begin{turn}{90} 
\textbf{top}
\end{turn}
\begin{subfigure}{0.23\textwidth}
  \centering
  \includegraphics[width=1\linewidth]{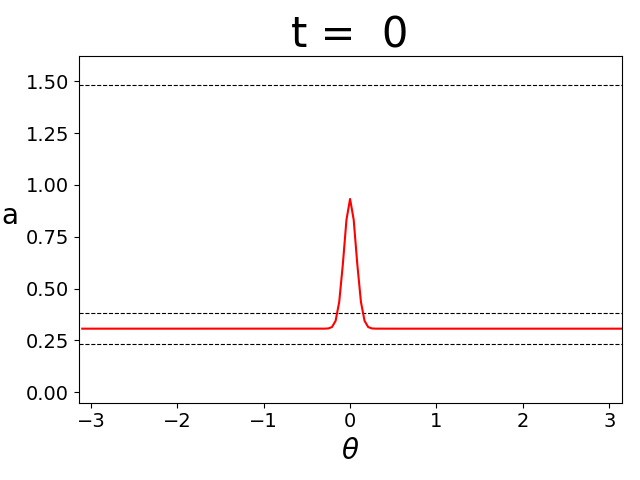}
  \caption{}
  \label{fig:asymptotic_analysis_a_1}
\end{subfigure}
\begin{subfigure}{0.23\textwidth}
  \centering
  \includegraphics[width=1\linewidth]{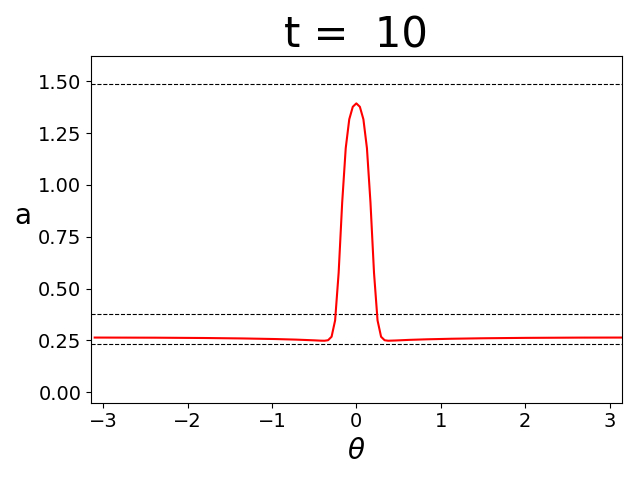}
  \caption{}
  \label{fig:asymptotic_analysis_a_2}
\end{subfigure}
\begin{subfigure}{0.23\textwidth}
  \centering
  \includegraphics[width=1\linewidth]{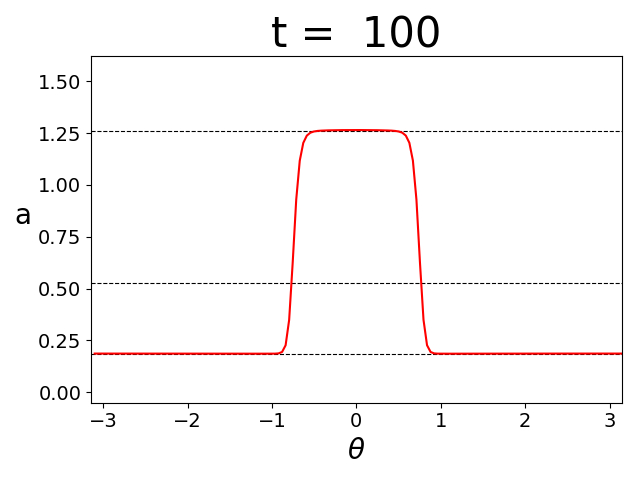}
  \caption{}
  \label{fig:asymptotic_analysis_a_3}
\end{subfigure}%
\begin{subfigure}{0.23\textwidth}
  \centering
  \includegraphics[width=1\linewidth]{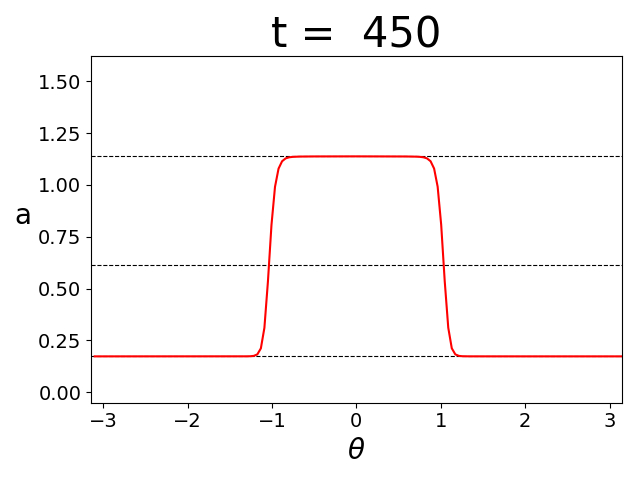}
  \caption{}
  \label{fig:asymptotic_analysis_a_4}
\end{subfigure}
\vspace{0.5cm}
\\
\begin{turn}{90} 
\textbf{bottom}
\end{turn}
\begin{subfigure}{0.23\textwidth}
  \centering
  \includegraphics[width=0.8\linewidth]{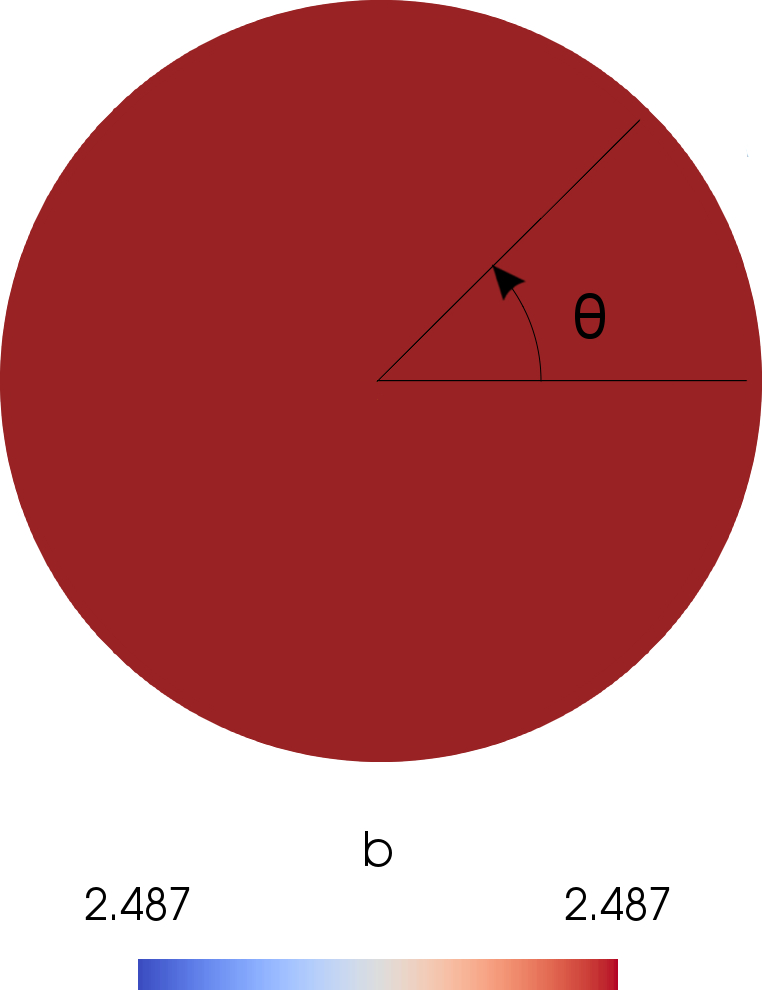}
  \label{fig:asymptotic_analysis_b_1}
\end{subfigure}
\begin{subfigure}{0.23\textwidth}
  \centering
  \includegraphics[width=0.8\linewidth]{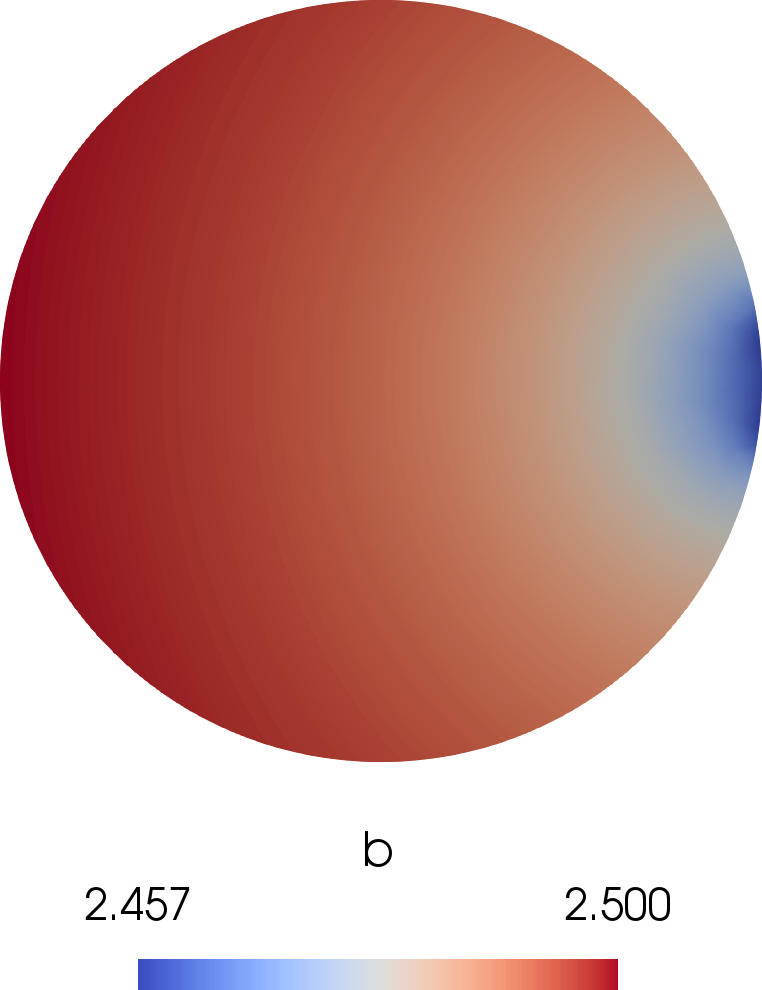}
  \label{fig:asymptotic_analysis_b_2}
\end{subfigure}
\begin{subfigure}{0.23\textwidth}
  \centering
  \includegraphics[width=0.8\linewidth]{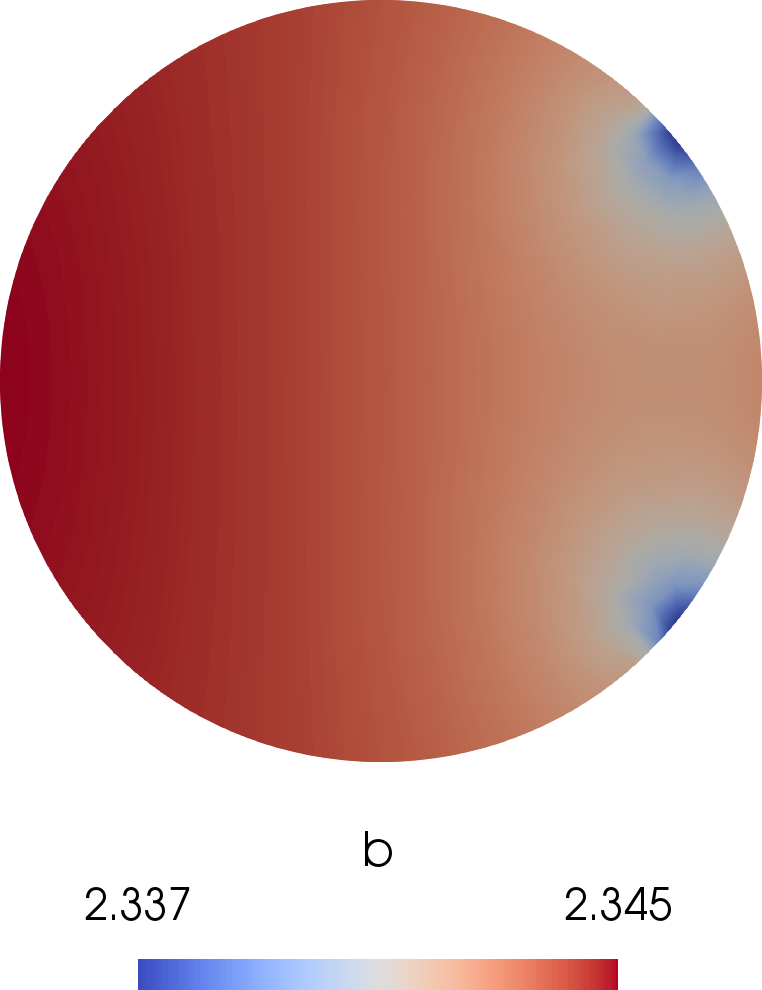}
  \label{fig:asymptotic_analysis_b_3}
\end{subfigure}%
\begin{subfigure}{0.23\textwidth}
  \centering
  \includegraphics[width=0.8\linewidth]{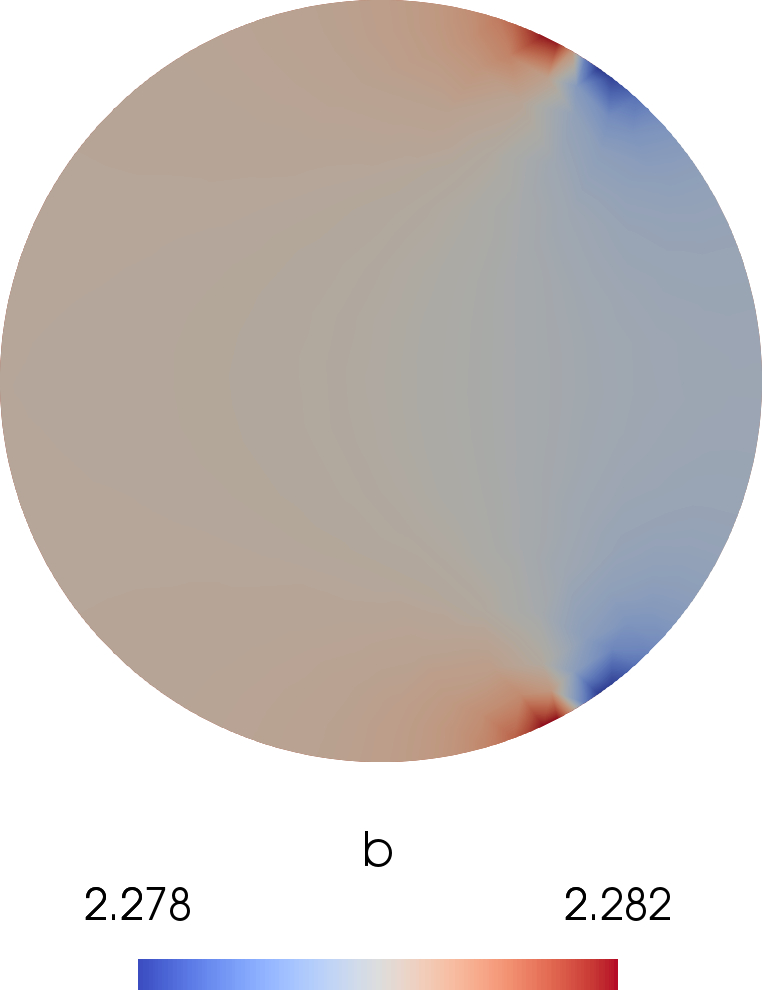}
  \label{fig:asymptotic_analysis_b_4}
\end{subfigure}
\caption{Numerical solutions of the BSWP model \eqref{eq:modelnondim_b}-\eqref{eq:f(a,b)_nondimensional} with $\varepsilon^2=0.001$ on a disc 
at different time steps. 
The parameter $\varepsilon$ plays a crucial role in sharpening the fronts of the solution $a$ \cite{Mori2011}  and smaller choices of $\varepsilon$ result in a clearer effect of the BSWP mechanism. 
On the top row we plot the solution $a$ (red line) over the circle at different time steps, whereas the horizontal dashed lines indicate the three solutions $a_1, a_2, a_3$ of $f\left(a, \overline{b} \right)=0$, where $\overline{b} =  \frac{1}{|\Omega|}\int_\Omega b\dx$ and $a_1<a_2<a_3$. On the bottom row we plot the solution $b$ inside the disk. It is important to note the scale values for $b$: at every time step, $b$ is approximately spatially constant. 
(a) (top) A narrow Gaussian function is summed over a spatially homogeneous initial condition. In most of its domain $a$ is initially smaller than $a_2$, except for the Gaussian peak. We use the centre of the peak as reference for the polar coordinate system.
(bottom) The initial condition for $b$ is spatially homogeneous.  
(b) (top) Attraction of $a$ towards the values $a_1$ and $a_3$ is well visible: the peak grows towards $a_3$, while the rest of the solution tends to the lower value $a_1$. (bottom) Depletion of $b$ starts from  the boundary of the disc at around $\theta=0$. 
(c) (top) At time $t=100$ $a$
overlaps $a_1$ and $a_3$ in most of the domain except in the two very small areas where the transition between the two states occurs very sharply. In addition, the peak of $a$ has visibly increased its width, as propagation has started. (bottom) $b$ is depleted in correspondence of the sharp moving fronts of active GTPase. 
(d) The steady states for $a$ and $b$. $b$ has reached its critical value and there is no more source of GTPase available for $a$, which therefore is pinned in an almost piece-wise constant shape. 
Details of the numerical methods and tools used for the simulation will be given in Section \ref{sec:BSFEM}.
}
\label{fig:asymptotic_analysis}
\end{figure}

We consider initial conditions of the following type 
\begin{align}
&b_{in}(\mathbf{x})=\overline{b_0}\in(b_1,b_2), &\mathbf{x}\in\Omega,\label{eq:asympt_analysis_initial_condition_b}\\
&a_{in}(\mathbf{x}) = a_{g} + a_p(\mathbf{x}), &\mathbf{x}\in\Gamma,\label{eq:asympt_analysis_initial_condition_a}
\end{align}
where $a_g\in \left[0, a_2(\overline{b_0})\right)$ and $a_p$ is a continuous function over $\Gamma$ such that
if 
\[
\Gamma_p:=\left\{\mathbf{x}\in\Gamma ~:~ a_p(\mathbf{x}) + a_g > a_2(\overline{b_0}) \right\}
\]
then 
\begin{align*}
0<\frac{|\Gamma_p|}{|\Gamma|} \ll 1
\quad \text{ and }\quad
\frac{1}{|\Gamma|}\int_{\Gamma_p} a_p \dx \ll 1.
\end{align*} 
In biological terms, the above describe that initially the inactive cytosolic protein is homogeneously constant, while the initial concentration of $a$ is less than the value $a_2(\overline{b_0})$ in most of its domain except for tiny regions in which its mass is negligible.
In the simulations we have represented $a_p$ with very narrow Gaussian functions. 

We consider a flat cell
$
\Omega = \{
(x,y) ~:~ x^2+y^2<r^2, \;  r>0
\}
$
 which, being a simple circular domain, makes the exposition clearer. We are also interested in a single peak for $a$, which means $\Gamma_p$ is connected, in other words, $a_{in}(\mathbf{x}) = a_2(\overline{b_0}) $ has two solutions $\mathbf{x}$.
{
\corr In our exposition we next show that the evolution of $a$ is strongly characterised by different time scales with the development of well defined spatial patterns and 
formation of boundary layers in which the solution drastically passes from one ``stable'' state to the other. This corresponds to a sudden large variation of the gradient of $a$, in very small regions, which is otherwise negligible elsewhere. This leads to the need of a spatial rescaling around these areas. 
A typical strategy for studying this class of equations is presented in \cite{Rubinstein1992}, where a mass conserved reaction-diffusion equation with a double-well potential is studied through  multiple temporal rescaling and matched asymptotic analysis.
}
Our analysis is described in four steps (see also Figure \ref{fig:asymptotic_analysis}) as outlined below, and it follows the asymptotic analysis done by Mori \textit{et al.} \cite{Mori2011} for the unidimensional model \eqref{eq:model_Mori_a}-\eqref{eq:model_Mori_f(a,b)}, which we have re-adapted to the BSWP model \eqref{eq:model_b}-\eqref{eq:model_a} thanks to the circular geometry.

\begin{enumerate}[(a)]
	\item At the initial time, $a$ evolves into a well defined profile with two fronts: over $\Gamma_p$ it is attracted by $a_3(b)$, while on the rest of the domain  it is attracted by $a_1(b)$. On the other hand, $b$ is approximately spatially homogeneous. We study this evolution over the zoomed time scale $\tau=t/\varepsilon$.
	
	\item In the intermediate time scale {\corr $t$} we observe the movement of the fronts in the $a$ profile, in particular we are interested in the expansion of the high concentration peak. In order to achieve this, we need to show that
	
	\begin{itemize}
		\item The speed of the propagating fronts is strictly related to the sign of the function defined by 
		\begin{equation}\label{eq:I(b)}
		I(b):=\int_{a_1(b)}^{a_3(b)}f(\xi,b) \text{d}\xi, \quad b\in(b_1,b_2).
		\end{equation}
		\item $I(b)$ is an increasing function in $(b_1, b_2)$ and there exists $b_c\in(b_1,b_2)$ such that $I(b_c)=0$.
	\end{itemize}

	\item {\corr The propagation of $a$ coincides with the depletion of $b$, which is always approximately spatially homogeneous (note the color scale in Fig \ref{fig:asymptotic_analysis} bottom). }
	
	\item Under particular conditions on the initial concentrations, the propagation stops before the whole boundary is activated. This occurs when $b$ has decreased to its critical value $b_c$.    
\end{enumerate}

We are now in a position to discuss the steps (a)-(d) in more detail.\\

{\corr \textbf{Step a)}
We first study the initial evolution of the system \eqref{eq:modelnondim_b}-\eqref{eq:f(a,b)_nondimensional} by introducing the fast time scale $\tau=t/\varepsilon$. Temporal rescaling results in the following coupled bulk-surface system
}
\begin{align*}
\frac{\partial {b}}{\partial {\tau}} &={\Delta}{b},\quad &\mathbf{x}\in\Omega, \\
\frac{\partial {a}}{\partial {\tau}} &=\varepsilon^2{\Delta_\Gamma} {a}+ \; f(a, b),\quad &\mathbf{x}\in\Gamma,\\
-(\mathbf{n}\cdot{\nabla}{b}) &=f\;({a}, {b})\quad &\mathbf{x}\in\Gamma.
\end{align*}
Looking for solutions of the form $a=a_0+a_1\varepsilon+a_2\varepsilon^2+\cdots$ and $b=b_0+b_1\varepsilon+a_2\varepsilon^2+\cdots$ we find, at the leading order
\begin{align*}
\frac{\partial {b_0}}{\partial {\tau}} &={\Delta}{b_0},\quad &\mathbf{x}\in\Omega, \\
\frac{\partial {a_0}}{\partial {\tau}} &= f(a_0, b_0),\quad &\mathbf{x}\in\Gamma,\\
-\mathbf{n}\cdot{\nabla}{b_0} &=f({a_0}, {b_0}),\quad &\mathbf{x}\in\Gamma.
\end{align*}
The equation for $a_0$ is an ordinary differential equation and, at each $\mathbf{x}$, the solution will tend to the stable stationary point $a_3(b)$ for $\mathbf{x}\in\Gamma_p$ or  
$a_1(b)$ elsewhere: at the end of this time scale we will have $\frac{\partial {a_0}}{\partial {\tau}}\approx 0$. This means that over $\Gamma$, $f(a_0,b_0)\approx 0$.

The equation for $b_0$ is the heat equation with Neumann boundary conditions that will become approximately homogeneous at the end of the time scale. Then $b_0(\mathbf{x},\tau)$ will tend to reach a spatially homogeneous profile over the domain $\Omega$.

{\corr \textbf{Step b)}} In the intermediate time scale $t$, we again look for solutions of the form  $a=a_0+a_1\varepsilon+a_2\varepsilon^2+ \cdots$ and $b=b_0+b_1\varepsilon+a_2\varepsilon^2+ \cdots$. At the leading order we have
\begin{align*}
{\Delta}{b_0} &= 0,\quad &\mathbf{x}\in\Omega, \\
f(a_0, b_0) &=0, \quad &\mathbf{x}\in\Gamma,\\
-\mathbf{n}\cdot{\nabla}{b_0} &=f({a_0}, {b_0}),\quad &\mathbf{x}\in\Gamma.
\end{align*}
We see that the flux condition is actually $-\mathbf{n}\cdot{\nabla}{b_0}=0$, consistent with the Laplace equation in $\Omega$. $b_0(\mathbf{x}, t)$ is now at equilibrium all over the domain. On the other hand, $a_0(\mathbf{x}, t)$ remains at its low and high values, either $a_1(b)$ or $a_3(b)$. This is valid far from the two front layers where the solution passes from $a_1$ to $a_3$ and vice versa. Our goal is to see if these front layers move in time over the boundary $\Gamma$. 
We take advantage of the circular geometry of the domain and re-write the model \eqref{eq:modelnondim_b}-\eqref{eq:f(a,b)_nondimensional} in polar coordinates
\begin{align*}
\varepsilon \frac{\partial {b}}{\partial {t}} &=\frac{\partial^2 {b}}{\partial {\rho^2}}+\frac{1}{\corr \rho}\frac{\partial {b}}{\partial {\rho}}+\frac{1}{\corr \rho^2}\frac{\partial^2 {b}}{\partial {\theta^2}},\quad &\rho\in(0,r), \theta\in(-\pi,\pi], \\
\varepsilon\frac{\partial {a}}{\partial {t}} &= \frac{\varepsilon^2}{\corr r^2}\frac{\partial^2 {a}}{\partial {\theta^2}} + f(a,b), \quad &\rho=r, \theta\in(-\pi,\pi],\\
-\frac{\partial {b}}{\partial {\rho}}&=f(a,b), \quad &\rho=r, \theta\in(-\pi,\pi],
\end{align*}
where $r$ is the radius of the disk.
In this coordinate system it becomes easier to define the positions of the front layers. Indeed, an angle $\theta$ is enough to uniquely identify a point on $\Gamma$. Let us set $\theta=0$ at the centre of the boundary subset $\Gamma_p$, so that there exist a value $\theta_1<\pi$ such that $\Gamma_p = (-\theta_1, \theta_1)$, see also  Figure \ref{fig:asymptotic_analysis_a_1} (top and bottom).

The positions of the two fronts of $a$ are therefore initially defined by $-\theta_1$ and $\theta_1$ and our goal is to show that these positions can change in time subject to \eqref{eq:modelnondim_b}-\eqref{eq:f(a,b)_nondimensional}. We will consider $\theta_1(t)$, which is initially small. We define the variable
\[
\varphi_1(t):=\frac{\theta-\theta_1(t)}{\varepsilon}, 
\]
such that 
\[
\lim_{\varepsilon\to 0}\varphi_1=\begin{cases}
+\infty \quad \text{ if } \theta>\theta_1\\
-\infty \quad \text{ if } \theta<\theta_1
\end{cases}
\] and
\[
\lim_{\varphi_1\to-\infty}a(\varphi_1)=a_3(b),
\quad
\lim_{\varphi_1\to+\infty}a(\varphi_1)=a_1(b),
\]
\textit{i.e.} the wave front connects the high and low plateau values of $a$. {\corr We remark that for $\theta<0$ the situation reverses:  the solution is close to $a_1(b)$ for values of $\theta<-\theta_1$ and to $a_3(b)$ for $\theta>-\theta_1$. More generally, the periodicity of the two-dimensional domain requires an even number of fronts in $(-\pi, \pi]$, which was not necessary in previous works on the wave pinning mechanism.}
The equation for $a$ in the new coordinate $\hat{a}(\varphi_1(t),t)=a\left(\frac{\theta-\theta_1(t)}{\varepsilon},t\right)$ is
\[
\varepsilon\frac{d \hat{a}}{dt} 
-
\theta_1'(t)\frac{\partial \hat{a}}{\partial \varphi_1}
=
\frac{1}{r^2}\frac{\partial^2 \hat{a}}{\partial {\varphi_1^2}} 
+
f(\hat{a},b),
\quad 
\varphi_1\in(-\infty, + \infty).
\]
The term $\theta_1'(t)$ in the left hand side of the above equation describes the speed of the front, which we want now to investigate.  
Using again asymptotic expansion $\hat{a}=\sum \hat{a}_i\varepsilon^i$ we get, at the leading order 
\[
\frac{1}{r^2}\frac{\partial^2 {\hat{a}_0}}{\partial {\varphi_1^2}} 
+ \theta_1'(t)\frac{\partial \hat{a}_0}{\partial \varphi_1}
+ f(\hat{a}_0,b)
=0,
\quad \varphi_1\in(-\infty, + \infty).
\]
Multiplying the above by $\frac{\partial \hat{a}_0}{\partial \varphi_1}$ and integrating in $\varphi_1\in(-\infty,\infty)$ leads to
\[
\frac{1}{2r^2}\int_{-\infty}^{+\infty}\frac{\partial}{\partial \varphi_1} \left(\frac{\partial \hat{a}_0}{\partial \varphi_1}\right)^2~\text{d}\varphi_1
+\theta_1'(t)\int_{-\infty}^{+\infty}\left(\frac{\partial \hat{a}_0}{\partial \varphi_1}\right)^2~\text{d}\varphi_1
+ \int_{-\infty}^{+\infty}f(\hat{a}_0,b_0)\frac{\partial \hat{a}_0}{\partial \varphi_1} ~\text{d}\varphi_1
= 0.
\]
The first integral is zero 
\[
\frac{1}{2r^2}\int_{-\infty}^{+\infty}\frac{\partial}{\partial \varphi_1} \left(\frac{\partial \hat{a}_0}{\partial \varphi_1}\right)^2~\text{d}\varphi_1
=\frac{1}{2r^2} \left(\frac{\partial \hat{a}_0}{\partial \varphi_1}\right)^2\Big|_{\varphi_1=-\infty}^{\varphi_1=+\infty}=0,
\]
since $\hat{a}_0$ is constant at the limits of $\varphi_1$. Applying a change of variable  $s=\hat{a}_0(\varphi_1,t)$
the last integral can be written as
\[
\int_{-\infty}^{+\infty}f(\hat{a}_0,\bar{b}_0)\frac{\partial \hat{a}_0}{\partial \varphi_1} ~\text{d}\varphi_1
=
-\int_{a_1(\bar b_0)}^{a_3(\bar b_0)}f(\xi,\bar{b}_0) ~\text{d}\xi.
\]
Hence, finally the following equality holds
\begin{equation}\label{eq:theta'}
   \theta_1'(t)
=
\frac{\int_{a_1(\bar b_0)}^{a_3(\bar b_0)}f(\xi,\bar{b}_0) ~\text{d}\xi}
{\int_{-\infty}^{+\infty}\left(\frac{\partial \hat{a}_0}{\partial \varphi_1}\right)^2~\text{d}\varphi_1}.
\end{equation}

\begin{figure}[ht!]
    \centering
    \includegraphics[scale=0.5]{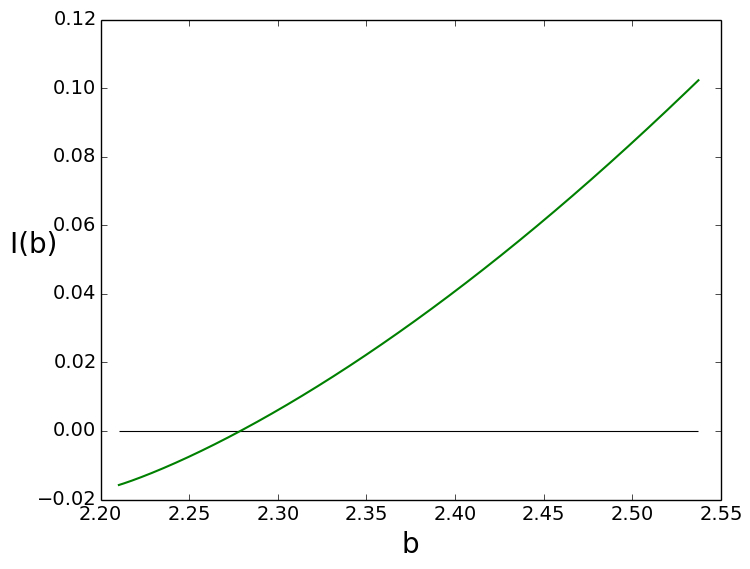}
    \caption{The integral $I(b)=\int_{a_1(b)}^{a_3(b)} f(a,b) \text{d}a $ with $f$ defined in \eqref{eq:f(a,b)_nondimensional} and parameter values in Table \ref{tab:parameters}. $I(b)$ increases in $[b_1, b_2]$ and has one zero $b_c\approx 2.28$, obtained by numerical estimation. The speed of the propagation of $a$ is related to a decreasing of the bulk component and stops when $b$ reaches the critical value $b_c$.}
    \label{fig:I(b)}
\end{figure}

As ${\int_{-\infty}^{+\infty}\left(\frac{\partial \hat{a}_0}{\partial \varphi_1}\right)^2d\varphi_1}>0$, the previous equality gives us an important information about the speed of the front, which moves with the same sign of the function 
\begin{equation}
I(b)={\int_{a_1(b)}^{a_3(b)}f(s,b) ~\text{d}s},
\end{equation}
which is represented in Figure \ref{fig:I(b)}.

{\corr We remark that $I(b)$ is an increasing function, since
\begin{align*}
I'(b) &= f(a_3(b), b)a_3'(b) - f(a_1(b), b)  a_1'(b) + \int_{a_1(b)}^{a_3(b)}\frac{\partial f(a,b)}{\partial b}~\text{d}a \\ &
=\int_{a_1(b)}^{a_3(b)}\left(k_0+\frac{\gamma a^2}{1 + a^2}\right)~\text{d}a>0
\end{align*}
given that $a_1(b)<a_3(b)$ and the parameters $k_0$ and $\gamma$ are positive. }The existence of a critical value $b_c$ such that $I(b_c)=0$ can be proven by showing that, for some $\varepsilon>0$, $I(b_1+\varepsilon)<0$ and  $I(b_2-\varepsilon)>0$, where $b_1$ and $b_2$ are the extremal values for the existence of three zeros $a(b)$ of $f(a,b)$. In fact when $b=b_1$ or $b=b_2$ the function {\corr $f(a,b)$} has only two roots, i.e. between the roots it is either entirely negative or entirely positive. If $b=b_1-\varepsilon$ or $b=b_2+\varepsilon$ then the integral is infinite. However as $I(b)$ is an increasing function, by continuity it follows that $I(b_1+\varepsilon)<0$ and  $I(b_2-\varepsilon)>0$. This shows the existence of the critical value $b_c$ and for \eqref{eq:theta'} we know that for $b>b_c$ then $a$ increases its high concentration region.

{\corr \textbf{Step c)}} {\corr We now prove that if $\theta_1$ 
increases, i.e. the high concentration peak for $a$ expands, then the quantity $b$ decreases all over the domain. Since  $a_1$ and $a_3$ are not constant, in principle propagation of $a$ does not necessarily imply an increment of its overall amount (which, by conservation of total mass \eqref{eq:conservation_mass} would have implied depletion of $b$). Therefore, we start rewriting \eqref{eq:conservation_mass} at the leading order of the asymptotic expansion as}
\[
M_0=\int_\Omega b_0 \dx + \int_{\Gamma} a_0~ \text{d}s +O(\varepsilon).
\]
At the previous step we have seen that $b_0$ is spatially homogeneously distributed and $a_0$ is approximately $a_3(b_0)$ if $|\theta| <\theta_1$ or $a_1(b_0)$ otherwise. Therefore we can rewrite the previous equation as
\begin{equation}\label{eq:asymptoticanalysis_conservation}
\pi r^2b_0(t) + 2\theta_1(t)r\; a_3(b_0) +2r\left( \pi-\theta_1(t)\right)a_1(b_0) + O(\varepsilon)=M_0.
\end{equation}
Discarding terms $O(\varepsilon)$ and differentiating (\ref{eq:asymptoticanalysis_conservation}) with respect to $t$ results in 
\[
\pi r^2b_0'(t) + 2r\theta_1'(t)a_3(b_0) + 2r\theta_1(t)a_3'(b_0)b_0'(t) + 2 r(\pi- \theta_1(t))a_1'(b_0)b_0'(t) - 2r\theta_1'(t)a_1(b_0) = 0,
\]
from which, rearranging terms leads to 
\begin{equation}\label{eq:asymptoticanalysis_b0'}
 b_0'(t)  =- 2\frac{ a_3(b_0)-a_1(b_0)}{\pi r^2 + 2\theta_1(t)r\;a_3'(b_0) + 2 r\left(\pi  - \theta_1(t)\right)a_1'(b_0)}\theta_1'(t){\corr r}.
\end{equation}

We now prove that the denominator in (\ref{eq:asymptoticanalysis_b0'}) is positive.
Let us differentiate with respect to $b$ the equation $f(a_{i}(b),b)=0$ for $i=1,3$
\begin{equation}\label{eq:dfdb}
0=\frac{d}{db}f(a_{i}(b),b)=  a_{i}'(b)\frac{\partial f}{\partial a}\Big|_{(a,b)=(a_{i}(b),b)}+ \frac{\partial f}{\partial b}\Big|_{(a,b)=(a_{i}(b),b)}.
\end{equation}
From which we get, if $\frac{\partial f}{\partial a}\Big|_{(a,b)=(a_{i}(b),b)}\neq 0$, that 
\begin{equation}\label{eq:a'(b)_sign}
a_{i}'(b)= - \left(\frac{\partial f}{\partial a}\Big|_{(a,b)=(a_{i}(b),b)}\right)^{-1}\frac{\partial f}{\partial b}\Big|_{(a,b)=(a_{i}(b),b)}.
\end{equation}
However, if the term in braces in \eqref{eq:a'(b)_sign} vanishes,  then from (\ref{eq:dfdb}), it needs to be that
\[
\frac{\partial f}{\partial b}\Big|_{(a,b)=(a_{i}(b),b)}=0,
\]
but this is not possible since
\[
\frac{\partial f}{\partial b}=\left(k_0+\frac{\gamma a^2}{1+a^2}\right)>0, \quad \forall a.
\]
Hence, using (\ref{eq:f(a,b)_bistability}) 
in (\ref{eq:a'(b)_sign}), we conclude that
\begin{equation}\label{eq:asymptoticanalysis_a>0}
a_1'(b)>0, \quad \text{and} \quad a_3'(b)>0.
\end{equation}
From \eqref{eq:asymptoticanalysis_a>0}, it is now clear the positiveness of the denominator in (\ref{eq:asymptoticanalysis_b0'}), while the sign of the numerator of \eqref{eq:asymptoticanalysis_b0'} is the opposite of the sign of $\theta_1'(t)$: if $\theta_1'(t)>0$ then $b_0'(t)<0$ and vice-versa. This finally proves that the propagation of active GTPase $a$ over the boundary is related to a decreasing of the bulk component $b$.

{\corr \textbf{Step d)}} In order to achieve polarisation, the propagation needs to stop, i.e. at a certain time $\bar{t}$, $\theta_1'(\bar{t})=0$ and this happens when $b(t)$ reaches a minimum value $b_c$. Therefore, ignoring terms of order $O(\varepsilon)$ we have 
\[
M_0 = \pi r^2b_c + 2\theta_1(\bar{t}) r \;a_3(b_c)+2r(\pi -\theta_1(\bar{t}))a_1(b_c).
\]
We rewrite it in the form
\[
M_0=\pi r^2b_c + 2r\theta_1(\bar{t})\big(a_3(b_c)-a_1(b_c)\big)+2\pi r~ a_1(b_c).
\]
Since we require $0<\theta_1<\pi$ then
\[
M_0
<\pi r^2b_c + 2\pi r\big(a_3(b_c)-a_1(b_c)\big)+2\pi r ~a_1(b_c)
=\pi r^2b_c + 2\pi r ~a_3(b_c)
\]
and 
\[
M_0
>\pi r^2b_c +2\pi r ~a_1(b_c).
\]
We therefore have found a condition on $M_0$ equivalent to the classical wave pinning model \cite{Mori2011}. To have pinning we need to take an initial value $b_0>b_c$ and $a_0$ such that
\begin{equation}\label{eq:asympt_analysis_condition_M_0}
m_1
<{M_0}
<m_2,
\end{equation}
where
the quantity $m_1:=\pi r^2 b_c + {2 \pi r}~a_1(b_c)$ represents the total mass at the equilibrium with the lowest active GTPase, while the quantity 
$m_3:=\pi r^2 b_c + {2 \pi r}~a_3(b_c)$ represents the total mass at the equilibrium where the whole membrane has been activated, with no pinning taking place. 
In order to have a heterogeneous steady state for $a$, \textit{i.e.} obtain a pinned active GTPase propagation state, the total amount $M_0$ of GTPase should not be neither too low nor too high.

\section{Bistability and polarisation}\label{sec:bistabiliy_LPA}
In this section we are interested in mapping parameter regions for all possible different behaviors of the {\corr two- and three-dimensional} BSWP model \eqref{eq:model_b}-\eqref{eq:model_a} in order to get some insights on the role of geometry. Indeed, depending on the parameters, the model is able to generate different responses, for example it supports spatial homogeneous solutions. We will start from this point, analysing the role of the reactions in the system. In a second step, we will use an approximated nonlinear analysis in order to identify the spatial responses of the BSWP model with respect to small perturbations of the boundary component from the spatially homogeneous state. {\corr We remark that the following analysis is basically independent of the spatial dimension.}

\subsection{Well mixed model} Integrating equation (\ref{eq:model_b}) in $\Omega$ and applying the divergence theorem with (\ref{eq:model_b_Gamma}), we get 
\[
\int_\Omega \frac{\partial b}{\partial t} \dx = -\int_{\Gamma} f(a,b)\ds.
\]
Since we want to consider spatial homogeneous solutions, this corresponds to
\[
\frac{\partial b_g}{\partial t}\int_\Omega 1 \dx = -f(a_g,b_g)\int_{\Gamma} 1 \ds.
\]
Finally, we will analyse the so-called \textit{well mixed system} defined by
\begin{align}
&\frac{\text{d} a_g }{\text{d} t} = f(a_g,b_g),\label{eq:wellmixed_ag}\\
&{\omega}\frac{\text{d} b_g }{\text{d} t} = -  f(a_g, b_g), \label{eq:wellmixed_bg}
\end{align}
where we recall that $\omega=|\Omega|/|\Gamma|$ denotes a parameter describing the geometry of the domain. Given that $\omega$ has unit length makes the above system unit dimensionally consistent (see also Table \ref{tab:parameters}).
We  note that the following quantity is conserved
\[
\frac{a_g(t)}{\omega} + b_g(t)=  \frac{a_g(0)}{\omega} + b_g(0),
\]
which can be interpreted as a scaled total concentration. Indeed, it follows from \eqref{eq:conservation_mass} that
\[
M_0 =
\int_{\Gamma} a(\mathbf{x}, t) \ds + 
\int_{\Omega} b(\mathbf{x}, t) \dx = |\Gamma| a_g(t) + |\Omega| b_g(t) = 
|\Omega| \left( \frac{a_g(t)}{\omega} +  b_g(t)\right).
\]
The analysis of \eqref{eq:wellmixed_ag}-\eqref{eq:wellmixed_bg} reduces to the single equation
\begin{equation}\label{eq:wellmixed_singleODE}
\frac{\text{d} a_g }{\text{d} t} = f\left( a_g,m_0- \frac{a_g}{\omega}\right),
\end{equation}
where $m_0:=\frac{M_0}{|\Omega|}$. From the study of the steady states, $ f\left( a_g,m_0-\frac{a_g}{\omega}\right) = 0$ is a third degree polynomial in {\corr $a_g$} and, by the Descartes' rule of signs, it can be shown that it has either one or three positive real roots. Therefore{\corr, from the negativity of the leading order coefficient,} it follows that there exists either a single stable steady state or 3 steady states where the outer two are stable. Bistability corresponds to the co-existence of high and low GTPase activities at the cell membrane. When only a single steady state is possible, then the well mixed model admits only one response between low and high activities. 

The responses of the model for different values of the parameters $m_0$ and $\gamma$ are shown in Figure \ref{fig:bifurcations_LPA_well_mixed}, where the bistability region is indicated by the blue color, and the white and red areas indicate existence of a unique steady state for \eqref{eq:wellmixed_singleODE}. 

\subsection{Local perturbation analysis}
\begin{figure}[ht]
    \centering
    \includegraphics[scale=0.22]{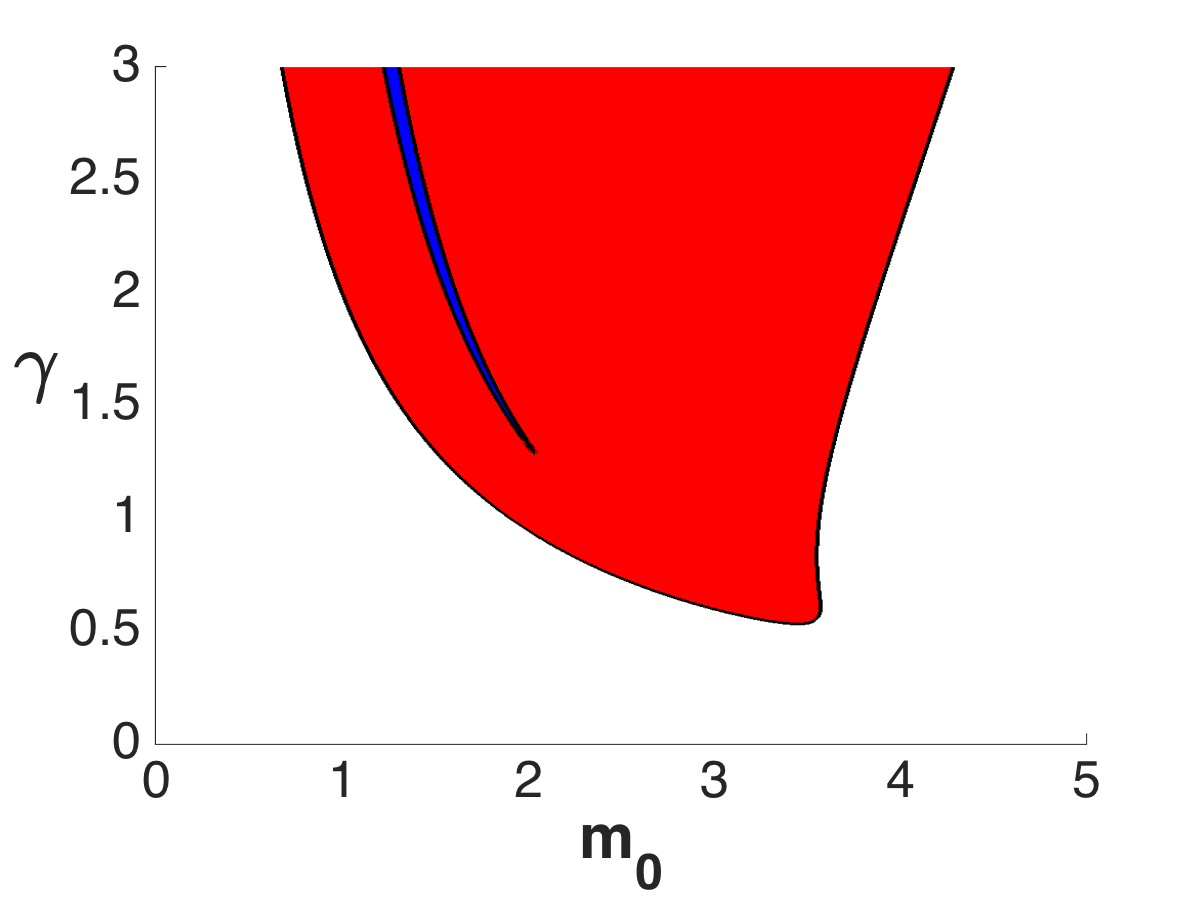}
    \includegraphics[scale=0.22]{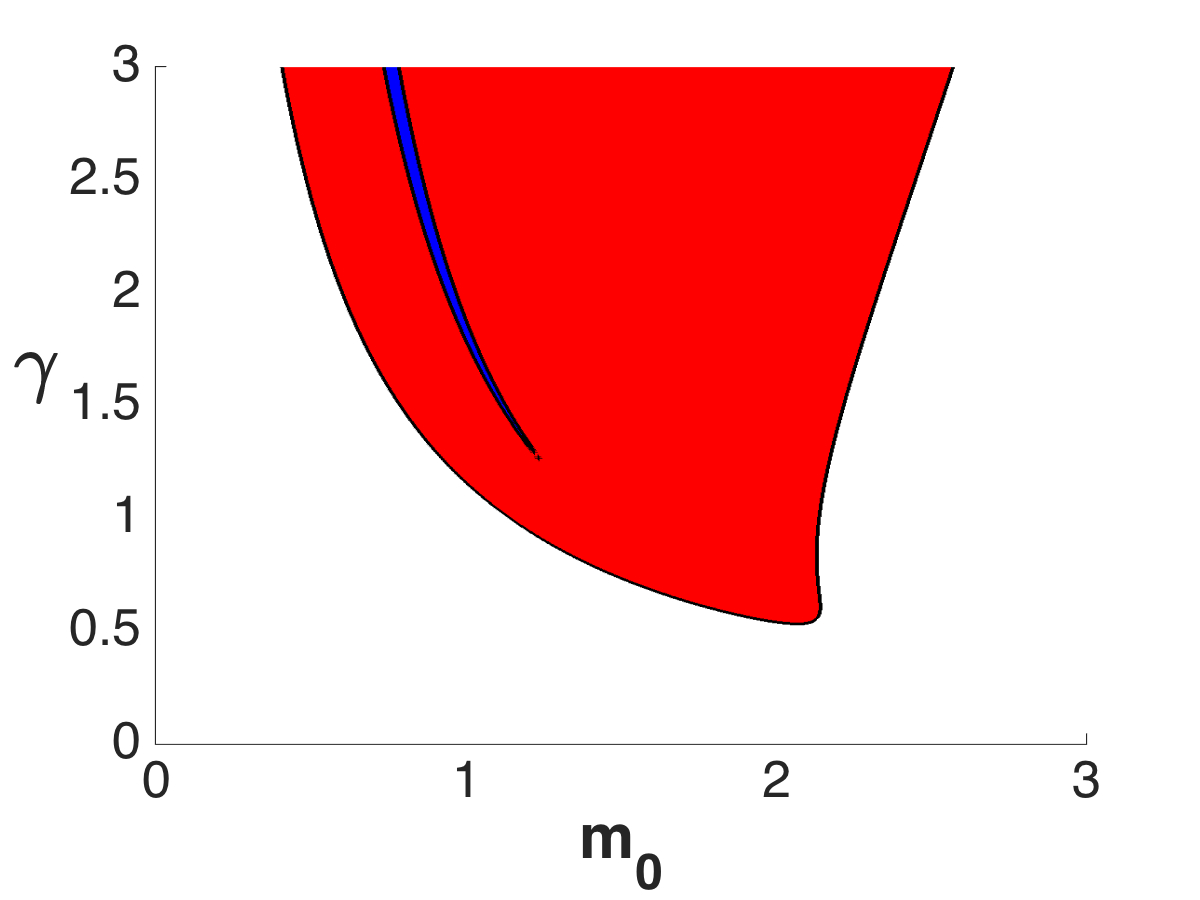}
    \includegraphics[scale=0.22]{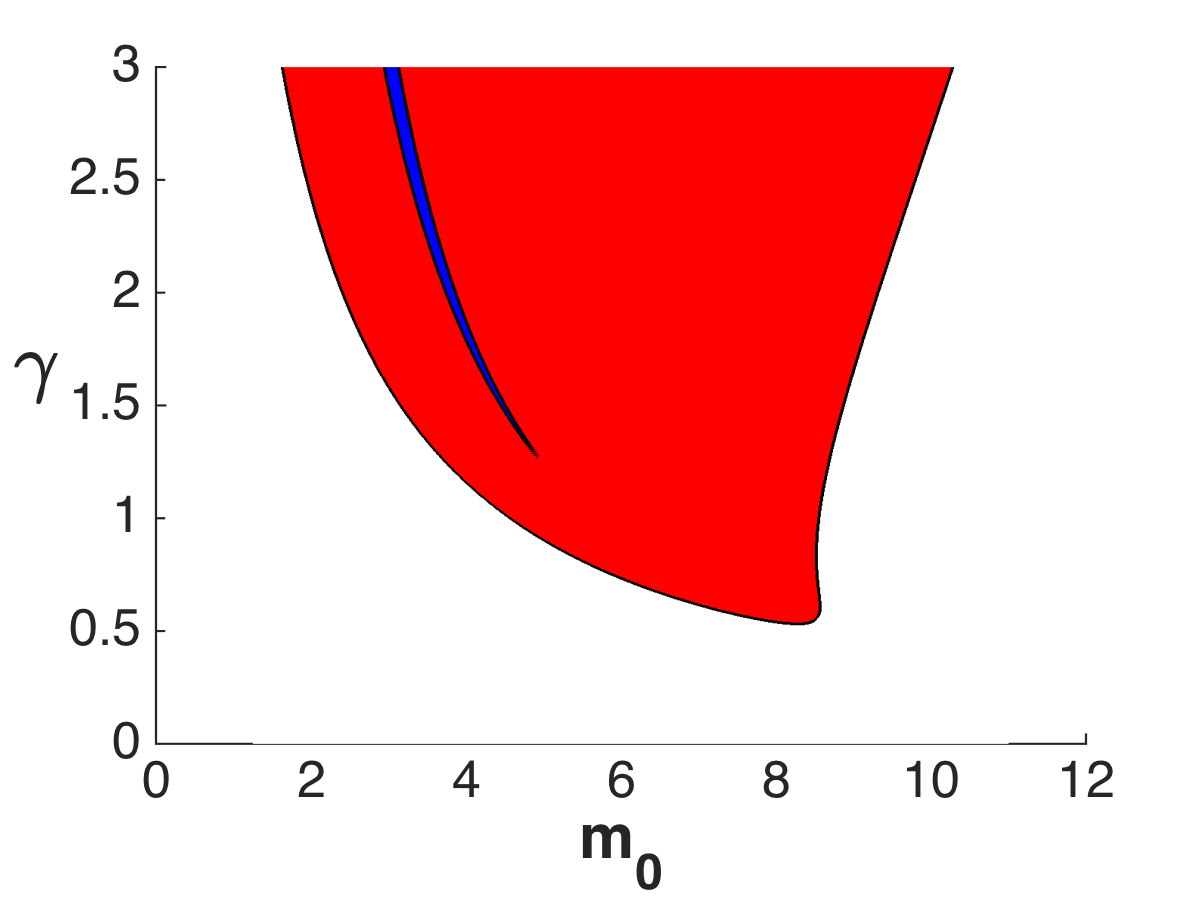}
    \caption{Bistability (blue) and polarity (red and blue) regions for different values of the parameter $\omega=|\Gamma|/|\Omega|$. On the $x$-axis we vary the total mass per unit volume $m_0=M_0/|\Omega|$. On the $y$-axis the activation rate $\gamma$ of Rho-GTPase positive feedback is varied. The blue region defines the parameter region in which all the possible responses (uniform high activity, uniform low activity or polarisation) can take place.
    Note that for small values of the positive-feedback rate $\gamma$ no polarisation is possible and the stronger the feedback, the bigger can be the total initial concentration. 
    From left to right: 
    $\omega = 1~\mu$m, which can be generated taking $\Omega$ as a sphere of radius $3~\mu$m; $\omega = 1.6~\mu$m which correspond to the choice of a sphere of radius of $5~\mu$m, as in \cite{Mori2008}; $\omega=0.42~\mu$m which corresponds to a non-spherical domain having a surface 4 times bigger than the one of a sphere of radius 5$~\mu$m but with  same volume. While we show qualitatively similar results, we also highlight the role of $\omega$: increments of the value $\omega$ (which might be due to an increase of the surface area) cause an increase of the bistability and polarisation areas. Although the three figures look almost the same, note the differences in the horizontal scales.
    }
    \label{fig:bifurcations_LPA_well_mixed}
\end{figure}

Local perturbation analysis (LPA) is a convenient tool  that can be very useful in understanding how a local perturbation might affect some classes of reaction-diffusion systems with fast and slow components. We refer the interested reader to \cite{Holmes2014,Holmes2016, Holmes2015} for more details and the LPA. The basic idea is the following: let system \eqref{eq:model_b}-\eqref{eq:model_a} possess a spatially homogeneous profile $(b_g(t), a_g(t))$ and apply a narrow and well localised perturbation to the slow-diffusive component $a$, such as defined by equation {\corr \eqref{eq:asympt_analysis_initial_condition_a}}. 
Based on the fact that we have a fast and a slow variable ($D_b>>D_a$) we consider the limits $D_b\to\infty$ and $D_a\to 0$. Therefore $b$ maintains a global spatial uniform profile $b_G(t)$.
On the other hand $a(\mathbf{x},t)$ has a global spatial uniform profile $a_g$ in most of the cell membrane, except in the narrow area where the perturbation $a_p$ is applied. Considering the limit $D_a\to 0$, the perturbation $a_p$ does not influence through diffusion the baseline level $a_g$ and, given its small mass, it does neither substantially influence $b$. 
In these terms it is possible to consider  $a_p(t)$ and $a_g(t)$ as different entities to obtain the following ODE system
\begin{align}
&\frac{\text{d} a_p  }{\text{d} t} = f\left(a_p,b_g\right),\label{eq:LPA_aP}\\
&\frac{\text{d} a_g  }{\text{d} t} = f\left(a_g,b_g\right),\label{eq:LPA_aG}\\
& \omega\frac{\text{d} b_g }{\text{d} t} = - f\left(a_g, b_g\right). \label{eq:LPA_b}
\end{align}
It can be easily shown using conservation that the above system can be reduced to the following system
\begin{align}
&\frac{\text{d} a_p  }{\text{d} t} = f\left(a_p, m_0 - \frac{a_g}{\omega}\right),\\
&\frac{\text{d} a_g  }{\text{d} t} = f\left(a_g, m_0 - \frac{a_g}{\omega}\right).
\end{align}
The above ODE system indicates that steady states for $a_p$ might differ from the steady states for $a_g$. Indeed, we interpret this case as the polarisation response: the perturbation has affected the system and two states on the boundary are simultaneously present, with a localised high activity and low activity elsewhere.
Using this analysis and numerical calculations, we obtain the polarisation region in the parameter plane $m_0-\gamma$, which is shown in red and blue color in Figure \ref{fig:bifurcations_LPA_well_mixed}. 

We have calculated the bistability and the polarisation regions for different values of $\omega$, obtaining qualitatively identical results. However, the regions increase their sizes with decreasing $\omega$. {\corr For the three-dimensional case,} for a given volume $|\Omega|$, $\max_{\Gamma}\omega=r/3$ where $r$ is the radius of the sphere enclosing that volume. Therefore, having a fixed volume, the more the surface increases, the smaller $\omega$ becomes.
This is an interesting result which suggests that changes in shapes and increases in the cell surface relative to its volume enhance the possibility of achieving polarisation. 
Indeed a key feature of cell migration is the change in cell shape \cite{Reig2014}. 

In \cite{Holmes2016}  the same analysis was done for the model \eqref{eq:model_Mori_a}-\eqref{eq:model_Mori_f(a,b)} where $a$ and $b$ are defined on the same unidimensional spatial domain. They derive a well mixed and LPA system which is a special case of our models \eqref{eq:wellmixed_bg}-\eqref{eq:wellmixed_ag} and \eqref{eq:LPA_aP}-\eqref{eq:LPA_b} when $\omega=1$. They initially use a sharp switch approximation for the reaction \eqref{eq:f(a,b)} (passing to the limit as $n\to\infty$) in order to be able to calculate the steady states analytically. Then they numerically calculate the bistability and polarisation regions for \eqref{eq:f(a,b)} with $n=4$. Our results, when $n=2$, are totally in line with their work and suggests that the bulk-surface framework maintains and extends the features of the original wave pinning model \eqref{eq:model_Mori_a}-\eqref{eq:model_Mori_b}.

\section{The bulk-surface finite element method} \label{sec:BSFEM}
Next, we present the bulk-surface finite element method (BS-FEM) \cite{Madzvamuse2016a} which we adopt to solve the BSWP model \eqref{eq:model_b}-\eqref{eq:model_a}. The basic idea is to describe the model numerically by systems of linear equations, which are easy to solve. In order to do this, we first describe the BSWP model using a weaker formulation, for which the regularity requirements are more flexible. 
In a second step we discretise the spatial and temporal domains. This allows us to finally derive the systems of linear equations. 

\subsection{Weak formulation}
We will use the following notation: for $D\subset\mathbb{R}^d$ we indicate with $H^1(D)$ and $H^{-1}(D)$ respectively the Sobolev space and its dual, see   \cite{Evans10} for definitions and theory.
$X$ being a Banach space we can define
\[
L^2([0,T];X):=
\Bigg\{u:[0,T]\to X \text{ s.t. } \int_0^T  ||u||_X^2 \;dt
<\infty  \Bigg\}.
\]  
In the following we will also use the dot notation to indicate the (temporal) derivative.
The weak formulation of the BSWP model \eqref{eq:model_b}-\eqref{eq:model_a}  reads: 
	{find $a\in L^2(0,T;H^1(\Gamma))$ with $\dot{a}\in L^2(0,T;H^{-1}(\Gamma))$ and $b\in L^2(0,T;H^1(\Omega))$ with $\dot{b}\in L^2(0,T;H^{-1}(\Omega))$
	such that
		\begin{align}
		&\int_\Gamma \dot{a} w \ds 
		+D_a\int_{\Gamma}\nabla_\Gamma a\cdot \nabla_\Gamma w \ds=
		\int_{\Gamma}f(a,b)w\ds, \label{eq:model_weakformulation_a}\\
		&\int_\Omega \dot{b} v  \;\text{d}\mathbf{x}
		+D_b\int_\Omega\nabla b\cdot\nabla v \;\dx=
		-\int_{\Gamma}g(a)b \;  v  \ds
		+\int_{\Gamma} \beta a\;  v  \ds
		,\label{eq:model_weakformulation_b}
		\end{align}
		$\forall w \in H^1(\Gamma)$ and $\forall  v \in H^1(\Omega)$. In equation \eqref{eq:model_weakformulation_b} we have introduced the function  $g(a):=\omega  \Big(k_0+\frac{\gamma a^2}{K^2+a^2}\Big)$.

\subsection{Spatial discretisation}
We consider a closed polyhedral approximation $\Omega_h$ of $\Omega$ and define a mesh over it, i.e. we find a suitable set $\Tau_h=\{T_1, ...,T_{N_\Tau} \}$ such that
$
\Omega_h = \bigcup_{i=1}^{N_\Tau} T_i,
$ 
where each $T_i$ is a tetrahedron, such that for any  $i\neq j$ we have
$\overset{\circ}{T}_i\cap\overset{\circ}{T}_j=\emptyset$ and if ${T}_i\cap{T}_j\neq\emptyset$ then the intersection is either a common face, side or vertex of the two elements.
As well, we approximate $\Gamma$ with $\Gamma_h:=\partial \Omega_h$. A natural mesh $\Sau_h$ for $\Gamma_h$ can be easily deduced from the bulk mesh $\Tau_h$. Indeed, the boundary of $\Omega_h$ is discretised by the external faces of some tetrahedra of $\Tau_h$. These faces, which are triangles, compose $\Sau_h$. We indicate with $N_h$ to represent the number of vertices in the mesh $\Tau_h$ and with $\hat{N}_h$ the number of vertices in $\Sau_h$.
The definition of the two meshes $\Tau_h$ and $\Sau_h$  and their compatibility is a crucial point for the bulk-surface finite element method.

Let now $\mathbb{P}_1(D)$ be the space of first degree polynomials over a set $D\subset\mathbb{R}^d$ and we define the following function spaces
\begin{align*}
V_h(\Omega_h) & := \left\{
v:\Omega_h\to\mathbb{R} ~:~v\in C^0(\Omega_h), v|_T \in \mathbb{P}_{1}(K), ~\forall T\in \Tau_h
\right\},\\
W_h(\Gamma_h) & := \left\{
w:\Gamma_h\to\mathbb{R} ~:~w\in C^0(\Gamma_h), w|_S \in \mathbb{P}_{1}(K), ~\forall S\in \Sau_h\right\},
\end{align*}
which are subsets, respectively, of $H^1(\Omega_h)$ and $H^1(\Gamma_h)$. The semi-discrete weak formulation therefore reads: find $a_h\in L^2([0,T];W_h(\Gamma_h))$ with $\dot{a_h}\in L^2([0,T];W_h(\Gamma_h))$ and $b_h\in L^2([0,T];V_h(\Omega_h))$ with $\dot{b}\in L^2([0,T];V_h(\Omega_h))$ such that
	\begin{align}
	&\int_{\Gamma_h}\dot{a_h} w_h \ds 
	+D_a\int_{\Gamma_h}\nabla_\Gamma a_h\cdot \nabla_\Gamma w_h \ds=
	\int_{\Gamma_h}f(a_h,b_h)w_h\ds  \label{eq:model_discreteweakformulation_a},\\
	&\int_{\Omega_h}\dot{b_h} v_h  \;\text{d}\mathbf{x}
	+D_b\int_{\Omega_h}\nabla b_h\cdot\nabla v_h \;\text{d}\mathbf{x}=
	-\int_{\Gamma_h}g(a_h)b_h \;  v_h  \ds
	+\int_{\Gamma_h} \beta a_h\;  v_h  \ds
	,\label{eq:model_discreteweakformulation_b}
	\end{align}
	$\forall w_h \in W_h(\Gamma_h)$ and $\forall  v_h \in V_h(\Omega_h)$. 
	
A basis for $W_h(\Gamma_h)$ is the set of the hat functions $\psi_i\in W_h(\Gamma_h)$ with the property that 
$\psi_i(\mathbf{x}_j)=\delta_{i,j}$ for any vertex $\mathbf{x}_j$ of $\Sau_h$ and $\forall i,j=1, ..., \hat{N}_h$.
As well, we denote with  $\varphi_1, \cdots, \varphi_{N_h}$ the hat functions on $\Tau_h$, which generate a basis of $V_h(\Omega_h)$.
Therefore we seek solutions of the form
\[
a_h(\mathbf{x},t) = \sum_{i=1}^{\hat{N}_h} a_h(\mathbf{x}_i, t) \psi_i(\mathbf{x})
\text{ and }
b_h(\mathbf{x},t) = \sum_{i=1}^{N_h} b_h(\mathbf{x}_i, t) \varphi_i(\mathbf{x}).
\]
In terms of the basis functions, the problem (\ref{eq:model_discreteweakformulation_a})-(\ref{eq:model_discreteweakformulation_b})
is equivalent to the following system of ODEs
\begin{align}
{\text M}_{\Gamma_h}\dot{\bm{a}} + D_a {\text K}_{\Gamma_h}\bm{a} = F(\bm{a}, \bm{b}), \label{eq:Galerkin_a}\\
{\text M}_{\Omega_h}\dot{\bm{b}} + D_b {\text K}_{\Omega_h}\bm{b} + G(\bm{a})\bm{b}= \beta H \bm{a}, \label{eq:Galerkin_b}
\end{align}
where
\begin{align*}
 \bm{a} & = \Bigg(a_h(\mathbf{x_i}, t)\Bigg)_{i=1, \cdots, \hat{N}_h},\; 
 \bm{b} =  \Bigg(b_h(\mathbf{x_i}, t)\Bigg)_{i=1, \cdots, N_h},\;
 {\text M}_{\Gamma_h}=\left(\int_{\Gamma_h} \psi_j\psi_i \ds\right)_{i,j=1, ..., \hat{N}_h}, \\
 {\text K}_{\Gamma_h}& =\left(\int_{\Gamma_h} \nabla_\Gamma \psi_j\cdot\nabla_\Gamma \psi_i \ds\right)_{i,j=1, ..., \hat{N}_h}, \; 
 F(\bm{a,b}) = \left(\int_{\Gamma_h} f({\corr a_h,b_h})\psi_i \ds
\right)_{i=1, ..., \hat{N}_h},\\
 {\text M}_{\Omega_h}& =\left(\int_{\Omega_h} \varphi_j\varphi_i \dx\right)_{i,j=1, ..., N_h}, \;
 {\text K}_{\Omega_h}=\left(\int_{\Omega_h}\nabla\varphi_j\cdot\nabla \varphi_i \dx\right)_{i,j=1, ..., N_h}, \\
 G(\bm{a}) & = \left(\int_{\Gamma_h}g({\corr a_h})
\varphi_j\varphi_i\; \ds
\right)_{i,j=1, \cdots, \hat{N}_h}\; \text{and} \; 
 H = \left(\int_{\Gamma_h}
\psi_j\psi_i\; \ds
\right)_{\substack{{i=1, \cdots, \hat{N}_h}\\{j = 1, \cdots, \hat{N}_h}}}.
\end{align*}

\subsection{Temporal discretisation}
We discretise the time interval $[0,T]$ uniformly with $N_t\in\mathbb{N}$ time points, corresponding to choosing a time step $\tau_h=\frac{T}{N_t}$. We  define 
\[
t^{n}= t^{n-1}+\tau_h, \text{ or equivalently } t^{n}=n\tau_h , \quad n=1,\cdots,N_t, 
\]
with $t^0=0$. We will indicate the solutions at discrete time $t^n$ with $\bm{a}^{n}$ and $\bm{b}^{n}$. We use a predictor-corrector finite difference method to approximate the time-derivatives (see for example \cite{Macdonald2016}). To calculate the solution at each time point, we follow the steps outlined below.
\begin{enumerate}
	\item We predict a solution  $\tilde{\bm a}^{n}$ for the surface component using the IMEX method (diffusion IMplicit, reaction EXplicit)
	\begin{align}
	\left(M_{\Gamma}+ {\corr \tau_h} D_a K_{\Gamma}\right)\tilde{\bm{a}}^{n} = M_{\Gamma}\bm{a}^{n-1} + {\corr \tau_h} F(\bm{a}^{n-1}, \bm{b}^{n-1}) 
	\label{eq:PC_prediction_a}.
	\end{align}
	
	\item We calculate the solution $\bm{b}^{n}$ using Crank-Nicholson time discretisation and the predicted solution $\tilde{\bm{a}}^{n}$ 
	\begin{align}
	&\left(M_{\Omega}+ \frac{1}{2}{\corr \tau_h} D_b K_{\Omega} 
	+ \frac{1}{2}{\corr \tau_h} G(\tilde{\bm{a}}^{n}) \right)\bm{b}^n \nonumber\\
	&= 
	M_{\Omega}\bm{b}^{n-1} 
	-\left( \frac{1}{2}{\corr \tau_h} D_b K_{\Omega} 
	- \frac{1}{2}{\corr \tau_h} {\corr G(\tilde{\bm{a}}^{n})} \right)\bm{b}^{n-1}
	+ \frac{1}{2}{\corr \tau_h}\beta H \tilde{\bm{a}}^{n} 
	+ \frac{1}{2}{\corr \tau_h}\beta H \bm{a}^{n-1}. \label{eq:PC_b}
	\end{align}
	
	\item Using the predicted $ \tilde{\bm{a}}^{n}$ and ${\bm b}^{n}$, we  correct the predicted solution for $\tilde{\bm a}^{n}$ using the Crank-Nicholson scheme
	\begin{align}
\hspace{-1cm}	\left(M_{\Gamma}+ \frac{1}{2}{\corr \tau_h} D_a K_{\Gamma}\right){\bm{a}}^{n} = M_{\Gamma}\bm{a}^{n-1} 
	- \frac{1}{2}{\corr \tau_h} D_a K_{\Gamma} \tilde{\bm{a}}^{n}
	+ \frac{1}{2}{\corr \tau_h} F(\tilde{\bm{a}}^{n}, \bm{b}^{n}) 
	+ \frac{1}{2}{\corr \tau_h} F(\bm{a}^{n-1}, \bm{b}^{n-1}).
	\label{eq:PC_correction_a}
	\end{align} 
\end{enumerate}
{\corr The method is second order accurate in time \cite{Quarteroni2010}, and moreover the following property holds.}
\begin{proposition}
 The numerical method (\ref{eq:PC_prediction_a})-(\ref{eq:PC_correction_a}) is conservative, i.e. 
 \[
 \int_{\Omega_h} b_h(\mathbf{x}, t^n) \dx + \int_{\Gamma_h} a_h(\mathbf{x}, t^n) \ds = \int_{\Omega_h} b_h(\mathbf{x}, 0) \dx + \int_{\Gamma_h} a_h(\mathbf{x}, 0) \ds, \quad \forall n = 1, ..., N_t.
 \]
\end{proposition}

\begin{proof}
It is sufficient to sum over the rows of each one of the three systems (\ref{eq:PC_prediction_a}), (\ref{eq:PC_b}) and (\ref{eq:PC_correction_a}). One obtains three different equations in which the property of the basis functions 
\[
\sum_{i=1}^{M_h}\psi_i(\mathbf{x})=1
\text{ and }
\sum_{i=1}^{N_h}\varphi_i(\mathbf{x})=1,
\]
is exploited to simplify the calculations. Summing the three equations together it is easy to see that
\[
\int_{\Omega_h} b_h(\mathbf{x}, t^n) \dx + \int_{\Gamma_h} a_h(\mathbf{x}, t^n) \ds =
\int_{\Omega_h} b_h(\mathbf{x}, t^{n-1}) \dx + \int_{\Gamma_h} a_h(\mathbf{x}, t^{n-1}) \ds.
\]
An iterative procedure leads  to the complete proof of the Proposition. 
\end{proof}
\noindent {\corr Details on the implementation of the numerical algorithm for the BS-FEM are given in \ref{Appendix:computations}.}

\section{Results}\label{sec:Results}
In this section we present some simulations on three different domains: a sphere, a capsule and a complex domain, caricature of a polarised fibroblast. In all the simulations except for last one, we set the initial conditions as follows: referring to Proposition \ref{prop:roots_f(a,b)},
the bulk component is spatially homogeneous with value 
\begin{equation}\label{eq:simulations_initial_condition_b}
\overline{b_0} = b_2 - \varepsilon_b(b_2-b_1),
\end{equation}
with $\varepsilon_b<1$ such that $\overline{b_0}>b_c$, where $b_c$ is the only zero of $I(b)$ in \eqref{eq:I(b)}.
For the surface component, we superimpose a narrow Gaussian function with magnitude $a_{p}=(a_2+a_3)/2$ on a spatially homogeneous profile with magnitude $a_{g}=(a_1+a_2)/2$, where $a_1$, $a_2$, $a_3$ are the solutions of $f(a,\overline{b_0})=0$, i.e.
\begin{equation}\label{eq:simulations_initial_condition_a}
a_{in} = a_{g}+a_{p}\, exp{\left(-\frac{(x-x_0)^2+(y-y_0)^2+(z-z_0)^2}{\sigma^2}\right)}
\end{equation}
where $(x_0,y_0,z_0)$ is the centre of the perturbation. In case of two perturbation peaks with centres $(x_0,y_0,z_0)$ and $(x_1,y_1,z_1)$, we impose the following initial condition
\begin{align}\label{eq:simulations_initial_condition_a_2peaks}
a_{in} = a_{g} & +a_{p}\, exp{\left(-\frac{(x-x_0)^2+(y-y_0)^2+(z-z_0)^2}{\sigma^2}\right)} \notag \\
& +a_{p}\, exp{\left(-\frac{(x-x_1)^2+(y-y_1)^2+(z-z_1)^2}{\sigma^2}\right)}.
\end{align}
{\corr  The following simulations present a variety of choices for the parameters $\varepsilon_b$, $\sigma^2$ as well as for the centre of the perturbations. Although these parameters do not play a fundamental role in the qualitative behavior of the solutions, here we show a selection of our most significant results.
}

\subsection{Sphere} Our first three-dimensional geometry on which we  solve the BSWP model \eqref{eq:model_b}-\eqref{eq:model_a} is the sphere which is the simplest possible three-dimensional shape. We consider a radius of 5$\mu$m, which is the radius used in the simulations of the WP model \cite{Mori2008}. We consider $\varepsilon_b=0.154$ in \eqref{eq:simulations_initial_condition_b} and $\sigma^2=0.5\mu$m$^2$ in \eqref{eq:simulations_initial_condition_a}. 
The perturbation of the homogeneous state is strong enough to trigger polarisation: from this small region, a propagative activation is started in all directions. This will be finally pinned in about 100 seconds, resulting in a stable active area. In Figure \ref{fig:simulation_sphere_solution} we show the evolution of $a$ and in Figure \ref{fig:simulation_sphere_masses} the temporal evolution of the masses of $a$ and $b$ which become constants when the propagation gets pinned.
\begin{figure}[ht]
	\includegraphics[width=.3\linewidth]{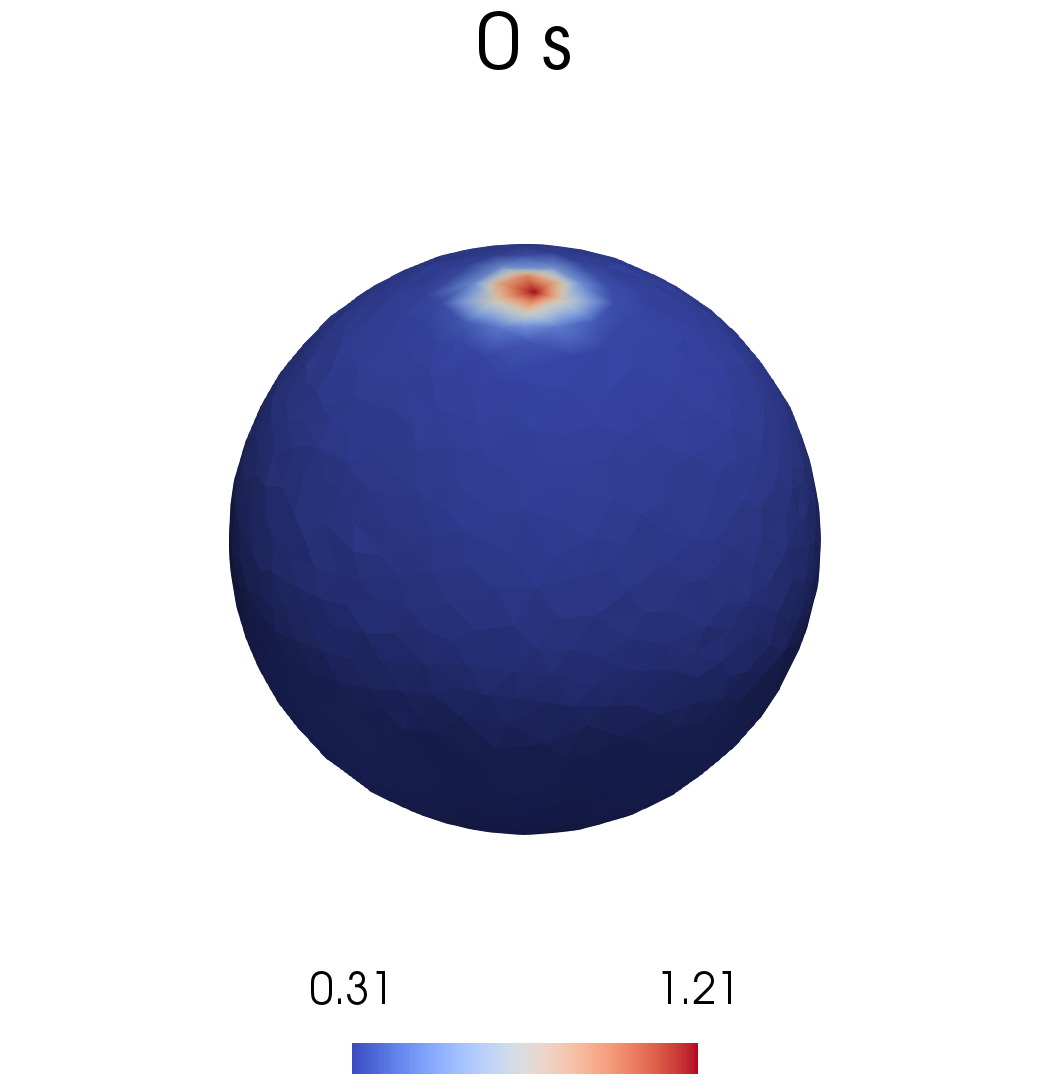}	\includegraphics[width=.3\linewidth]{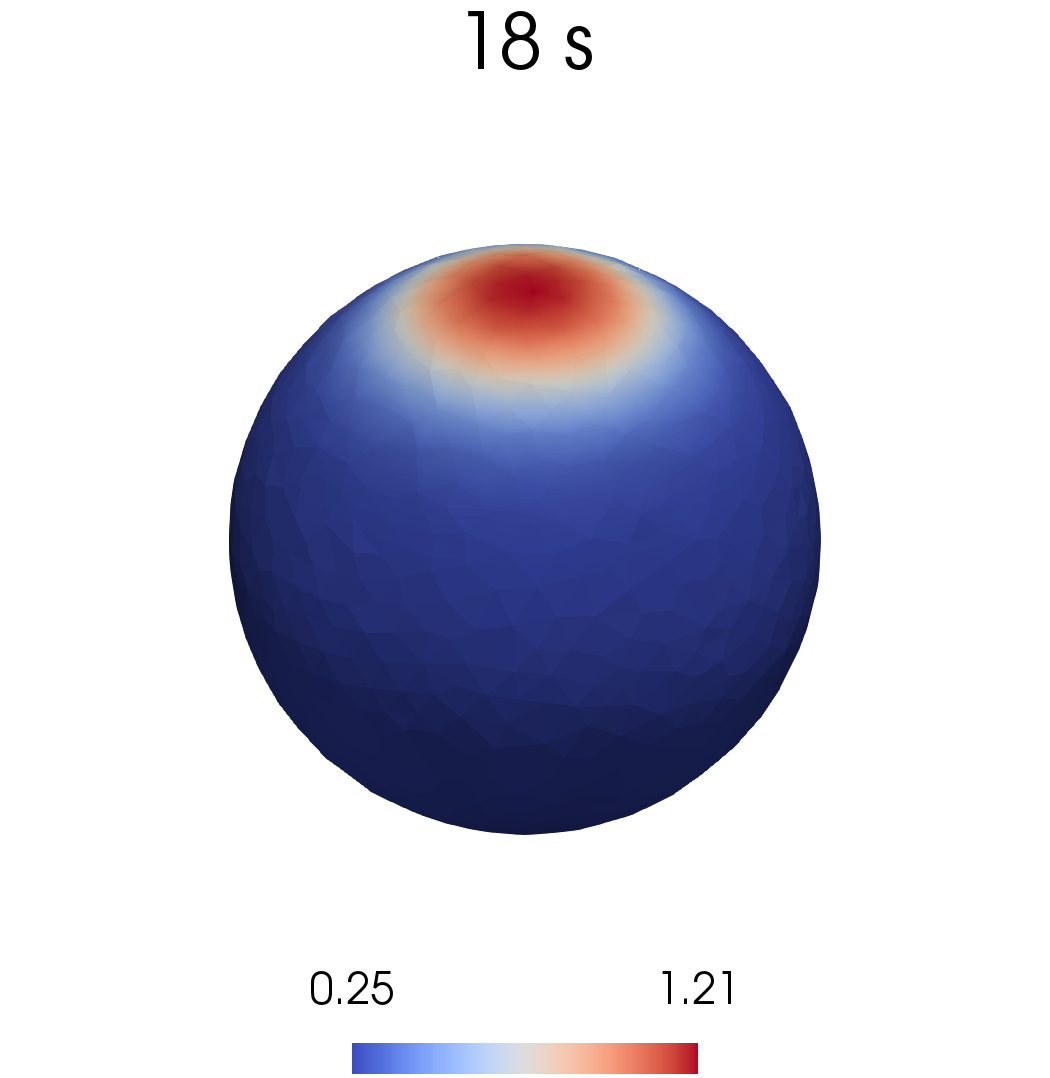}	\includegraphics[width=.3\linewidth]{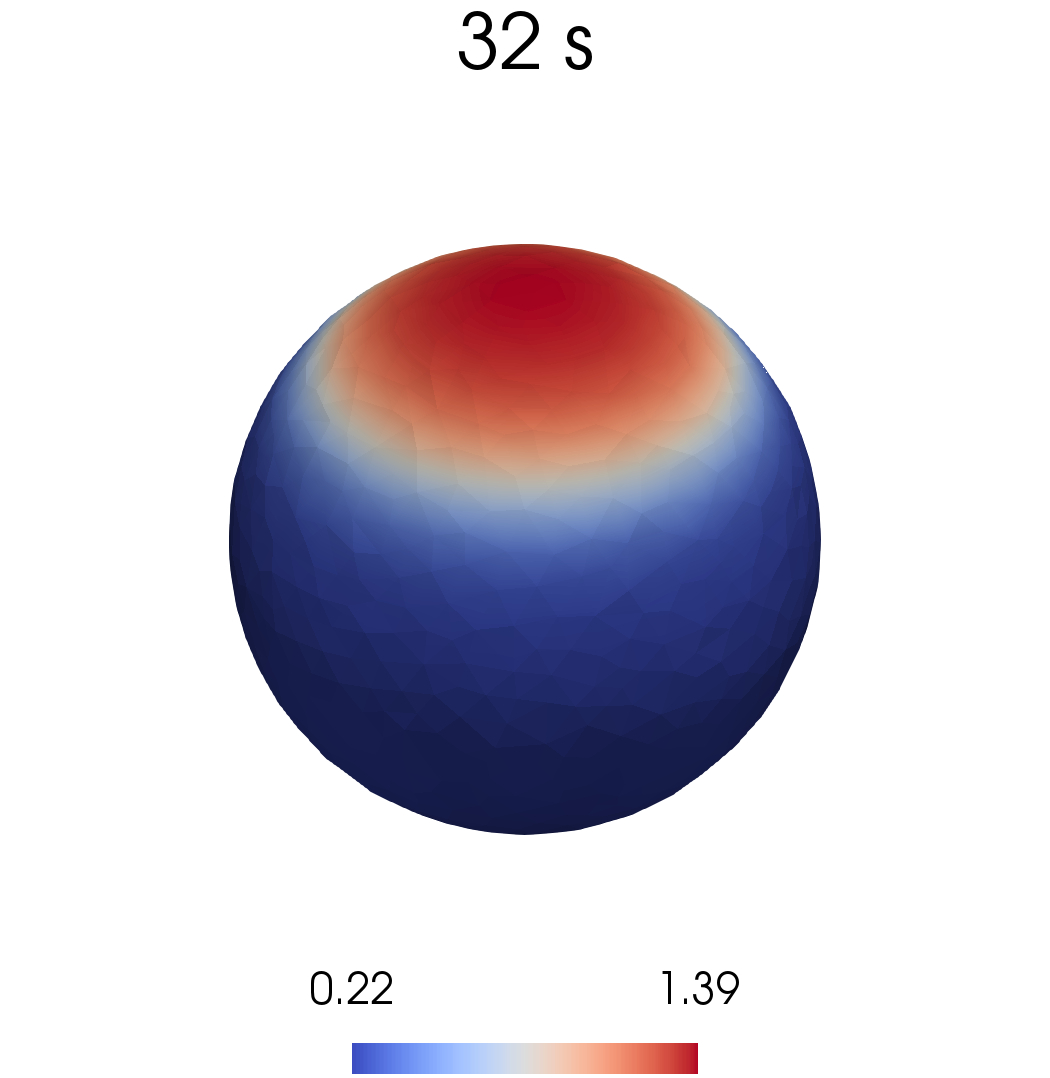}\\	\includegraphics[width=.3\linewidth]{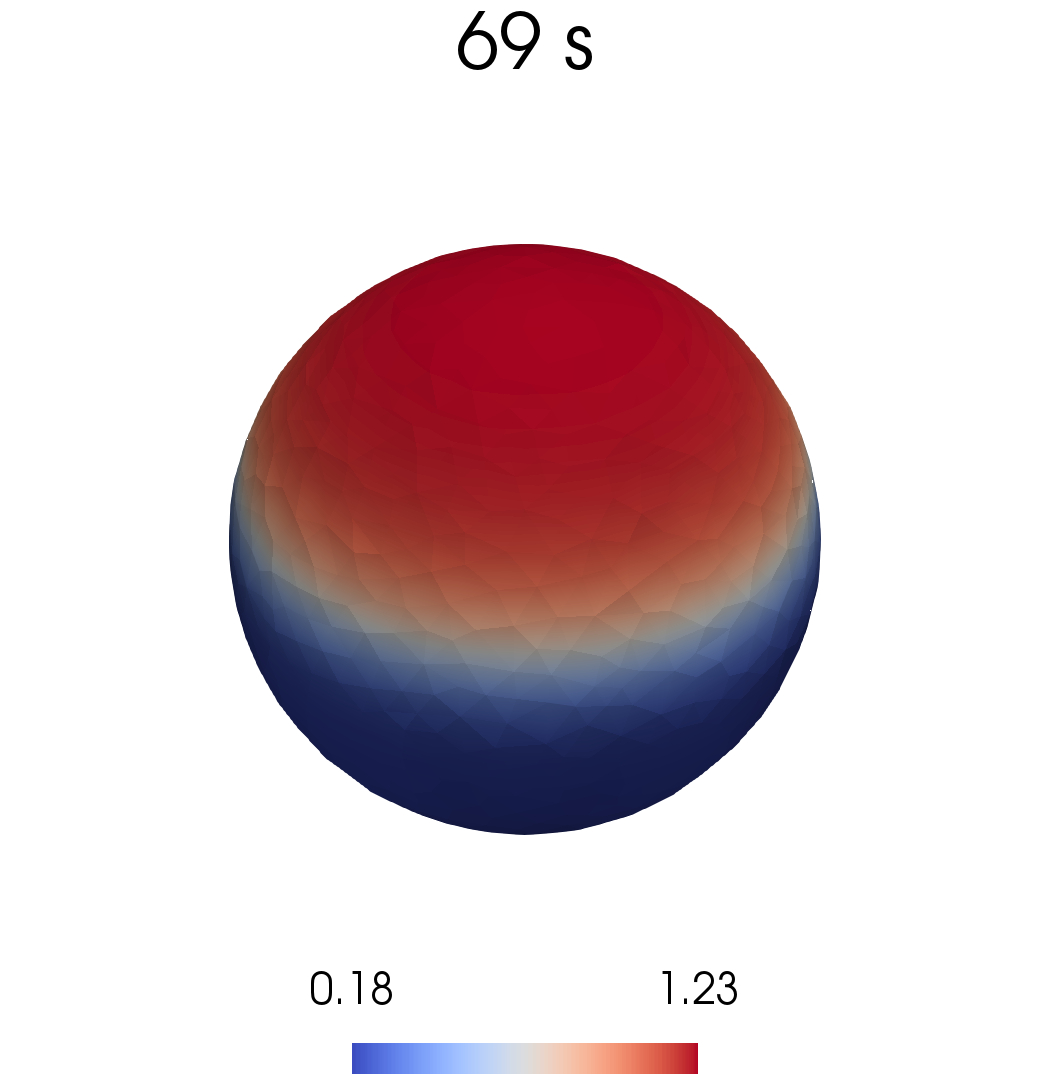}	\includegraphics[width=.3\linewidth]{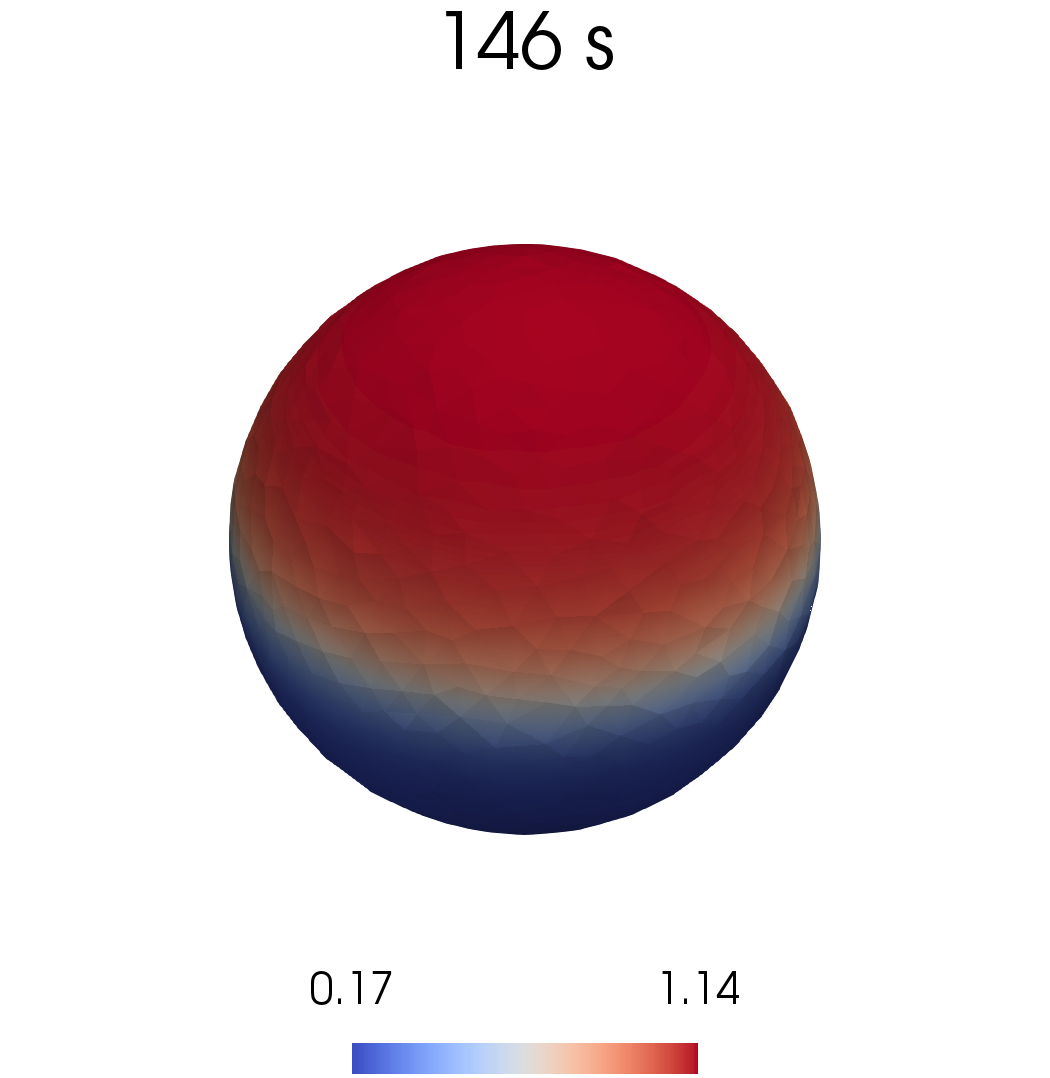}	\includegraphics[width=.3\linewidth]{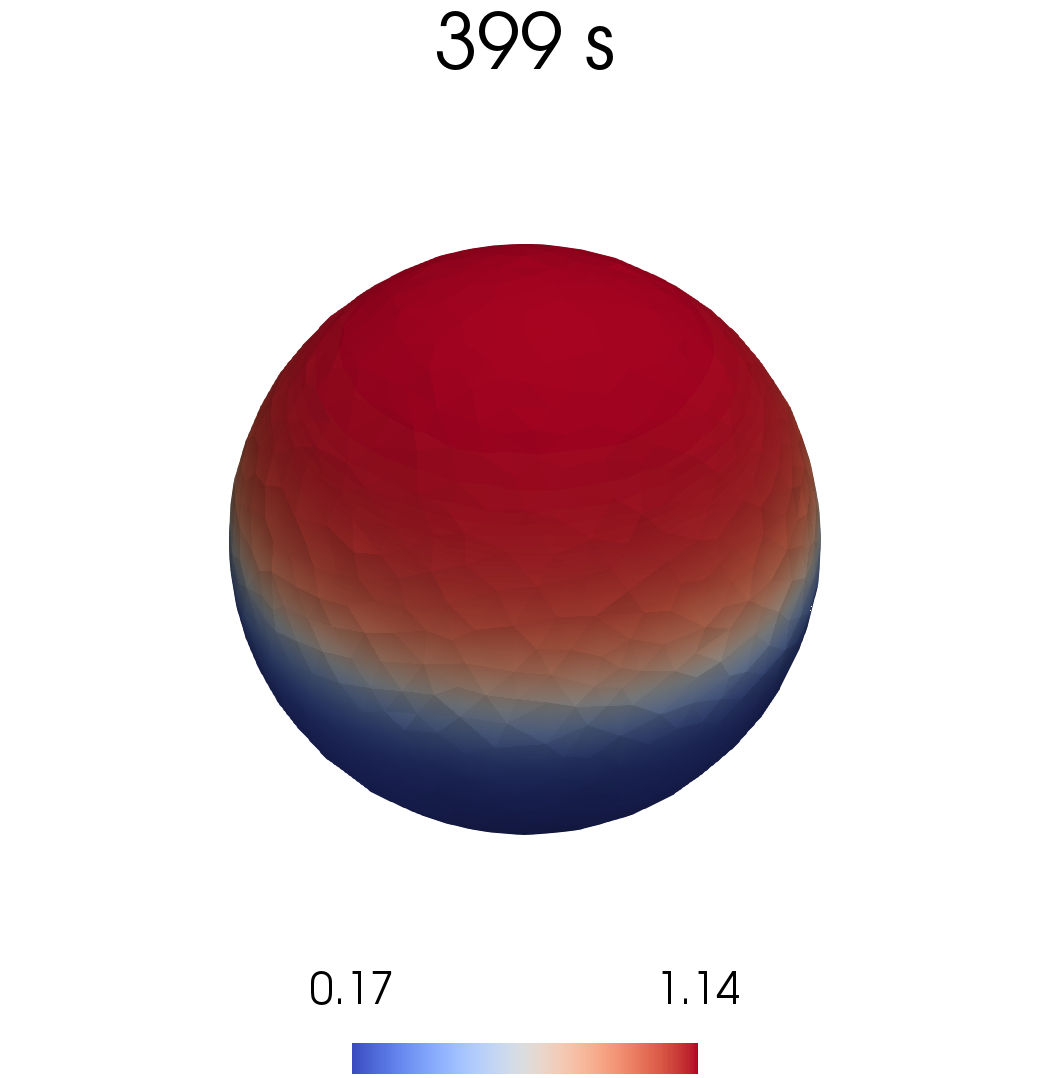}
\caption{Numerical simulations of the BSWP model \eqref{eq:model_b}-\eqref{eq:model_a} on a sphere: The active form of Rho GTPase $a$ propagating from a "stimulating" initial condition \eqref{eq:asympt_analysis_initial_condition_a} over a sphere. The numerical solution reaches a stable configuration after about 100 seconds. See text for further details.}\label{fig:simulation_sphere_solution}
\end{figure}

\begin{figure}[ht]
	\centering
	\includegraphics[width=.4\linewidth]{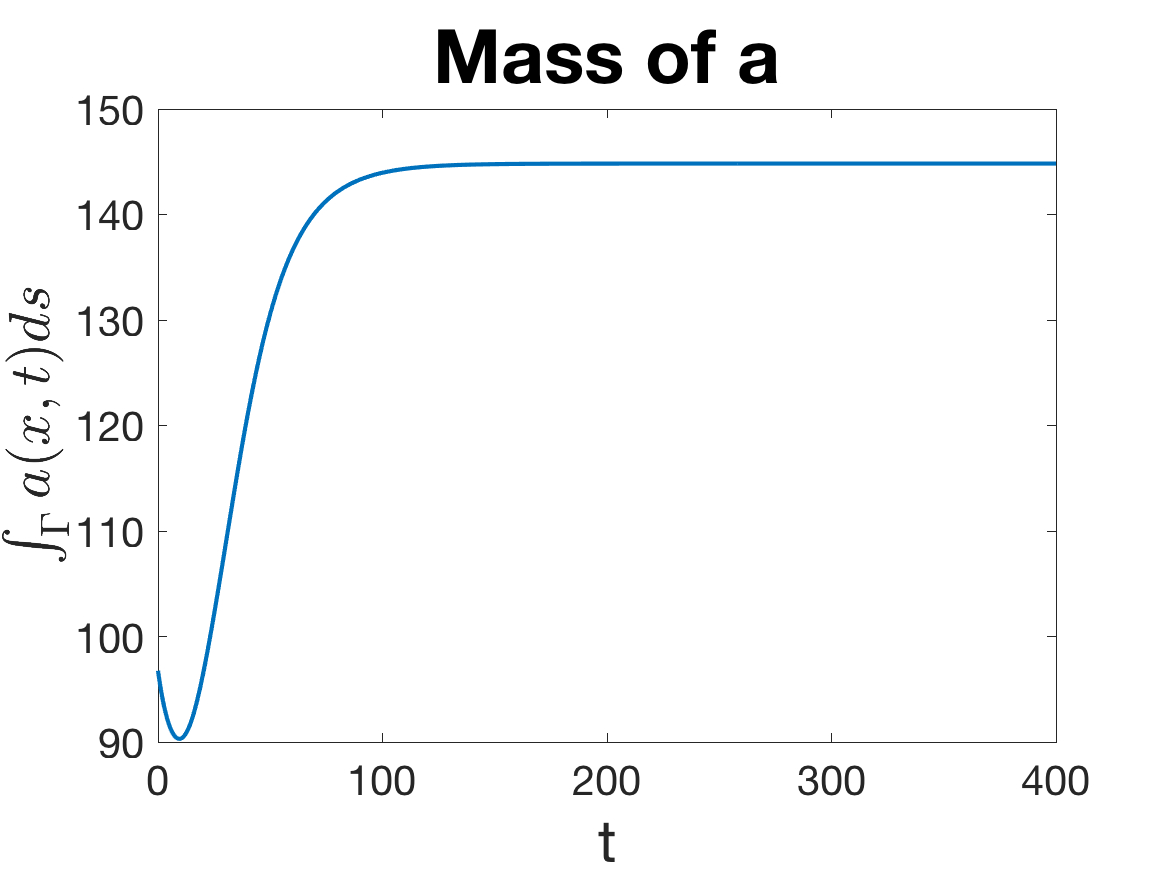}
	\includegraphics[width=.4\linewidth]{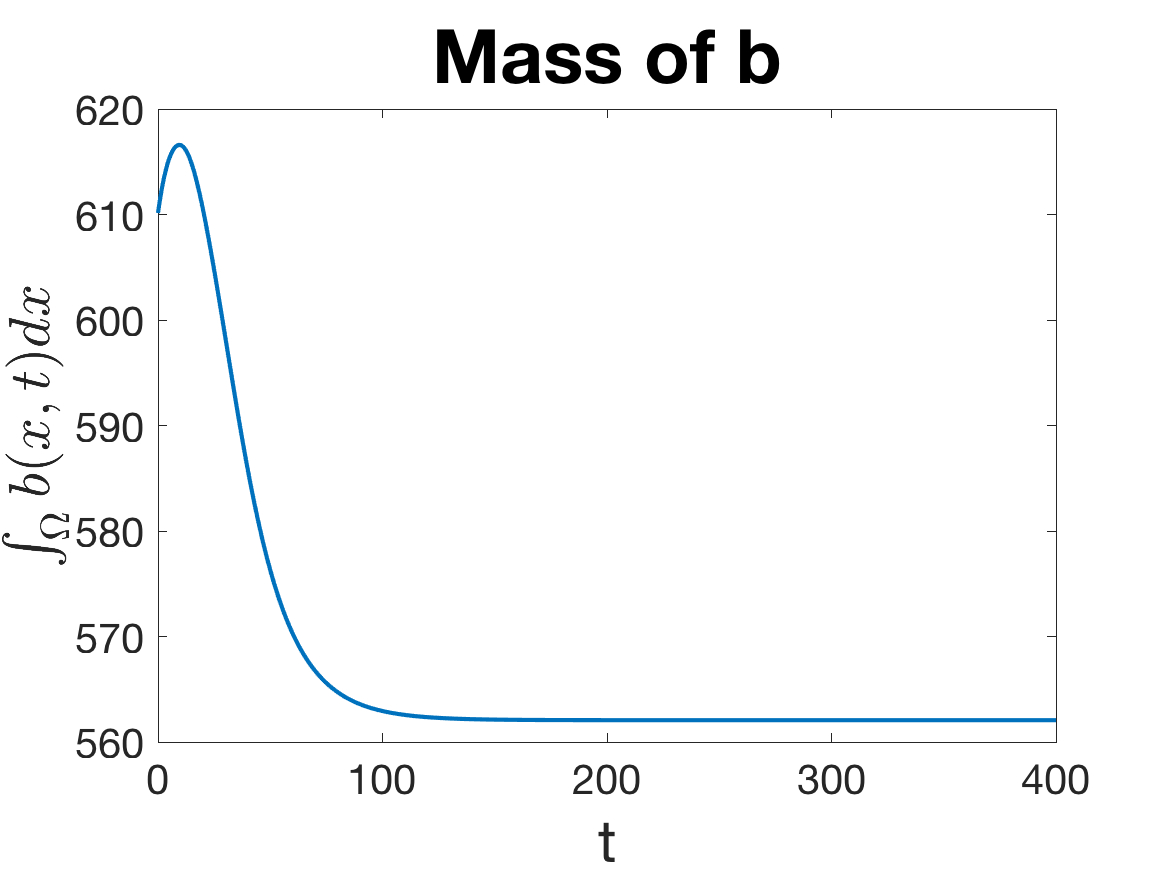}
	\caption{Variation in time of the total mass of active (left) and inactive (right) GTPases of the numerical solution shown in Figure \ref{fig:simulation_sphere_solution}. The initial decrease in the mass of $a$ is due to the attraction of the solution towards the smaller value $a_1(b)$ in most of its domain. Consequently we observe an initial increase of the mass of $b$. The mass of $a$ starts increasing with the spreading of its activity over the surface, which reduces the mass of $b$.
	After about 100 seconds the two components approach the equilibrium in mass.}\label{fig:simulation_sphere_masses}
\end{figure}

\subsection{Capsule}\label{sec:Capsule} As a second example, we compute numerical solutions of the BSWP model \eqref{eq:model_b}-\eqref{eq:model_a} on a capsule composed of cylinder of radius 5 $\mu$m and height 4 $\mu$m and two spherical caps at its extremities.
{\corr The results shown in Figure \ref{fig:simulation_pill_solution} are obtained with parameter values $\varepsilon_b=0.006$ in \eqref{eq:simulations_initial_condition_b}
and {\corr $\sigma^2=0.2\mu$m$^2$} in \eqref{eq:simulations_initial_condition_a}. A very small value of $\varepsilon_b$ is chosen in order to have an initial total quantity of $b$ very close to its possible maximal value $b_2$, therefore increasing the available source for the activation. The small value for $\sigma^2$ narrows the initial activated area, but it is still big enough to maintain the ability to propagate.}
As expected, the initial condition triggers the activation process, which apparently reaches the steady state in about 120 seconds, see Figure \ref{fig:simulation_pill_solution}. Eventually, we compute and observe the behavior of the numerical solutions for very long times for the BSWP model and notice that the activated region is moving very slowly from its ``apparent" steady state, towards one of the caps of the capsule, which is finally covered in more than 3 hours. }
Vanderlei \textit{et al.} showed the same property for the classical wave pinning model \eqref{eq:model_Mori_a}-\eqref{eq:model_Mori_f(a,b)}: on two-dimensional geometries the ``steady state" active concentration has the tendency to move very slowly towards more rounded regions of the domain \cite{Camley2017,Vanderlei2011}. It is interesting to note that in our case, the slow motion requires a much bigger time, which in \cite{Vanderlei2011} was of only around 200 seconds.

\begin{figure}[ht!]
	\includegraphics[width=.3\linewidth]{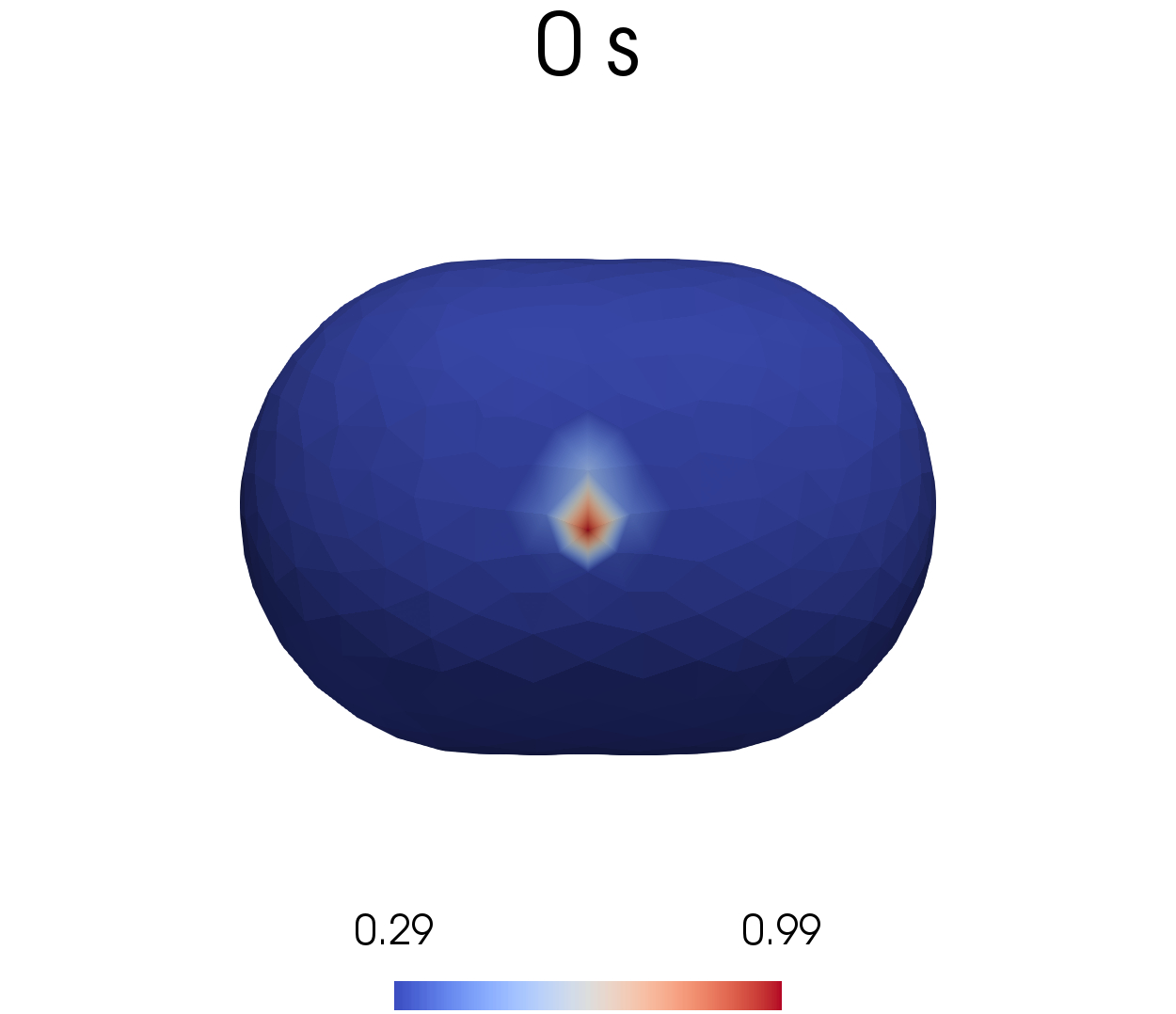}
	\includegraphics[width=.3\linewidth]{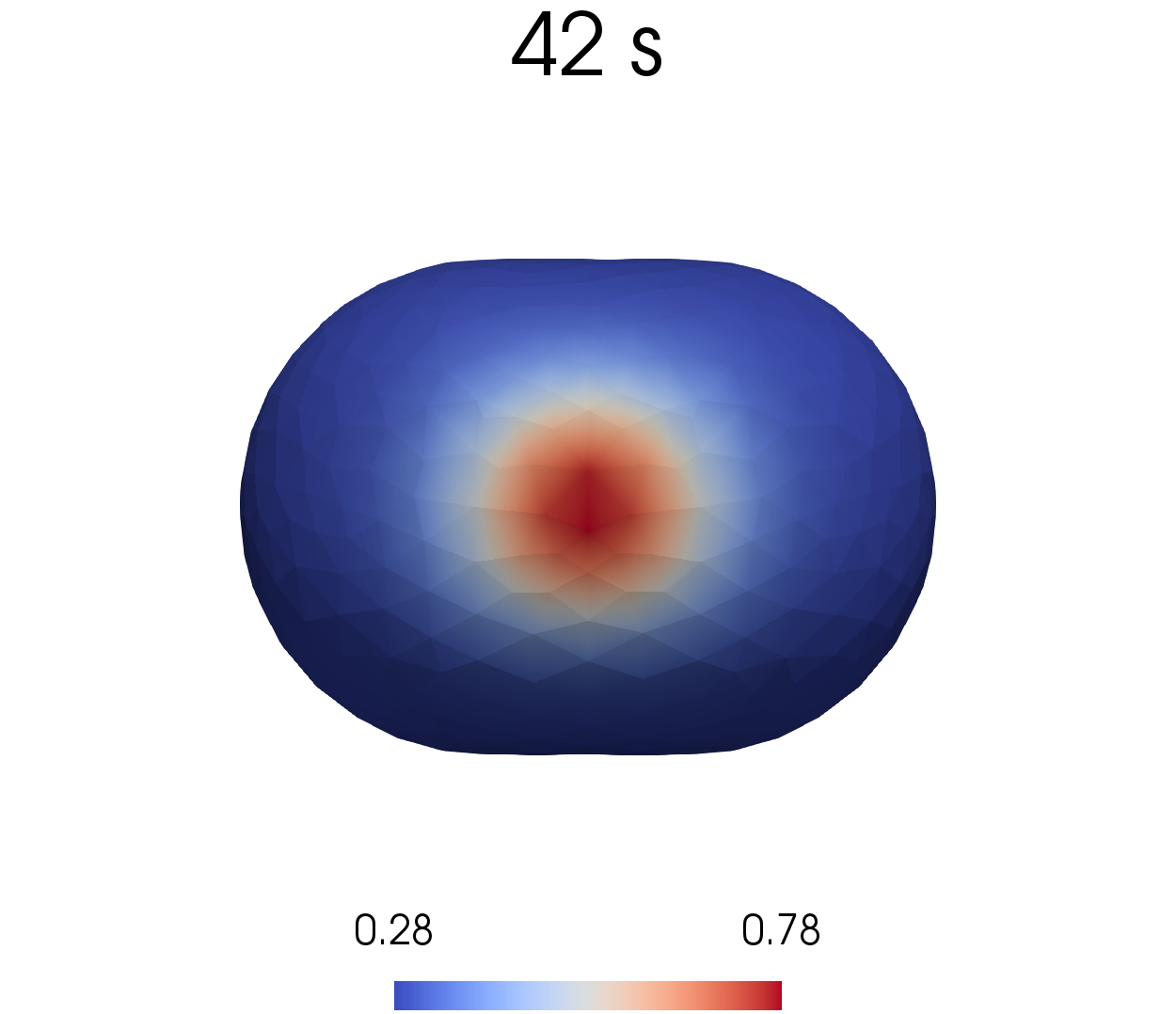}	\includegraphics[width=.3\linewidth]{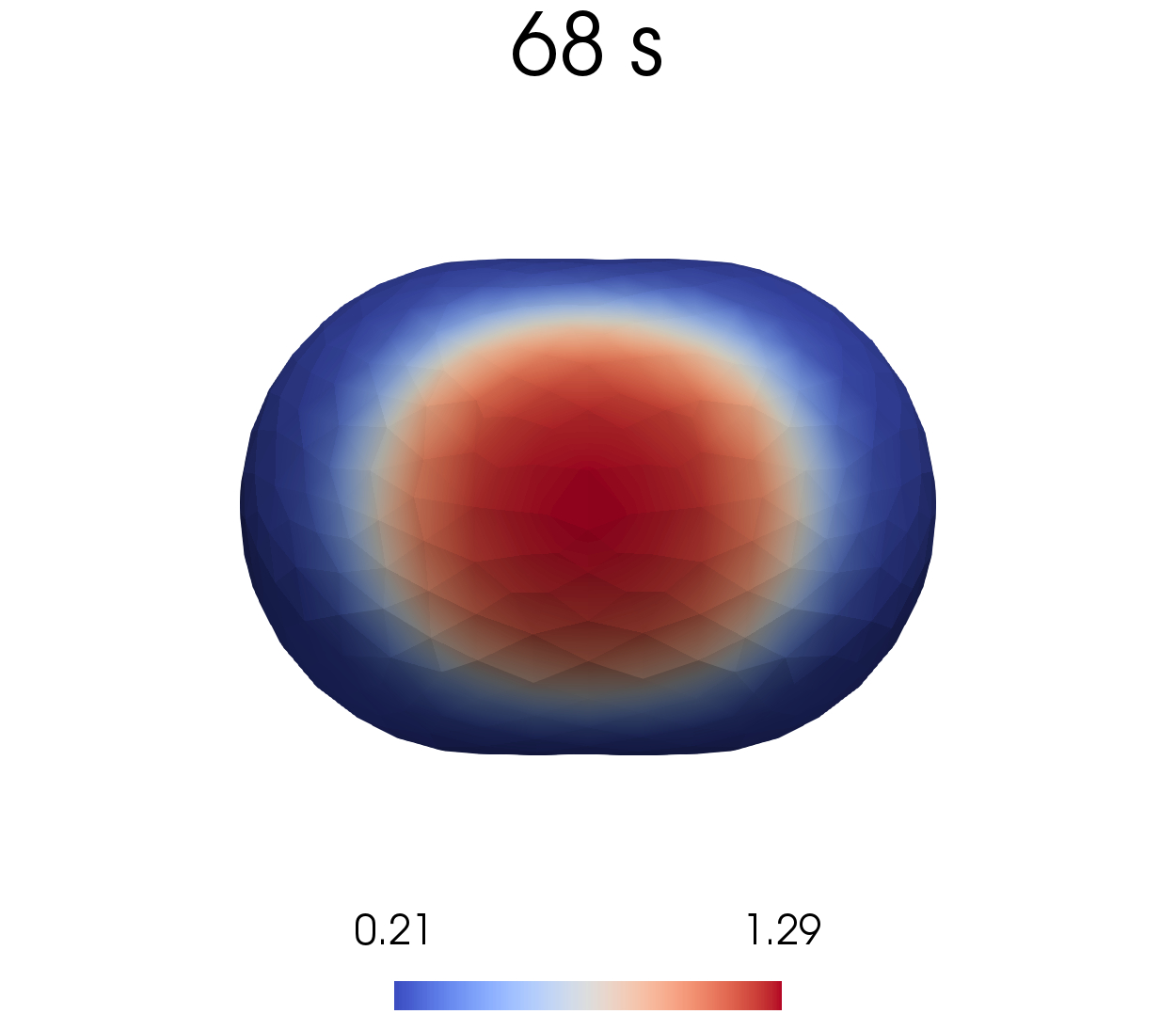}\vspace{10pt}\\	\includegraphics[width=.3\linewidth]{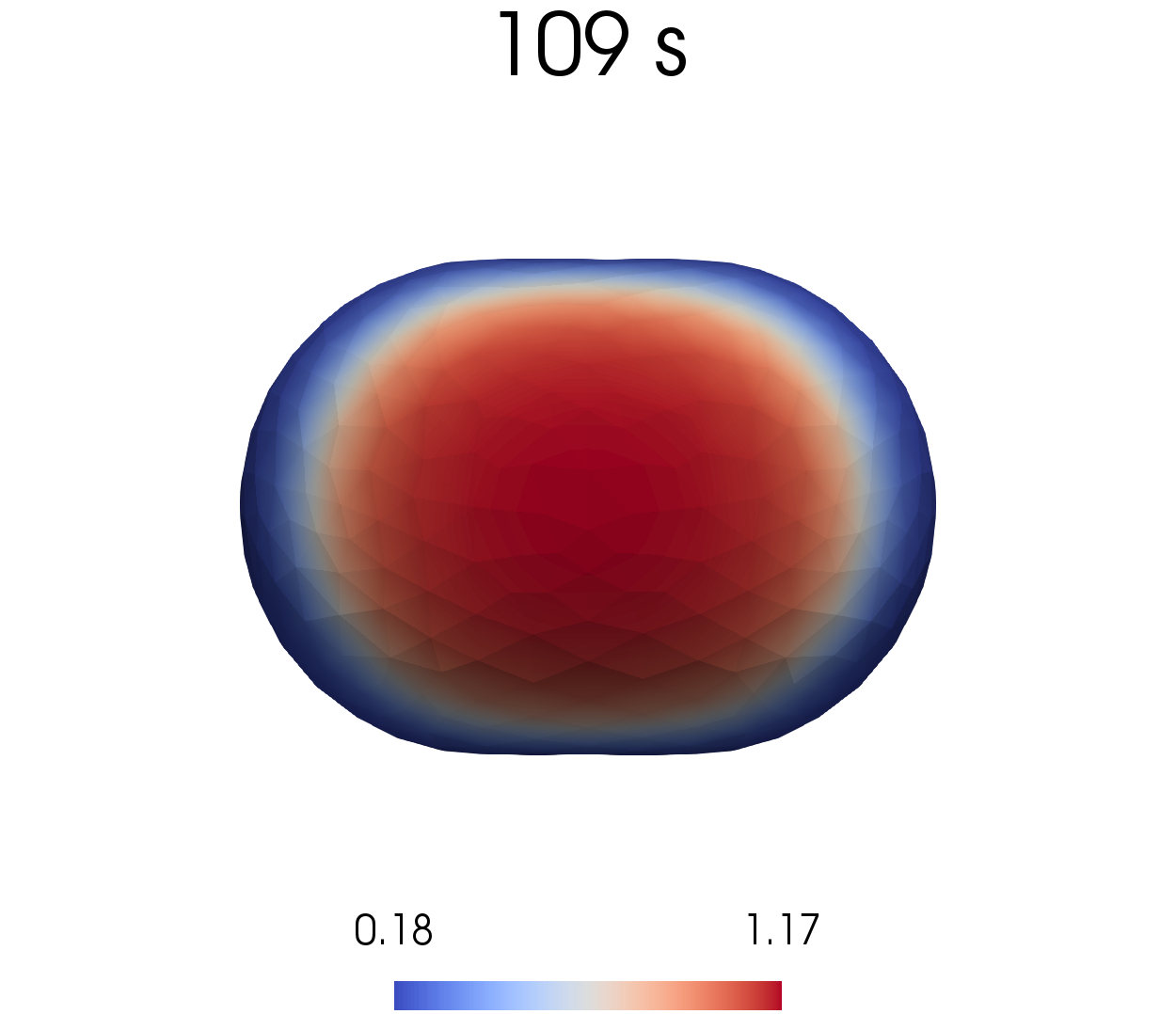}	\includegraphics[width=.3\linewidth]{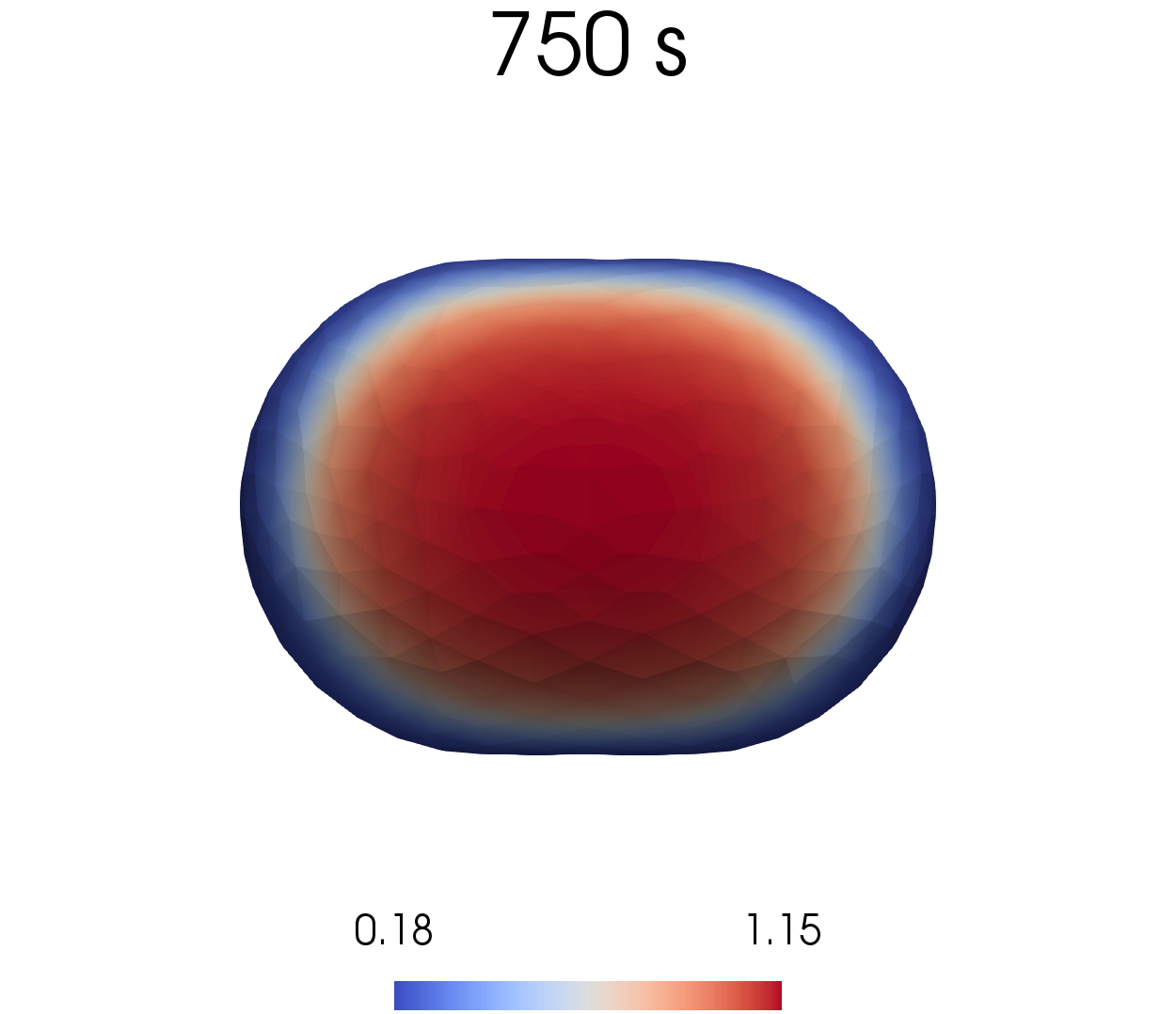}	\includegraphics[width=.3\linewidth]{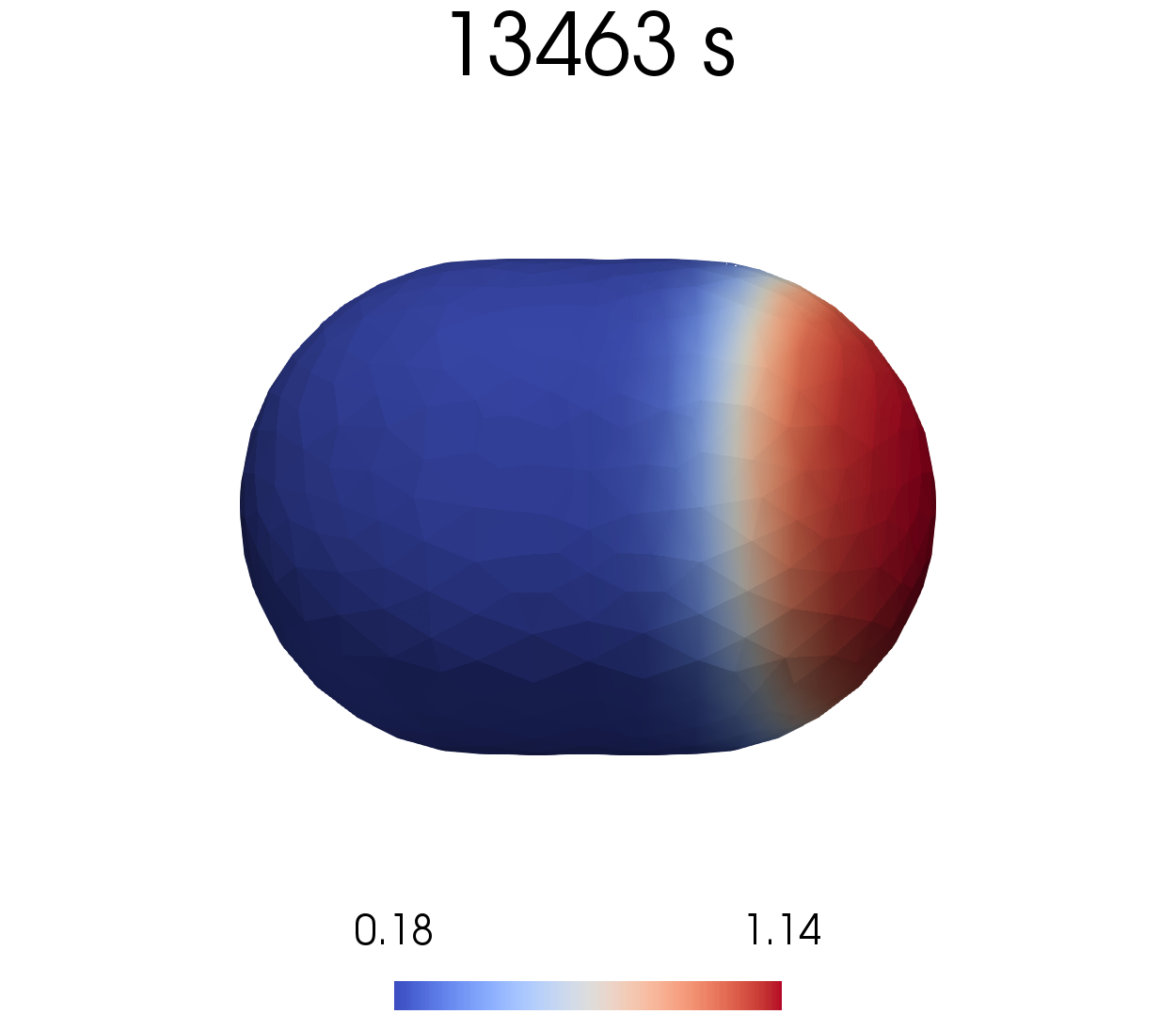}
	\caption{Numerical simulations of the BSWP model \eqref{eq:model_b}-\eqref{eq:model_a} on a capsule. The solution $a$ is here reported at several time steps: a small area in the lateral side of the capsule is activated, causing the activation of the entire lateral section which reaches its maximal size after around 120 seconds. Eventually, after a very long time, the activated area moves towards one of the  spherical caps of the domain.}\label{fig:simulation_pill_solution}
\end{figure}

\subsection{Polarised cell shape} 
\begin{figure}[ht!]\centering
\includegraphics[trim=60mm 90mm 60mm 100mm, clip,scale=0.20]{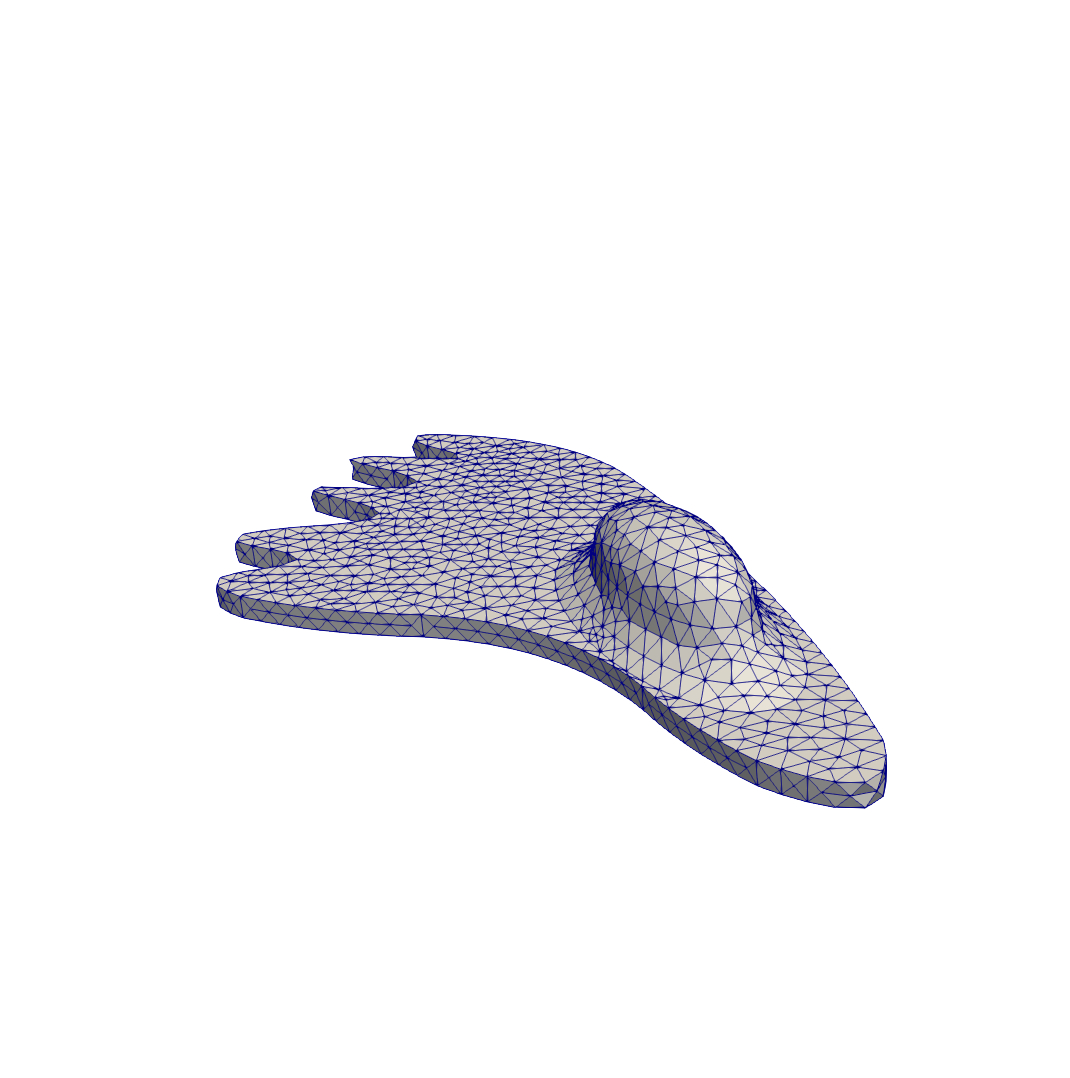}
\caption{The surface of a polarised cell shaped domain. The domain has been discretised with 5362 tetrahedrons which induced a surface discretisation with 3044 triangles.}
\label{fig:Migrating_cell}
\end{figure}

\begin{figure}[ht!]
	\includegraphics[width=.3\linewidth]{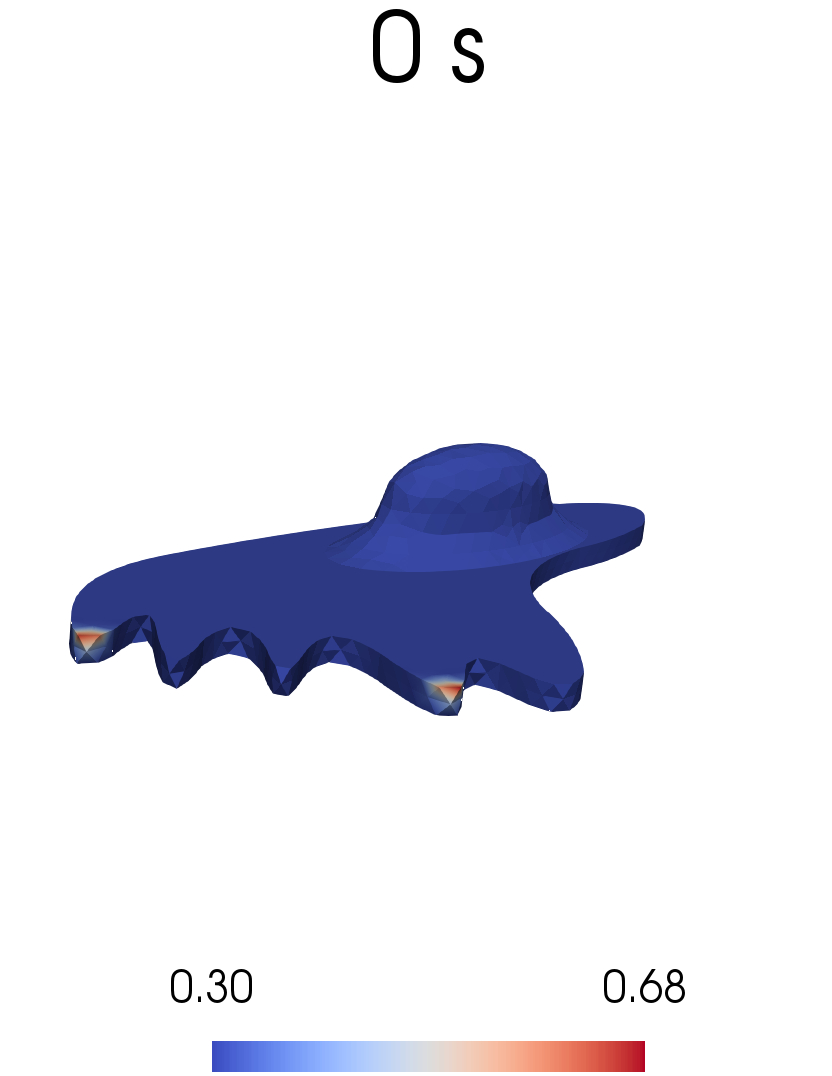}
	\includegraphics[width=.3\linewidth]{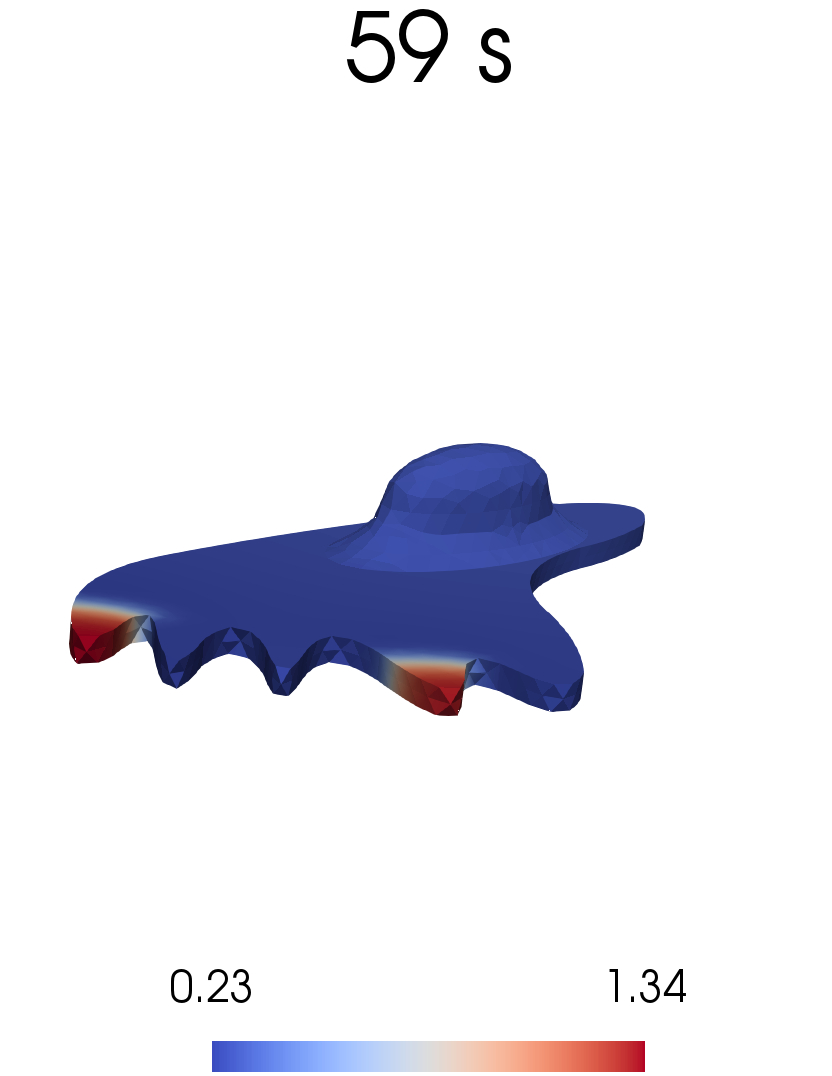}	\includegraphics[width=.3\linewidth]{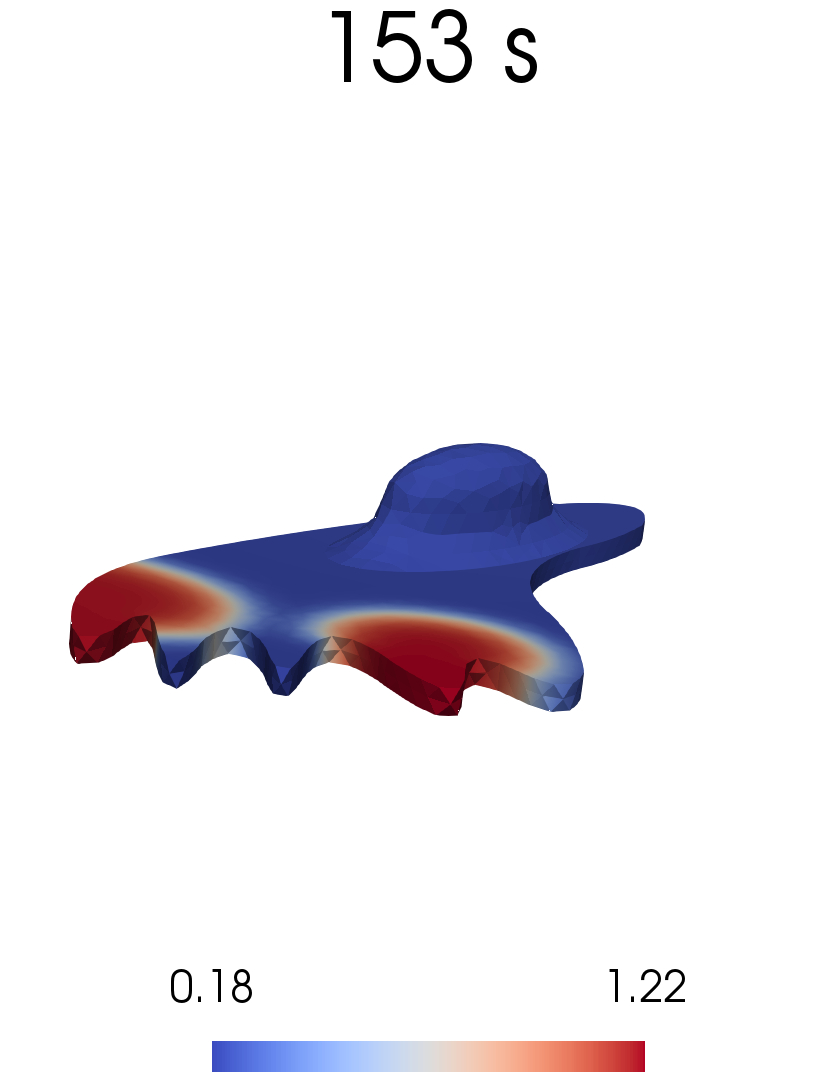}\vspace{10pt}\\	\includegraphics[width=.3\linewidth]{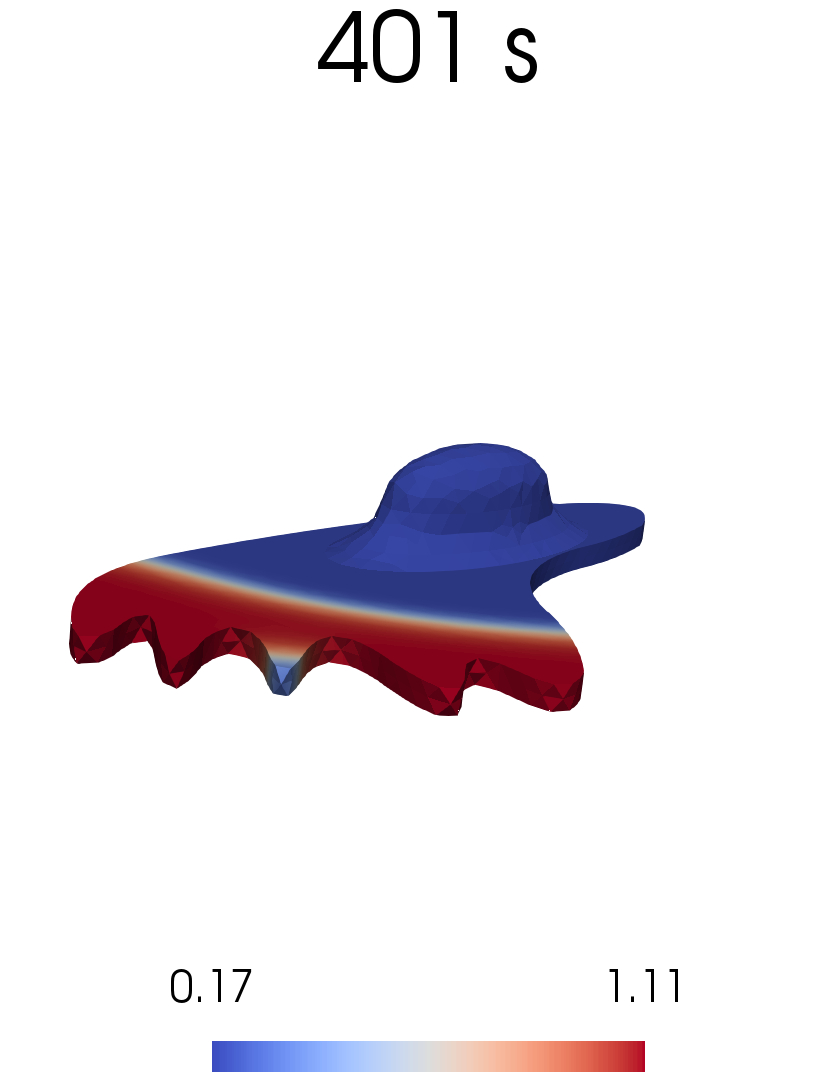}	\includegraphics[width=.3\linewidth]{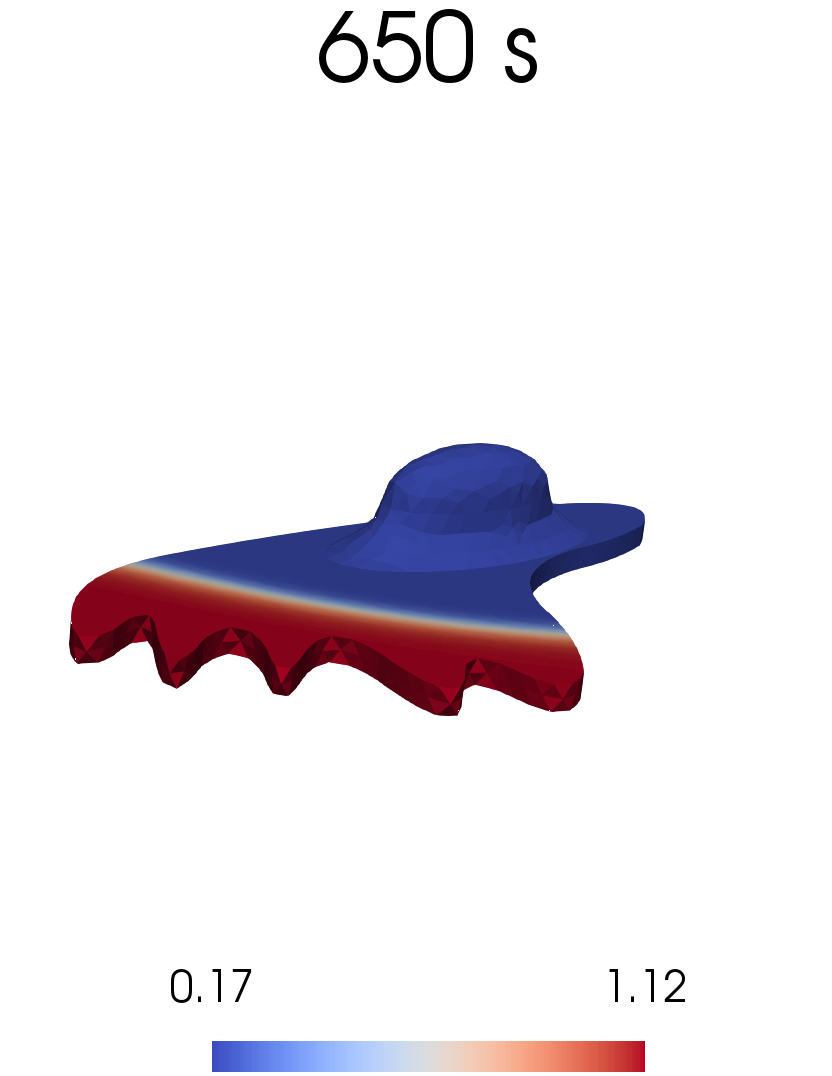}	\includegraphics[width=.3\linewidth]{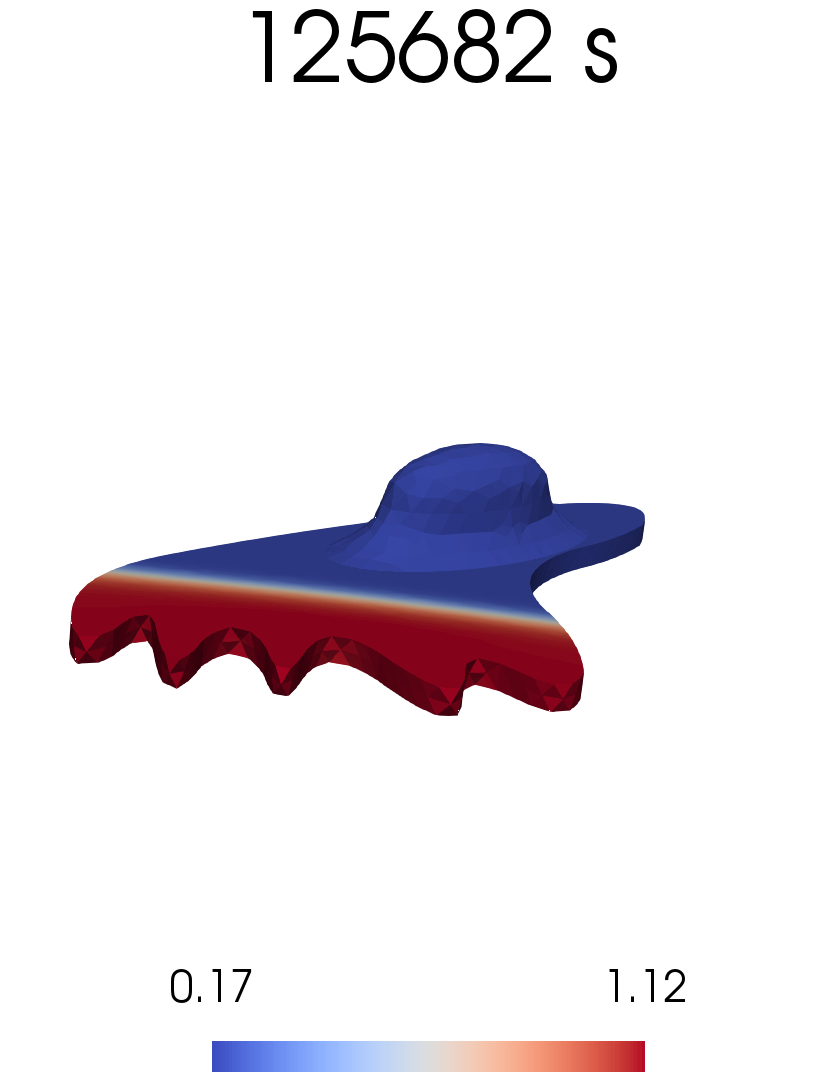}
	\caption{Numerical simulation of  the BSWP model \eqref{eq:model_b}-\eqref{eq:model_a} on a more complex domain, caricature of a polarised cell. The numerical results show the solution $a$ at different stages of the polarisation process: a small area in two of the five tips of the domain is activated and this generates two propagating fronts which merge together in about 4 minutes. The activated area stabilises covering the whole front of the domain in about 10 minutes. {\corr A video illustrating the wave pinning process is provided in the supplementary material.}}\label{fig:Migrating_cell_2_tips}
\end{figure}

\begin{figure}[ht!]
	\includegraphics[width=.3\linewidth]{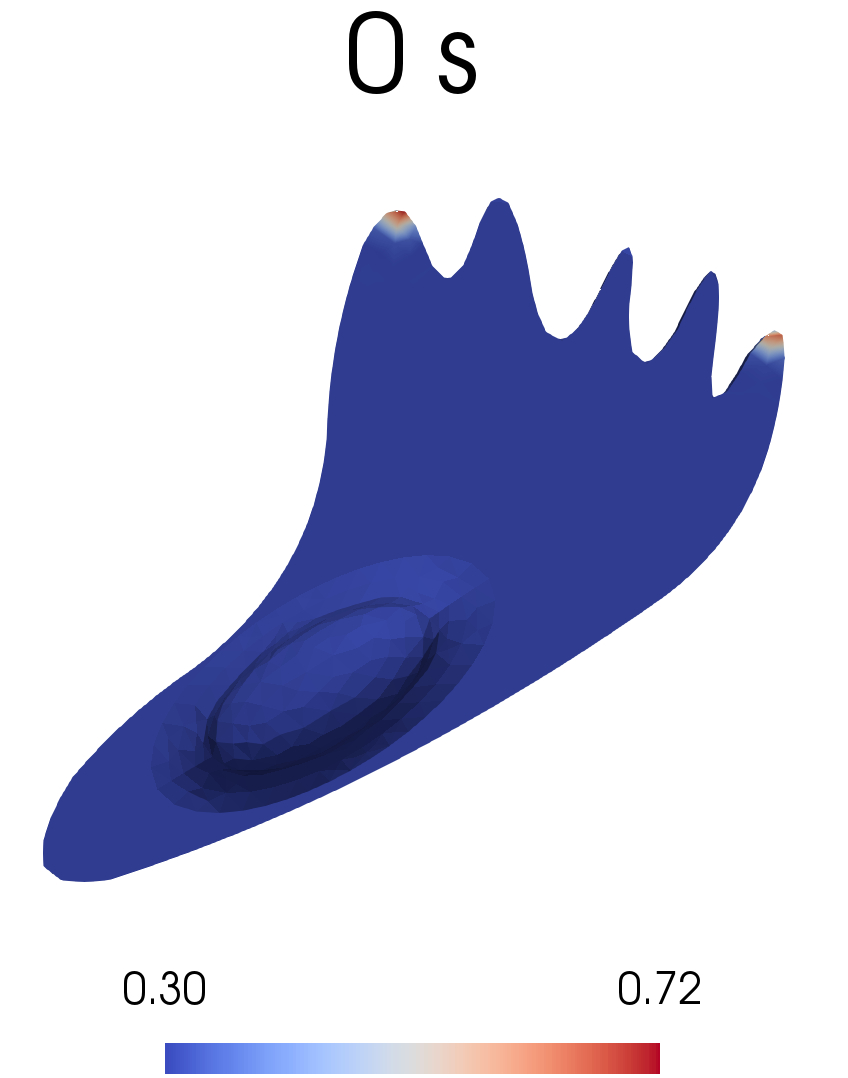}
	\includegraphics[width=.3\linewidth]{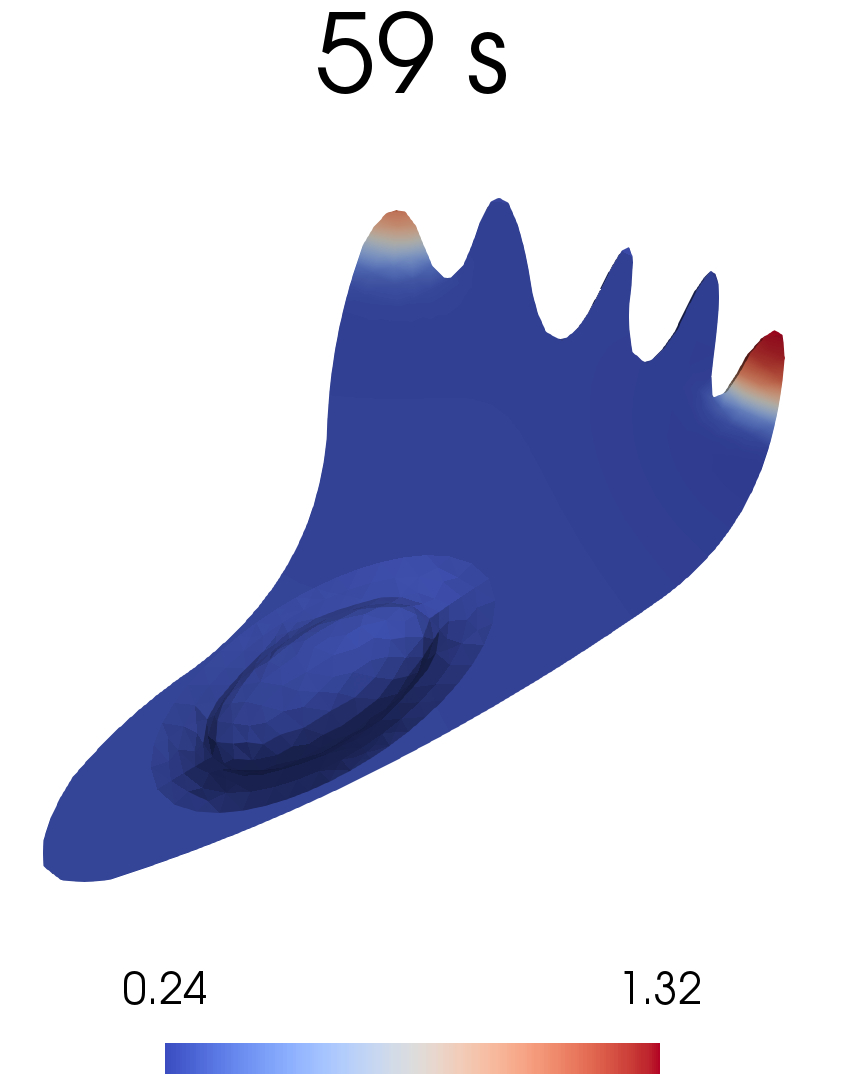}	\includegraphics[width=.3\linewidth]{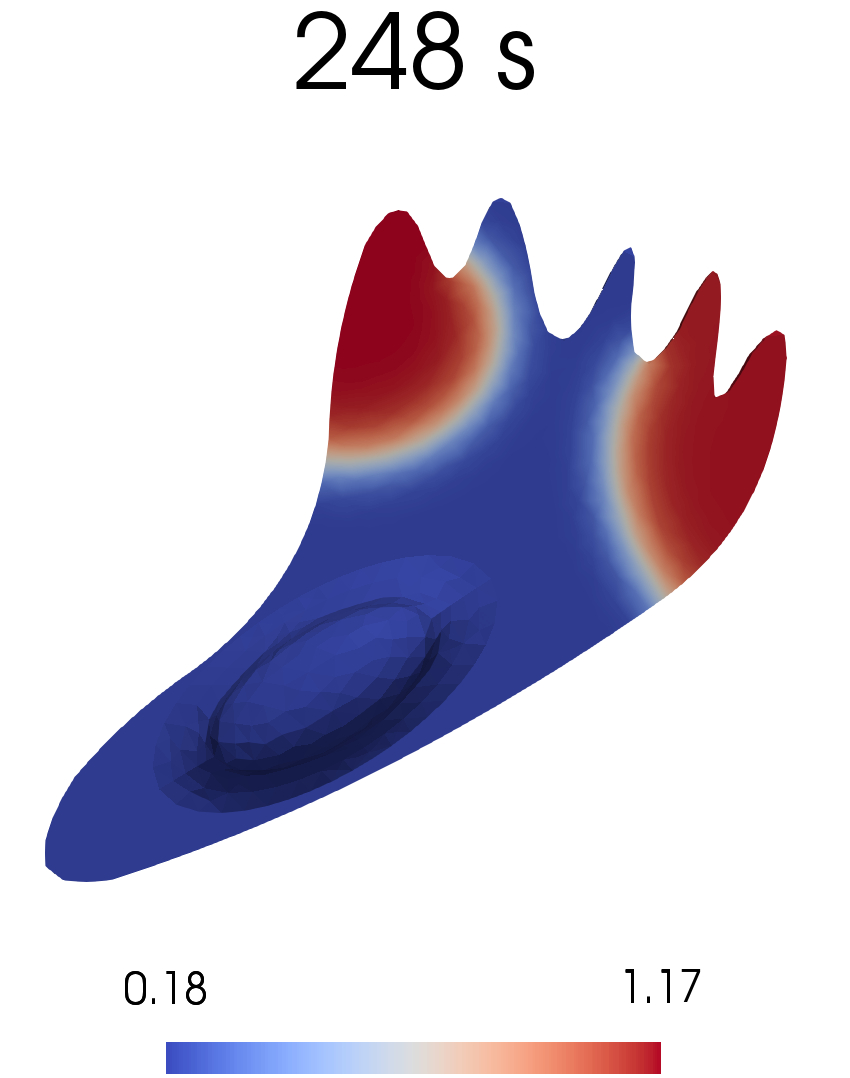}\vspace{10pt}\\	\includegraphics[width=.3\linewidth]{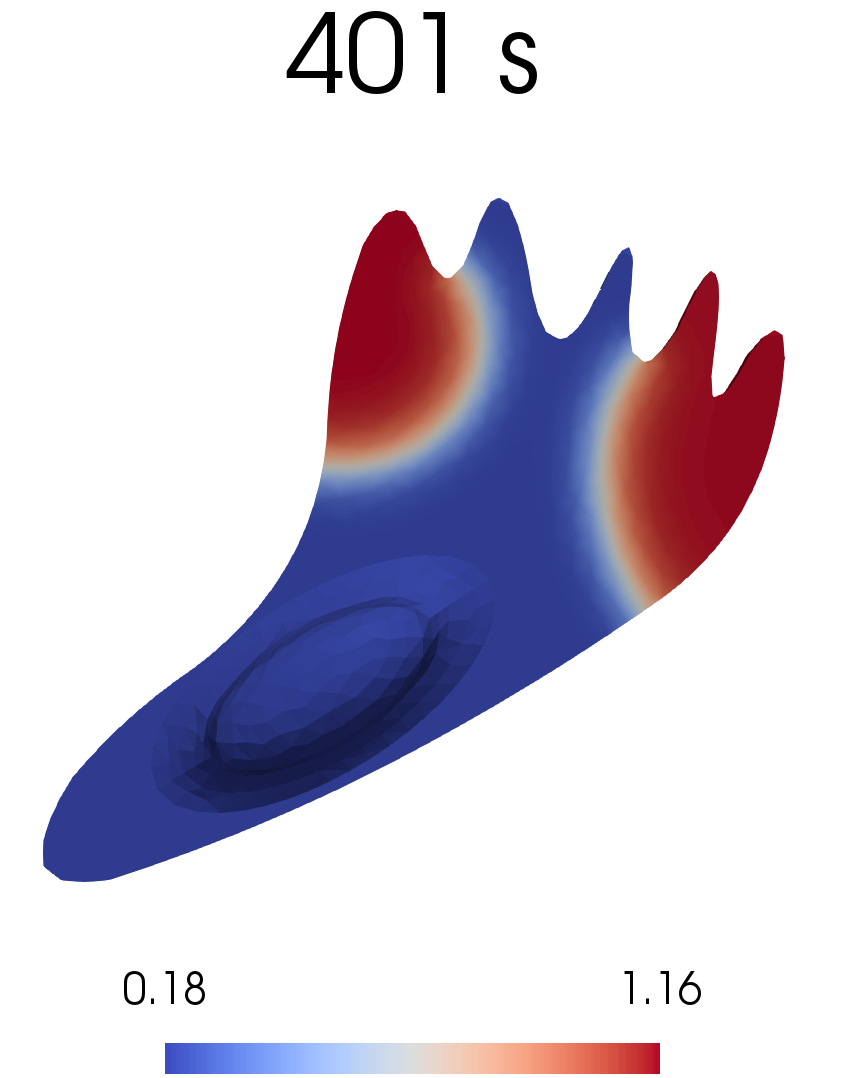}	\includegraphics[width=.3\linewidth]{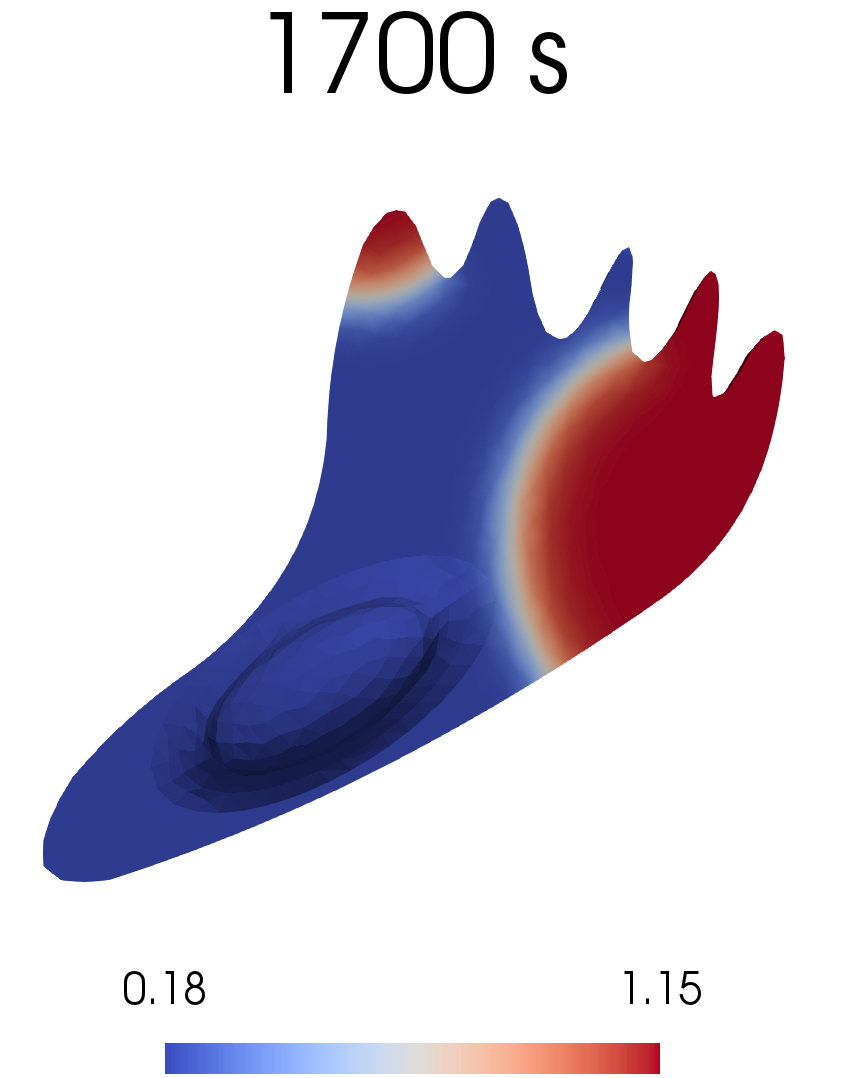}	\includegraphics[width=.3\linewidth]{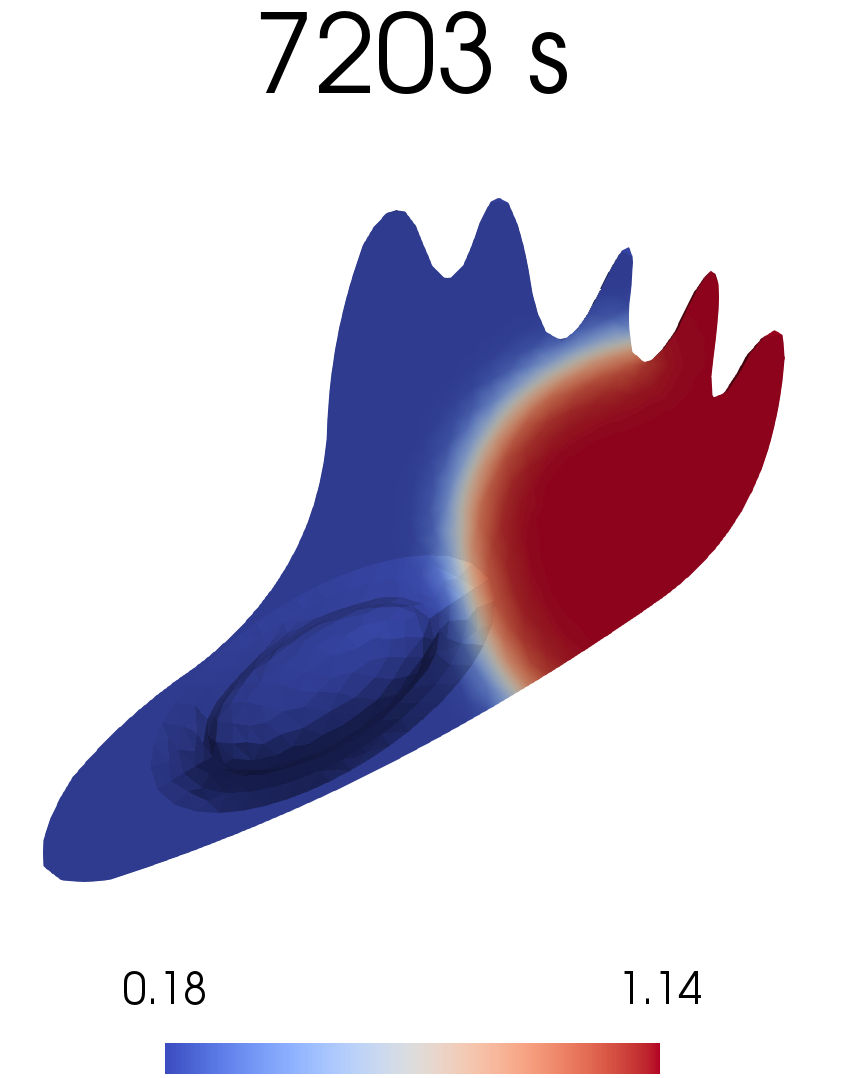}
	\caption{Numerical simulation of  the BSWP model \eqref{eq:model_b}-\eqref{eq:model_a}  on a domain mimicking a polarised cell. In this simulation the external tips are activated. Two active  waves are generated and start propagating on the surface. After about 5 minutes the competition effects between the two active patches start being visible, and one active region (left tip) starts reducing its size, until it disappears completely. Disappearing and stabilisation of the remaining active area occur after more than 30 minutes.
	Competition in two-dimensional wave pinning model was very recently investigated in \cite{Chiou2018}.}\label{fig:migrating_cell_competition}
\end{figure}

Next, we consider a more complex geometry whose shape mimics that of a polarised cell \emph{in vitro}, see Figure \ref{fig:Migrating_cell}. The domain has a volume of 538 $\mu$m$^3$ and surface area of 911 $\mu$m$^2$, almost three times more than the surface area of a sphere with the same volume. 
The front of the domain presents some protrusions with five tips. We set $\varepsilon_b=0.154$ in \eqref{eq:simulations_initial_condition_b} and $\sigma^2=0.5$ in \eqref{eq:simulations_initial_condition_a_2peaks}. In Figure \ref{fig:Migrating_cell_2_tips} we activate one external tip and one internal tip, while in Figure \ref{fig:migrating_cell_competition} activation starts from the external tips. Both perturbations are strong enough to trigger the polarisation process, which starts the enlargement of the polarity patches. In the first simulation shown in Figure \ref{fig:Migrating_cell_2_tips}, in about four minutes the two activated spots merge together into a unique stable active region which enlarges over the whole front of the domain and gets pinned in about 10 minutes.

In the second simulation shown in Figure \ref{fig:migrating_cell_competition}, the cell needs a much longer time to stabilise as it has to deal with two competitive polarity patches. Initially, propagation occurs normally with two different enlarging areas. After about five minutes one active region inverts its behavior and starts disappearing. This leads to a winning tip, which continues enlarging on its side, until final stabilisation. 

\begin{figure}[ht!]
	\centering{
	\includegraphics[width=.2\linewidth]{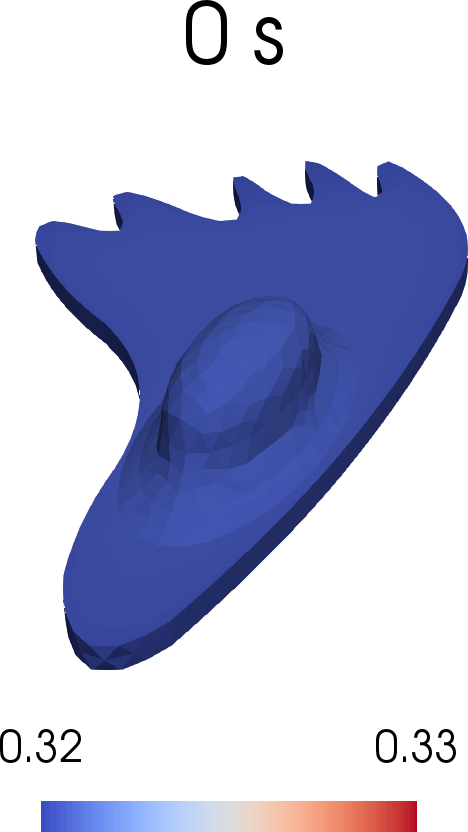}\hspace{10pt}
	\includegraphics[width=.2\linewidth]{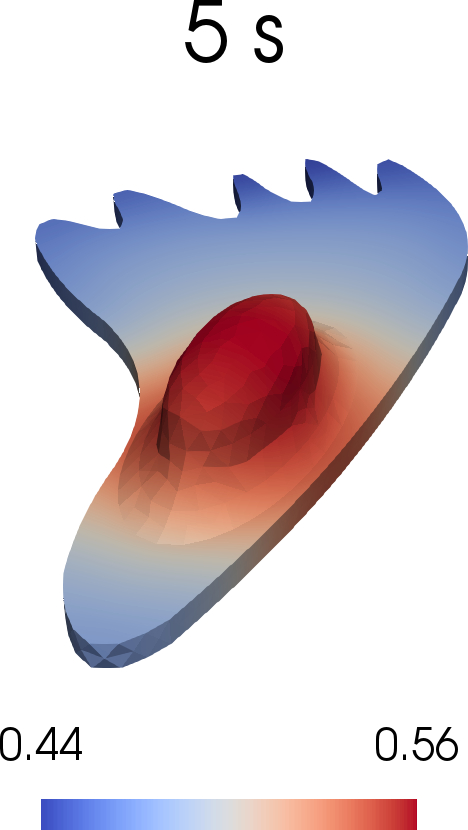}\hspace{10pt}	\includegraphics[width=.2\linewidth]{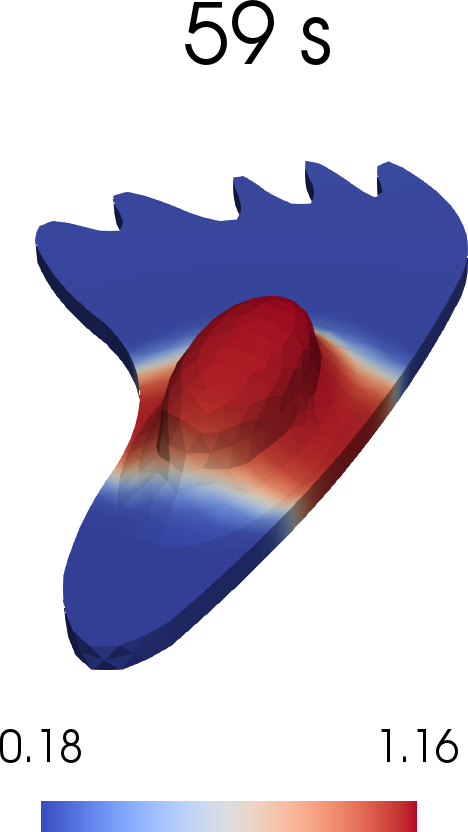}\vspace{15pt}\\	\includegraphics[width=.2\linewidth]{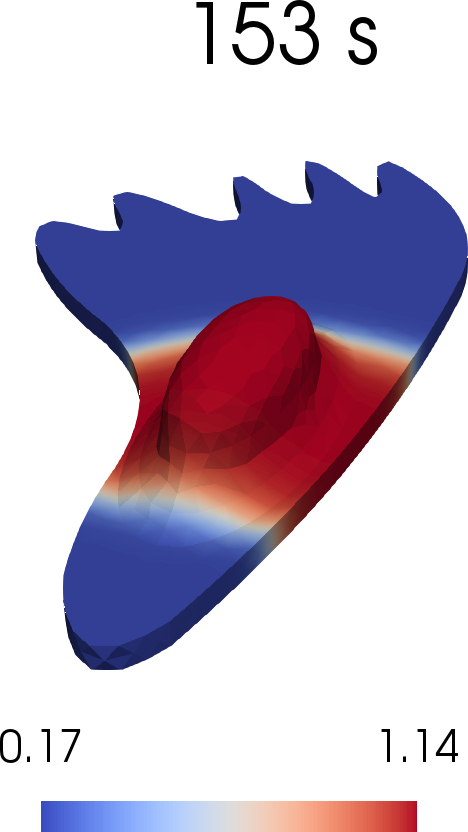}\hspace{10pt}	\includegraphics[width=.2\linewidth]{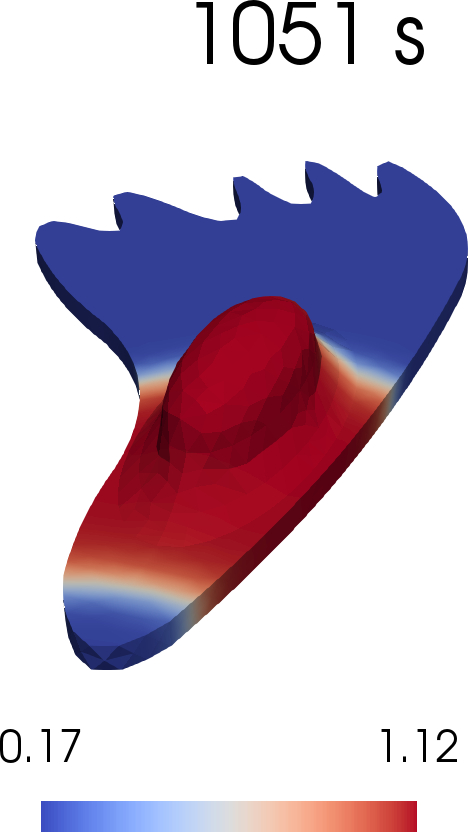}\hspace{10pt}	\includegraphics[width=.2\linewidth]{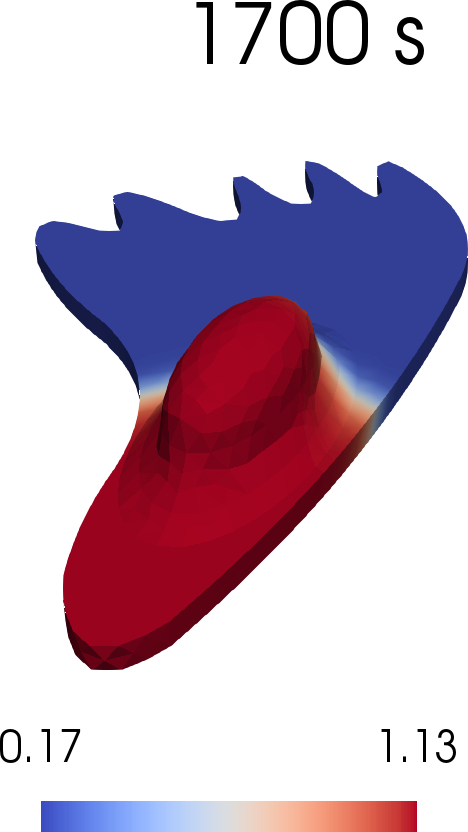}}
	\caption{Numerical simulation of  the BSWP model \eqref{eq:model_b}-\eqref{eq:model_b_Gamma} and \eqref{eq:model_a} with reaction kinetics (\ref{eq:f(a,b)+fs}) on a domain mimicking a polarised cell. We apply a constant stimulus $k^s(\mathbf{x},t)=0.03$ s$^{-1}$ until time $t_{S}$=20 s which induces an activation at the level of the nucleus-shaped volume. From here, a wave starts, covering the whole rear. In about 20 minutes the BSWP model has reached its steady state, with the rear having high levels of active GTPase.}\label{fig:migrating_cell_stimulus}
\end{figure}

In all previous simulations, we have used suitable initial conditions in the form of perturbations of the spatially homogeneous profile of $a$. This has been shown to be enough to give rise to polarisation,
in the numerical results, as well as for the asymptotic and local perturbation analysis. However, similar perturbations can be induced by perturbing the reaction \eqref{eq:f(a,b)}. Indeed, in most of the papers simulating the WP model, polarisation was initiated from a stimulus included in the reaction function, rather than a stimulus in the initial conditions, which were, in turn, spatially homogeneous.
Following this latter approach, the BSWP model is given by equations \eqref{eq:model_b}-\eqref{eq:model_a} with reaction
\begin{align}\label{eq:f(a,b)+fs}
f(a,b)&=\omega  \Big(k_0+\frac{\gamma a^2}{K^2+a^2}\Big)b-\beta a + \omega k^s b, & \mathbf{x}\in\Gamma.
\end{align}
where $k^s=k^s(\mathbf{x},t)$ is an arbitrary function, generally non-negative until a certain time $t_s$ and zero afterwards \cite{Mori2008}.
Appropriate choices of $k^s$ can lead to the formation of local peaks in the solutions, which trigger the propagation of $a$ over the surface. An interesting result of the two-dimensional BSWP model \eqref{eq:model_b}-\eqref{eq:model_a} was its ability to self polarise from homogeneous initial conditions in asymmetric geometries when a spatially homogeneous stimulus was applied in an initial time interval $[0, t_{s}]$ \cite{Giese2015}. 
In Figure \ref{fig:migrating_cell_stimulus} we present the same experiment on our three-dimensional domain in which we apply a homogeneous stimulus of $0.03$ s$^{-1}$ for 20 seconds. This induces a rapid local activation of the ellipsoidal volume on the top of the cell, with noticiable effects within the first 5 seconds. The high $a$ concentration starts increasing and sharpening the fronts, and successively it spreads towards the rear of the domain. Our simulation confirms the interesting geometry-induced self-polarisation ability also for the three-dimensional case. 

\section{Discussion}\label{sec:conclusions}
In this paper, we have presented a three-dimensional extension of the wave pinning model in a bulk-surface setting, in which membrane-bound GTPase and cytosolic GTPase are spatially localised and their interactions occur on the cell surface. The model describes cell polarisation through a minimal circuit of GTPase switching between active and inactive forms as well as between the membrane and the cytosol.
In our work we were able to show many analogies to the classical wave pinning model \cite{Mori2008,Mori2011,Vanderlei2011} not previously shown in three-dimensional domains. 

In this framework, the bulk-surface wave pinning (BSWP) model \eqref{eq:model_b}-\eqref{eq:model_a} maintains the three key properties (conservation of total mass, different diffusivities and bistability of the reaction) which are again necessary to achieve polarisation. 
Different techniques and methods have been used to get a good understanding of the behavior of the bulk-surface wave pinning model. By employing asymptotic analysis in Section \ref{sec:asymptotic_analysis} we show how a perturbation of the homogeneous initial condition can trigger a propagation of the high level of active GTPase over the cell membrane. Effects of the geometry and parameters mapping have been investigated in Section \ref{sec:bistabiliy_LPA}, where we have highlighted how polarisation behavior is more probable in complex domains. {\corr This has been done using local perturbation analysis which allows a reduction to a ODE system.}
{\corr Finally,} using the bulk-surface finite element method, presented in Section \ref{sec:BSFEM}, we  computed numerical solutions of the BSWP model on different domains. An interesting result has been obtained from the model over a capsule-shape domain, where long time behavior of the model has been simulated, showing another common property of the classical wave pinning model derived by Mori \textit{et al.} \cite{Mori2008}, noted in \cite{Vanderlei2011}: the high active concentration region moves very slowly from its apparent stable steady state towards more rounded areas, until it covers one of the spherical caps of the capsule. 

Simulations have been done also on a more complex geometry mimicking a polarised cell-like shape. We  showed competition between different highly active areas, as previously reported for the classical wave pinning mechanism \cite{Chiou2018}. {\corr In addition}, we show how geometry plays a crucial role on the spontaneous polarisation in our three-dimensional BSWP model, as  reported in the two-dimensional case by Giese \textit{et al.} \cite{Giese2015}. {\corr In the latter case, the asymmetric geometry of the domain plays a crucial role in enhancing activation of the GTPases. Indeed, activation was induced by a spatial homogeneous stimulus, but its effects appear well localised in specific areas of the surface. }

Positive feedback, known to be a biological feature of Rho GTPases \cite{Graessl2017}, has been confirmed as a key player also in the new formulation of the model. It is  represented by the Hill function in  \eqref{eq:f(a,b)}, but many other nonlinear choices are possible. Identification of Rho GTPase feedback is an extremely interesting task and hopefully coordinated efforts between biologists and mathematicians can lead the way to a more complete understanding of cell polarisation and migration.

We expect the BSWP  \eqref{eq:model_b}-\eqref{eq:model_a}  to be a starting point for a more complete work, in which the biochemical mechanisms shown above are coupled with mechanical properties of the cell, such as membrane tension and migration. Indeed, in real cells, GTPase concentration would lead to shape changes, through cytoskeleton interactions. The classical wave pinning model has been already coupled to mechanistic models for membrane tension \cite{Wang2017} and cell migration \cite{Camley2017,Vanderlei2011}. 
In these latter works the migrating cell, instead of keeping a straight direction, was turning over one side. This corresponded to the slow motion of the polarised area, as discussed in the Section \ref{sec:Capsule} and in Figure \ref{fig:simulation_pill_solution}. 
In view of this and taking into account the influence of the geometry of the domain,
it can be of interest to extend these results and investigate how the bulk-surface approach influences
the mechanical properties. Indeed, as reported in Figure \ref{fig:simulation_pill_solution}, the slow motion appears to be much slower with respect to the one reported in the literature \cite{Vanderlei2011} and, in a reasonable amount of time, the turning effect might not be noticeable. As well, the effects of the geometry reported in Section \ref{sec:bistabiliy_LPA} might play an important role on evolving domains describing more accurately migrating cells, in which the parameter $\omega=|\Omega|/|\Gamma|$ is subject to changes in time.

Another interesting extension of this study is whether it is possible to achieve similar mechanisms in a bulk-surface model with three species, when membrane recruitment of cytosolic GTPase is taken into account. This idea of GTPase model has been presented in \cite{Ratz2014}, but the polarisation mechanisms were Turing-type. 

\section*{Data accessibility}
The authors declare no use of primary data as a result there is no supporting material to present in association to the results pertained by the current manuscript.

\section*{Acknowledgments}
 DC would like to thank Dr. Christopher Rowlatt for the useful discussions during his visit to University of Strathclyde. As well, many thanks to Mr Farzad Fatehi Chenar and Mr Victor Ogesa Juma for the helpful conversations about bifurcation diagrams.
This project has received funding from the European Union's Horizon 2020 research and innovation programme under the Marie Skłodowska-Curie grant agreement no 642866. The authors would like to thank the Isaac Newton Institute for Mathematical Sciences for its hospitality during the programme \textit{Coupling Geometric PDEs with Physics for Cell Morphology, Motility and Pattern Formation}, supported by EPSRC Grant Number EP/K032208/1. LEK is supported by an NSERC Discovery Grant. SP is supported in part by an NSERC Discovery Grant. AM was partially supported by a fellowship from the Simons Foundation. AM is a Royal Society Wolfson Research Merit Award Holder, generously supported by the Wolfson Foundation.

\bibliographystyle{acm}
\bibliography{references.bib}

\appendix
\section{Non dimensionalisation}
Let $A$, and $B$ be some dimensional concentration quantities with {\corr $[A]=$mol $\mu$m$^{-(d-1)}$, and  $[B]=$mol $\mu$m$^{-d}$ where $d$ is the dimension of the domain. Let $L$ be a typical length in the cell ($[L]=\mu$m), representing for example its radius,} and $T$ a temporal quantity ($[T]=$s). Then we can define the non-dimensional variables 
\[
\hat{a}=a/A,\quad \hat{b}=b/B,\quad \hat{t}=t/T,\quad \hat{\mathbf{x}}=\mathbf{x}/L. 
\]
Using the chain rule in the derivation we have
\[
\frac{\partial }{\partial t}=\frac{1}{T}\frac{\partial }{\partial \hat{t}}, \quad \nabla=\frac{1}{L}\hat{\nabla}, \quad \Delta=\frac{1}{L^2}\hat{\Delta}.
\]
Therefore, from \eqref{eq:model_b}-\eqref{eq:model_a}, 
we obtain 
\begin{align*}
\frac{B}{T}\frac{\partial \hat{b}}{\partial \hat{t}} &=D_b\frac{B}{L^2}\hat{\Delta} \hat{b},\quad &\mathbf{x}\in\hat{\Omega}, \\
\frac{A}{T}\frac{\partial \hat{a}}{\partial \hat{t}} &=D_a\frac{A}{L^2}\hat{\Delta_\Gamma} \hat{a}+f(A\hat{a}, B\hat{b}),\quad &\mathbf{x}\in\partial\hat{\Omega},\\
-D_b\frac{B}{L}(\mathbf{n}\cdot\nabla \hat{b}) &=f(A\hat{a}, B\hat{b}),\quad &\mathbf{x}\in\partial\hat{\Omega},\\
f(A\hat{a}, B\hat{b})&=\Big(k_0+\frac{\gamma \hat{a}^2}{\left(\frac{K}{A}\right)^2+\hat{a}^2}\Big)\omega  B\hat{b}-\beta A\hat{a}.
\end{align*}
We now set $\hat{K}=K/A$ so we can write
\[
f(A\hat{a}, B\hat{b})=\beta A \left[\frac{\omega  B}{\beta A}  \left(k_0+\frac{\gamma\hat{a}^2}{\hat{K}^2+\hat{a}^2}\right)\hat{b}- \hat{a}\right].
\]
In the system we get
\begin{align*}
\frac{\partial \hat{b}}{\partial \hat{t}} &=\frac{D_b T}{L^2}\hat{\Delta} \hat{b},\quad &\mathbf{x}\in\hat{\Omega}, \\
\frac{1}{T}\frac{\partial \hat{a}}{\partial \hat{t}} &=\frac{D_a}{L^2}\hat{\Delta_\Gamma} \hat{a}
+\beta  \left[\frac{\omega  B}{\beta A}  \left(k_0+\frac{\gamma\hat{a}^2}{\hat{K}^2+\hat{a}^2}\right)\hat{b}- \hat{a}\right],
\quad &\mathbf{x}\in\partial\hat{\Omega},\\
-(\mathbf{n}\cdot\hat{\nabla} \hat{b}) &=
\frac{A L}{D_b B} \beta\left[\frac{\omega  B}{\beta A}  \left(k_0+\frac{\gamma\hat{a}^2}{\hat{K}^2+\hat{a}^2}\right)\hat{b}- \hat{a}\right],
\quad &\mathbf{x}\in\partial\hat{\Omega},\\
\end{align*}
{\corr
As in \cite{Mori2011}, we make the assumption
\[
L=\sqrt{\frac{D_b}{\beta}}
\]
i.e. $L$ is approximately the length that the diffusing protein $b$ covers in its biochemical activation time scale. With this choice, using the parameters in Table \ref{tab:parameters} we have $L=10\mu$m.
We also define $A=K$ and $B=K/L$, so $A$ and $B$ are related to the quantity $K$ of active component needed to reach half of the maximal activation rate induced by the positive feedback. 
For the time we use 
\[
T=\frac{1}{\beta}\sqrt{\frac{D_b}{D_a}}=\frac{L}{\sqrt{\beta D_a}}
\]
This choice is particularly convenient for the analysis of the model at different time scales in Section \ref{sec:asymptotic_analysis}. For comparison with the previous works, we remark that the same expressions for $L$ and $T$ were used in \cite{Mori2011}.
Finally 
}
we get
\begin{align*}
\frac{\partial \hat{b}}{\partial \hat{t}} &=\sqrt{\frac{D_b}{D_a}}\hat{\Delta} \hat{b},\quad &\mathbf{x}\in\hat{\Omega}, \\
\cancel{\beta}\sqrt{\frac{D_a}{D_b}}\frac{\partial \hat{a}}{\partial \hat{t}} &=\cancel{\beta}\frac{D_a}{D_b}\hat{\Delta_\Gamma} \hat{a}+\cancel{\beta}\left[  \left(\hat{k_0}+\frac{\hat{\gamma}\hat{a}^2}{1+\hat{a}^2}\right)\hat{b}- \hat{a}\right],\quad &\mathbf{x}\in\partial\hat{\Omega},\\
-(\mathbf{n}\cdot\hat{\nabla} \hat{b}) &= \left(\hat{k_0}+\frac{\hat{\gamma}\hat{a}^2}{1+\hat{a}^2}\right)\hat{b}- \hat{a},\quad &\mathbf{x}\in\partial\hat{\Omega},\\
\end{align*}
where 
\[
\hat{k_0} = \frac{\omega  B}{\beta A}k_0 =  \frac{\omega  }{\sqrt{\beta D_b}} k_0 \quad \text{ and } \quad
\hat{\gamma} = \frac{\omega  B}{\beta A}\gamma =  \frac{\omega  }{\sqrt{\beta D_b}} \gamma.
\] 
One of the main assumption of the model is that $a$ diffuses much slower than $b$, so we set 
{\corr
\[
\varepsilon^2={\frac{D_a}{D_b}}\ll1.
\]
}
Dropping all the hats, we finally have the system
\begin{align*}
\varepsilon\frac{\partial {b}}{\partial {t}} &={\Delta}{b},\quad &\mathbf{x}\in\Omega,  \\
\varepsilon\frac{\partial {a}}{\partial {t}} &=\varepsilon^2{\Delta_\Gamma} {a}+ \; f(a, b),\quad &\mathbf{x}\in\Gamma, \\ 
-(\mathbf{n}\cdot{\nabla}{b}) &=f\;({a}, {b}),\quad &\mathbf{x}\in\Gamma,
\end{align*}
with 
\begin{equation}\notag
f({a}, {b}):= \left({k_0}+\frac{{\gamma}{a}^2}{1+{a}^2}\right){b}-{a}.
\end{equation}
\section{Numerical code details}\label{Appendix:computations}
The numerical code we used to solve the model was written in Python 2.7 and the three systems of linear equations (\ref{eq:PC_prediction_a})-(\ref{eq:PC_correction_a}) were assembled using  FEniCS, which is an open source finite element software package for solving partial differential equations \cite{Alnaes2015}. We presented simulations on three different domains: a sphere, a capsule and a complex domain, caricature of a polarised fibroblast. The geometries, with the respective meshes, were created using the FEniCS mesh generator \textit{mshr} \cite{Alnaes2015} for the first case, Gmsh for the latter cases \cite{Geuzaine2009}.

\end{document}